%% file: PTFOrionOverview.tex
\documentclass[preprint]{aastex}

\usepackage{amsmath,url,afterpage,natbib}
\usepackage[hang,small,bf,nooneline]{caption}
\input{usersetcommands}

\shorttitle{The PTF Orion Project}
\shortauthors{van~Eyken et al.}

\begin{document}


\title{The Palomar Transient Factory Orion Project: Eclipsing Binaries and Young Stellar Objects}


\author{Julian~C.~van Eyken, David~R.~Ciardi}
\affil{NASA Exoplanet Science Institute, California Institute of
  Technology, 770 South Wilson Avenue, M/S~100-22, Pasadena, CA
  91125, USA \email{vaneyken@ipac.caltech.edu}}

\author{Luisa~M.~Rebull, John~R.~Stauffer}
\affil{Spitzer Science Center/California Institute of Technology, M/S
  220-6, 1200 East California Boulevard, Pasadena, CA 91125, USA}

\author{Rachel~L.~Akeson, Charles~A.~Beichman}
\affil{NASA Exoplanet Science Institute, California Institute of
  Technology, 770 South Wilson Avenue, M/S~100-22, Pasadena, CA
  91125, USA}

\author{Andrew~F.~Boden} \affil{Department of Astronomy, California
  Institute of Technology, Pasadena, CA 91125, USA}

\author{Kaspar~von~Braun, Dawn~M.~Gelino}
\affil{NASA Exoplanet Science Institute, California Institute of
  Technology, 770 South Wilson Avenue, M/S~100-22, Pasadena, CA
  91125, USA}

\author{D.~W.~Hoard} \affil{Spitzer Science Center/California
  Institute of Technology, M/S 220-6, 1200 East California Boulevard,
  Pasadena, CA 91125, USA}

\author{Steve~B.~Howell} \affil{National Optical Astronomy
  Observatory, 950 North Cherry Avenue, Tucson, AZ 85719, USA / NASA Ames Research Center, PO Box 1, M/S 244-30, Moffett Field, CA
    94035, USA}

\author{Stephen~R.~Kane, Peter~Plavchan, Solange~V.~Ram\'{i}rez} 
\affil{NASA Exoplanet Science Institute, California Institute of
  Technology, 770 South Wilson Avenue, M/S~100-22, Pasadena, CA
  91125, USA}


\author{Joshua S. Bloom, S. Bradley Cenko}
\affil{Department of Astronomy, University of California, Berkeley, CA
94720-3411, USA}

\author{Mansi M. Kasliwal, Shrinivas R. Kulkarni}
\affil{Cahill Center for Astrophysics, California Institute of
Technology, Pasadena, CA 91125, USA}

\author{Nicholas M. Law}
\affil{Dunlap Institute for Astronomy and Astrophysics, University of
Toronto, 50 St. George Street, Toronto M5S 3H4, Ontario, Canada}

\author{Peter E. Nugent}
\affil{Computational Cosmology Center, Lawrence Berkeley National
Laboratory, 1 Cyclotron Road, Berkeley, CA 94720, USA}

\author{Eran O. Ofek\altaffilmark{1}}
\affil{Cahill Center for Astrophysics, California Institute of
Technology, Pasadena, CA 91125, USA}

\author{Dovi Poznanski\altaffilmark{1}}
\affil{Department of Astronomy, University of California, Berkeley, CA
94720-3411, USA; Computational Cosmology Center, Lawrence Berkeley National
Laboratory, 1 Cyclotron Road, Berkeley, CA 94720, USA}

\author{Robert M. Quimby}
\affil{Cahill Center for Astrophysics, California Institute of
Technology, Pasadena, CA 91125, USA}


\author{Carl~J.~Grillmair, Russ~Laher}
\affil{Spitzer Science Center, M/S 220-6, California Institute of
  Technology, Jet Propulsion Laboratory, Pasadena, CA 91125,
  USA}

\author{David~Levitan} \affil{Department of Physics, California
  Institute of Technology, Pasadena, CA 91125, USA}

\author{Sean~Mattingly} \affil{Department of Physics and Astronomy,
  The University of Iowa, Iowa City, IA 52242, USA}

\and

\author{Jason~A.~Surace}
\affil{Spitzer Science Center, M/S 220-6, California Institute of
  Technology, Jet Propulsion Laboratory, Pasadena, CA 91125,
  USA}

\altaffiltext{1}{Einstein Fellow}



\begin{abstract}
\input{abstract}
\end{abstract}





\input{introduction}
\input{survey}

\input{datareduction}

\input{dataquality}

\input{binaries}

\input{ctts}
\input{conclusions}

\acknowledgments
\input{acknowledgments}


\bibliographystyle{apj} 
\bibliography{ptfbibliography}





\end{document}

%% file: usersetcommands.tex
\newcommand{\kms}{\,{\rm km}\,{\rm s}^{-1}}

\newcommand{\pc}{\,{\rm pc}}

\newcommand{\Msun}{\,M_{\odot}}

\newcommand{\Rsun}{\,R_{\odot}}

\newcommand{\Myr}{\,{\rm Myr}}

\newcommand{\hr}{\,{\rm hr}}
\newcommand{\vmag}{\,{\rm mag}}
\newcommand{\mmag}{\,{\rm mmag}}

\newcommand{\masyr}{\,{\rm mas\, yr^{-1}}}
\newcommand{\pix}{\,{\rm pixel}}
\newcommand{\um}{\,{\rm \mu m}}

%% file: abstract.tex
The Palomar Transient Factory (PTF) Orion project is one of the
experiments within the broader PTF survey, a systematic
automated exploration of the sky for optical transients. Taking
advantage of the wide ($3.5\degr\times 2.3\degr$) field of view
available using the PTF camera installed at the Palomar 48-inch
telescope, 40 nights were dedicated in 2009 December to 2010 January to
perform continuous high-cadence differential photometry on a single
field containing the young (7--$10\Myr$) 25 Ori association. Little is
known empirically about the formation of planets at these young ages,
and the primary motivation for the project is to search for planets
around young stars in this region. The unique data set also provides
for much ancillary science. In this first paper we describe the survey
and the data reduction pipeline, and present some initial results from
an inspection of the most clearly varying stars relating to two of the
ancillary science objectives: detection of eclipsing binaries and
young stellar objects. We find 82 new eclipsing binary systems, 9 of
which are good candidate 25~Ori or Orion~OB1a association
members. Of these, two are potential young W~UMa type systems. We report on the
possible low-mass (M-dwarf primary) eclipsing systems in the sample,
which include six of the candidate young systems. Forty-five of the
binary systems are close (mainly contact) systems, and one of these
shows an orbital period among the shortest known for W~UMa binaries,
at $0.2156509 \pm 0.0000071$ days, with flat-bottomed primary
eclipses, and a derived distance that appears consistent with
membership in the general Orion association. One of the candidate
young systems presents an unusual light curve, perhaps representing a
semi-detached binary system with an inflated low-mass primary or a
star with a warped disk, and may represent an additional young Orion
member. Finally, we identify 14 probable new classical T-Tauri stars
in our data, along with one previously known (CVSO 35) and one
previously reported as a candidate weak-line T-Tauri star (SDSS
J052700.12+010136.8).

%% file: introduction.tex
\section{Introduction}

The Palomar Transient Factory (PTF) is a survey built around a wide-field
mosaic camera installed on the Palomar 48-inch Samuel Oschin telescope
dedicated to surveying the sky to find photometric transient and
variable objects with variability timescales of minutes to years. The
camera consists of a mosaic of two rows of six 2k$\times$4k-CCDs (of
which one is non-functional), with $1\arcsec$ pixels, giving a total
field size of $\approx 3.5\degr\times 2.3\degr$. With a $5\sigma$
detection limit of $R\approx 21.0\vmag$ for 60\,s exposures
\citep{Law2010}, the main survey is fully automated, and is designed
to allow for very rapid followup of transient events using the Palomar
60-inch telescope, as well as a further global network of telescopes for
additional characterization. A technical summary of the system is
given by \citet{PTFtechnical}, and an overview of the science goals by
\citet{PTFsurvey}. A summary of the first year's performance is
given by \citet{Law2010}.

Among the programs within the survey is the PTF Orion project. The
primary goal of the Orion project is to search for transiting
exoplanets around stars that are 5--$10\Myr$ old. Little is known
empirically about exoplanets during the first few millions of years of
their lives -- the timescales on which they are expected to form and
migrate -- and the aim is to try and fill this gap in order to provide
constraints on theories of planet formation and evolution
\citep[e.g.,][]{ArmitagePlanetFormation}. Typical young circumstellar
disk lifetimes are on the order of $5-10\Myr$ \citep{Hillenbrand2008},
and by trying to catch stars at or just after the point of disk
dissipation, we can hope to find planets at the point where they may
first become observable without their photometric signatures being
swamped by that of the disk, and where disk--planet interaction has
ceased.

In addition to the planet search, the observations provide a unique
data set to study a variety of stellar astrophysics: eclipsing binary
systems enable tests of star formation and evolution models; stellar
activity and rotation at these young ages can be characterized; and
previously unknown pre-main-sequence (PMS) stars can be identified and
characterized within the region. (A similar young transit and eclipse
search, the Monitor project, is described by \citet{Monitor} and
\citet{Miller2008}.)

The \object{25 Ori} association, with an estimated age of around
7--$10\Myr$ \citep{Briceno2007}, provides an ideal target to achieve
the goals of the PTF Orion survey, being the most populous known
sample in its age range within $500\pc$, the right age, and with a
spatial extent well matched to the PTF field of view. The field has
already been well studied and characterized by \citet{Briceno2005,
  Briceno2007}. In the winter of 2009/2010, we obtained time-series
data centered on the 25 Ori association, taking continuous $R$-band
observations and obtaining $\approx$110,000 individual high-cadence
light curves.

The unique capacity of eclipsing binary systems for yielding accurate
measurements of fundamental stellar parameters, such as mass, radius, and
effective-temperature ratio, makes them useful for constraining
stellar models. This is especially true of young PMS
systems, where evolutionary models are not as well constrained
observationally \citep{MathieuPPV}. At primary masses below
$1.5\Msun$, \citet{Hebb2010} list in addition to their newly reported
system, MML 53, only six known PMS eclipsing
binaries (RXJ 0529.4+0041A, \citealt{Covino2000, Covino2004}; V1174 Ori,
\citealt{Stassun2004}; 2MASS J05352184-0546085, \citealt{Stassun2006,
  Stassun2007}; JW 380, \citealt{Irwin2007}; Par 1802, \citealt{Cargile2008,
  Stassun2008}; and ASAS J052821+0338.5, \citealt{Stempels2008}), with a
further four candidates in the Orion Nebula cluster more recently
reported by \citet{Morales2011}. Evolutionary models for stars at the
late end of the stellar mass range ($\gtrsim$M0) are also currently
not well constrained by empirical measurements, owing to their
intrinsic faintness, and the consequent difficulty in observing them. To
date, fewer than 20 such systems are known \citep[see][and references
therein]{Kraus2011}. It is therefore of particular interest to
identify binary systems which are young Orion members and/or which lie
in the low-mass regime in our data set.

Observations of contact binary systems, commonly known as W~UMa
systems, are also of interest both because they are potentially useful
distance indicators \citep{Rucinski2004WUmaDistances,
  Eker2009WUMaCalibration}, and because their formation and evolution
is currently not well understood \citep[see, e.g.,][]{Bilir2005,
  Eker2006, Pribulla2006}. It has been debated whether they may form
as the result of the merger of previously detached binary systems through
angular momentum loss due to tidal interaction and magnetic braking
\citep[e.g.,][]{Stepien1995, Stepien2006EvStat}, or as a result of
initially detached binary orbits decaying by a combination of the
Kozai mechanism and tidal friction due to the presence of a third more
distant body \citep{Tokovinin2006,Rucinski2007,Eggleton2006}, or
instead directly as contact systems
\citep[e.g.,][]{Bilir2005,Roxburgh1966}. The 7-10$\Myr$ age of the 25
Ori region, for example, is too short for the expected Gyr timescale
needed for in-spiral via angular momentum loss
\citep[][]{Stepien1995,Stepien2006EvStat,Bilir2005,Gazeas2008}, and
indeed, appears somewhat short even for merger by the
Kozai/tidal-friction mechanism, which operates on timescales of tens
of Myr or more, \citep{Eggleton2006}. Finding young contact binaries
within our sample is potentially useful for constraining these
formation mechanisms.

Single PMS T-Tauri stars (TTSs) are also useful for
understanding stellar formation and early evolution. At the age of the
25 Ori association, they are relatively unobscured and can be observed
in the optical regime. The advent of wide-field surveys like the
CIDA-QUEST survey \citep{Briceno2005} allows for broad, deep searches
that are sensitive to the more dispersed and more mature low-mass
populations, where previous studies have been more tightly focused on
the brighter OB stars in relatively compact regions. The PTF survey is
similarly sensitive, and identification of new young TTSs in
our data complements the sample reported by
\citet{Briceno2005,Briceno2007}.

In this first paper we present a description of the survey and the
data reduction procedure, and some of the initial results that relate
to these two stellar astrophysical science objectives of the program:
binary systems and new candidate PMS stars. Section
\ref{sec:survey} describes the survey; section \ref{sec:datareduction}
describes the data-reduction pipeline developed for the PTF Orion
project; section \ref{sec:dataquality} gives an overview of the data
obtained and an assessment of the current precision levels; section
\ref{sec:binaries} discusses the eclipsing binaries found, including
the candidate young binary systems, the low-mass binaries, and some of
the contact and near-contact binaries of interest (with some overlap
between these groups); and section \ref{sec:ctts} discusses the new
classical T-Tauri stars (CTTSs). A brief summary is given in section
\ref{sec:summary}.

%% file: survey.tex
\section{The PTF Orion Survey}\label{sec:survey}

\subsection{Field Selection}\label{sec:fieldselection}

Our target field around 25~Ori was chosen to balance a number of
requirements. Interstellar reddening is minimal along the line of
sight to the cluster \citep[mean $A_V = 0.29$;][]{Briceno2007}; the
source count density is optimal, being high but without overcrowding
($\approx 4.4$ sources per square arcminute at our detection threshold);
and the region is already relatively well characterized, with an
estimated age of $\approx 7$--$10\Myr$ \citep{Briceno2007}, corresponding
to the time at which the disks around young stars have mostly
dissipated \citep{Hillenbrand2008}. This allows us to probe the
photometric variability of young stars without it being completely
dominated by variations due to accretion or circumstellar
extinction. The majority of TTSs in the 25 Ori region are
weak-lined T Tauri stars (WTTSs), indicating that the disks have indeed
dissipated \citep{Briceno2007}.

\begin{figure*}[tbp] \epsscale{0.8}
\plotone{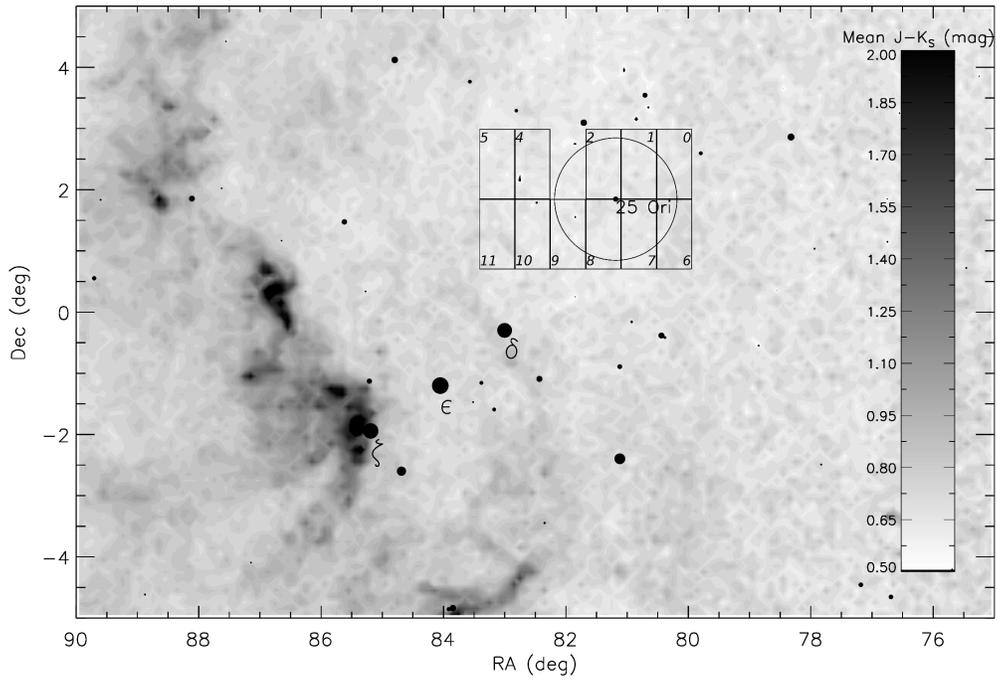}
\caption[Field location]{\label{fig:fieldlocation}Location of the
  field with respect to 25 Ori and the Orion belt stars, overlaid on
  grayscale contours representing mean 2MASS $J-K_s$ colors, which
  trace reddening. Bin size is $0.1\degr\times0.1\degr$. Filled
  circles mark bright stars ($V \le 6.0\vmag$), with size varying
  linearly with visual magnitude; the brightest of these are labeled,
  along with 25 Ori which is placed in a chip gap in the PTF field of
  view. The footprint for each chip in the field is marked and the
  chip number identified; the dead chip (NE of the central four) is
  omitted. The circle shows the $1\degr$ working radius used by
  \citet{Briceno2007} to define the 25 Ori group.}
\end{figure*}

Within these constraints, the field was aligned to maximize the number
of known TTS covered (\citealt{Briceno2005, Briceno2007}; C.~Brice\~no 2009,
private communication) whilst allowing for the
non-functioning chip, and to place 25~Ori itself on a gap between
chips, minimizing charge bleeding from what would otherwise be a
heavily saturated source. Figure \ref{fig:fieldlocation} shows the
positioning of the field with respect to 25 Ori and the brighter stars
in the region. The field is centered at approximately
$\alpha = 05\fh 26\fm 46\fs$, $\delta = +01\fdg 50\farcm 50\farcs$
(J2000), covering an area $3.5\degr$ in right ascension and $2.3\degr$
in declination. One of the chips in the array is non-functional (the
north-east of the central four, chip number 4) and is not shown.

The figure confirms that the amount of interstellar reddening is
apparently uniformly quite low in the 25~Ori field, and we adopt the
derived mean extinction value for the cluster from \citet[][$A_V = 0.29\vmag$]{Briceno2007}
as a single representative value. (In comparison,
typical values from the dust maps of \citet{SchlegelDustMaps} along
lines of sight to the sources in the field range from $\approx
0.3$ to $0.6\vmag$, and mostly $<1.0\vmag$. As estimates of the full
line-of-sight extinction through to the edge of the Galaxy, these
suggest an upper limit.) Following the extinction laws of
\citet{Cardelli1989}, and assuming $R_V=3.1$, the adopted value of
$A_V$ corresponds to extinctions in $R$, $J$, $H$, and $K_S$ of
$0.22\vmag$, $0.082\vmag$, $0.055\vmag$, and $0.033\vmag$,
respectively.

\subsection{PTF Observations}
Data were taken for the Orion project on the majority of the clear nights 
between 2009 December 1 and 2010 January 15. Weather allowing, the
chosen primary field was observed in the $R$ filter as near
continuously as possible, whenever it was higher in the sky than an
airmass of 2.0. Exposures were 30s long, with a cadence varying
generally between 70-90s, including readout time and depending on the
performance of the telescope control system and guiding control
loop (see below).

Since the Orion observing program could not easily be incorporated
into the normal PTF robotic scheduling, separate scripts for telescope
and camera control code were adapted from the standard PTF robot code for
the purposes of the Orion observations. Each night at the beginning of
Orion field observations, normal PTF operations were manually
interrupted to run the Orion observing program. When the field center
set once again below airmass 2.0, control was automatically returned
to the standard PTF robot scheduler and normal PTF operations resumed.

\begin{figure*}[tbp] \epsscale{0.9}
\plotone{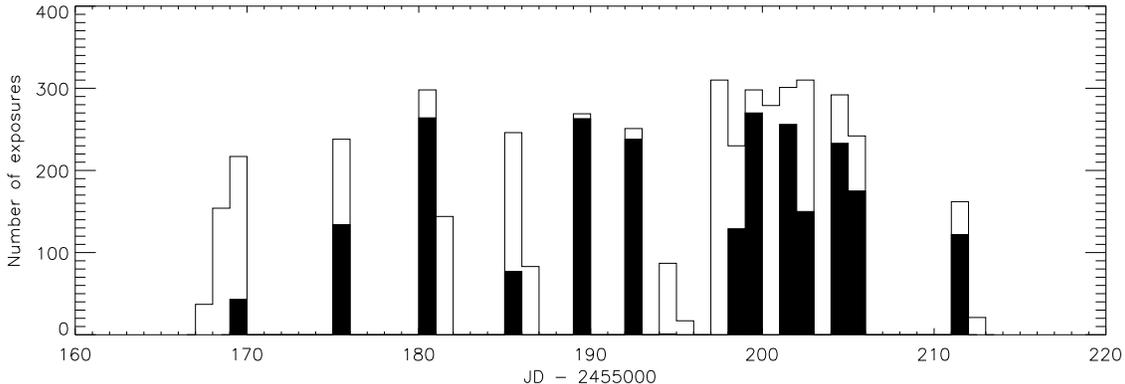}
\caption[Observations per night]{\label{fig:observations}Number of
  exposures taken per night for the PTF Orion project from December
  2009 -- January 2010. Solid black regions indicate the fraction
  of exposures which remained unflagged after processing by the data
  reduction pipeline, for chip 0 (similar fractions are seen for the
  other chips).}
\end{figure*}

Of the 40 nights dedicated to the Orion project, 14 yielded usable
data, the majority of the remainder being either of marginal quality
due to cloud or subject to telescope closure because of poor
weather. (One night also could not be processed due to an
insufficiency of non-Orion exposures to create a flat field for that
night, and can in principle be recovered.) A total of 4,486 exposures
were taken during the run. An average across the chips of $\approx
3,460$ exposures ran successfully through the pipeline after
rejection of those which could immediately be discarded at the outset
by visual inspection, and an average of $\approx 2,400$ of those
exposures passed the data reduction pipeline's automated data quality
assessment without flagging. There is some variation in these numbers
from chip to chip since each is processed completely
independently. Figure \ref{fig:observations} shows a histogram of the
number of exposures taken on each night, and the fraction of those
which passed unflagged for chip 0, which is used as a representative
example of all the chips.

In order to minimize the effects of flat-fielding errors for
high-precision photometry, it is desirable to minimize image shift on
the detector between frames. Since the Palomar $48\arcsec$ has no native
guiding facility, we developed software to guide on the science images
as they came from the telescope, providing feedback to the telescope
via offset pointing commands. Offsets were referenced to a single good
quality reference frame taken at the beginning of the run, and
telescope pointing updates were applied during readout time generally
every second frame. On initial pointing of the telescope, offsets of
typically $\approx 70$--90 pixels relative to the reference frame were
normally corrected within two exposures. With the guiding switched on,
the pointing toward the center of the field was stabilized with a
root-mean-square (rms) image shift of $\approx1.0$ and $\approx 0.5$ pixels
in right ascension and declination respectively, compared to unguided
linear tracking drifts of $\approx0.3\pix\min^{-1}$.

\subsection{Ancillary Data}\label{sec:ancillary}

In addition to the PTF data obtained, we also drew on data from
several other sources. The photographic US Naval Observatory USNO-B1.0
source catalog \citep{USNOB}
provided a reference data set for initial photometric zero-point
correction of the $R$-band PTF Orion data (section
\ref{sec:rawphot}). A further catalog of $\approx40,000$ reference
stars from the Centro de Investigaciones de Astronom\'ia--Quasar
Equatorial Survey Team (CIDA-QUEST), provided by C.~Brice\~{n}o
(2009, private communication), allowed for a final more accurate correction to
the $R$-band zero point after differential photometry had been
performed (section \ref{sec:diffphot}). This catalog had in turn been
calibrated against the Landolt system \citep{Landolt} with $R$
magnitudes ranging from $14\lesssim R \lesssim 19$ and measurement
errors varying with increasing magnitude from $\approx
0.015$ to $0.1\vmag$. The overall error for the brighter stars (down to
$R\approx 17$) including systematics is $\approx 0.02$--$0.03\vmag$. We
also made use of the PPMXL catalog \citep{PPMXL} to obtain proper
motions for our candidate binary systems. The PPMXL catalog provides a
determination of positions and proper motions by combining USNO-B1.0
and Two Micron All-Sky Survey (2MASS) astrometry, and aims to be complete to $V\approx 20\vmag$.

The 2MASS point-source
catalog \citep{2MASS} provides all-sky coverage of point sources in
the near-infrared $J$ ($1.25\micron$), $H$ ($1.65\micron$), and $K_s$
($2.16\micron$) bandpasses, with magnitude limits of 15.8, 15.1, and
$14.3\vmag$ respectively. PTF Orion source detections were matched
against this catalog to provide color information on each source in
combination with the PTF Orion $R$ magnitude. The 2MASS catalog
includes a `quality' flag for each measurement in each band, which can
take various values. In particular, designations from `A' to `F'
represent photometric quality in decreasing order (in the case of an
`F', error estimates could not be determined). In certain cases
throughout this paper we use these flags to restrict analysis to
sources with 2MASS data above a given quality level, as stated in the
text.

For certain target objects, anticipating that these young objects
might have discernible mid-infrared (MIR) excesses due to
circumstellar disks, we also examined the {\em Spitzer Space Telescope}
\citep{Werner2004} Heritage Archive for both Infrared Array Camera
\citep[IRAC,][]{Fazio2004} and Multiband Imaging Photometer for
{\em Spitzer} \citep[MIPS,][]{Rieke2004} data, discussed below. We found some
IRAC data (at 3.6, 4.5, 5.8, and $8 \micron$), but MIPS data (at
$24\micron$) were more common in our target region.  We found MIPS
and/or IRAC data for 15 objects, discussed in sections
\ref{sec:binariesspitzer} and \ref{sec:ctts}. All 15 of the objects
were found in at least one of 9 AORKEYs\footnote{AORKEYs are the
  unique eight-digit identifier for the Astronomical Observation Request
  (AOR), which can be used to retrieve these data from the {\em Spitzer}
  Heritage Archive.}: 10987776, 26066432, and/or 26067456 for IRAC,
and 10987008, 17052160, 26067712, 26067968, 26069760, and/or 26070016
for MIPS. The images for each object were examined for potential
source confusion, and none was found.

For IRAC, we started from the automatically-produced mosaics from the
Archive (S18.7 or greater), and performed aperture photometry using
the IDL\footnote{\url{http://www.ittvis.com/idl}} routine
aper.pro\footnote{Found in the IDL Astronomy User's Library,
  \url{http://idlastro.gsfc.nasa.gov/}; see also \citet{IDLastro} and
  \citet{DAOPHOT}.} on these mosaics at the designated positions. We
used an aperture radius of 3 native pixels ($1.2\,\mathrm{arcsec} \pix^{-1}$),
a sky annulus of radius 3--7 native pixels, and multiplicative aperture
corrections of 1.124, 1.127, 1.143, and 1.234 for the four channels,
respectively, as prescribed by the IRAC Instrument Handbook, available
on the Spitzer Science Center (SSC) Web site.\footnote{\url{http://ssc.spitzer.caltech.edu/}} The zero-points we used to convert between flux
density units (Jy) and magnitudes were, respectively, 280.9, 179.7,
115.0, and 64.13\,Jy.

For MIPS, we initially obtained the automatically-produced mosaics
from the Archive (S18.12 for MIPS), and on a case-by-case basis,
reprocessed the mosaics if necessary from the basic calibrated data
(BCD) level, to reduce the influence of instrumental artifacts.  We
also conducted aperture photometry using IDL's aper.pro on the mosaics
(either ours or the Archive's), using an aperture of 7$\arcsec$, an
annulus of $7\arcsec$--$13\arcsec$, and an aperture correction of 2.05, from
the MIPS Instrument Handbook, again available from the SSC
website. The MIPS-24 zero point we used was 7.14\,Jy.

In the ideal case, to determine whether or not there is infrared
excess due to a circumstellar disk, we need an estimate of a spectral
type so that we can fit a model to the spectral energy distribution
(SED) and estimate reddening ($A_V$), and then estimate the degree of
infrared excess due to a disk. In the absence of an accurate estimate
of spectral type, the degree of IR excess is not well defined. We can
instead attempt to estimate the degree of infrared excess by assuming
that these sources are all truly stars (as opposed to background
objects), and then comparing the longer {\em Spitzer} bands to a shorter
band. Where possible, even lacking a spectral type estimate, we can
use [3.6]$-$[24] to reasonably reliably indicate disk excess (where
the bracket notation denotes the magnitudes at the {\em Spitzer} bands),
since this range is within the Rayleigh-Jeans tail even for the latest
spectral types, giving them colors of zero. Reddening also does not
strongly affect this color unless $A_V$ is extreme, in which case the
extinction will be high enough to make a source detection unlikely in
our PTF data. MIPS-24 data are more widely available in our field than
the IRAC bands, however, and $3.6\micron$ data do not always exist. In
such cases, the 2MASS $K_s$ band is the most commonly available
alternative. $K_s - [24]$ therefore provides an alternative excess
estimate, although it is less reliable since late-type stars (such as
many of those in our sample) are not colorless at $K_s-[24]$
\citep{Gautier2007}, and the effects of reddening are stronger at
$K_s$ than at $3.6\micron$. For such cases, additional followup
observations will be needed to refine the estimate of the excess.

%% file: datareduction.tex
\section{The PTF Orion Data Reduction Pipeline}\label{sec:datareduction}

The initial PTF data reduction pipeline in place at the time the data
were taken was geared primarily towards absolute photometry and
extensive database operations for handling the vast quantities of data
coming from the broader survey. We developed a separate dedicated
relative-photometry pipeline, which took over from the standard
pipeline at the point where image processing was complete. Here, we
describe the two broad steps in the data reduction procedure:
obtaining the initial raw photometry, beginning with the standard PTF
image processing and the first steps of the Orion pipeline; and the
differential photometry process used to obtain the precisions needed
for the PTF Orion survey.

\subsection{Raw Photometry}\label{sec:rawphot}

The main PTF pipeline image processing consisted of standard bias
subtractions, flatfielding, and astrometric solutions. Each chip was
reduced separately and in parallel. Floating biases were measured as
the median of the overscan regions of the chips and subtracted along
with a nightly combined superbias frame; flatfields were
created by combining all (non-Orion-project) images taken with the
same filter during a night, all the bright pixels being first masked
out using SExtractor \citep{sextractor}, to leave a pure sky
flatfield. Astrometric information was added to the FITS image headers
using SCAMP \citep{SCAMP} to solve for the astrometry. The general PTF
pipeline is discussed in more detail in \citet{ptfpipeline} and
R.~Laher et al. (2011, in preparation).

The Orion pipeline ran separately and took as input the image data
from the standard pipeline after the astrometric solutions had been
added. Written in IDL, 
the main steps in the PTF Orion pipeline are as follows:

\begin{enumerate}

\item {\em Pixel masking}: Masks are initially applied to reject any
  badly behaved pixels on the detector using an algorithm loosely
  based on the IRAF\footnote{\url{http://iraf.noao.edu/}} `ccdmask'
  procedure \citep{IRAF1,IRAF2}. Masks are created from images made by
  dividing a 70\,s light-emitting diode (LED\footnote{See
    \citet{PTFtechnical}}) flatfield by a 35\,s LED flatfield; three
  independent such divided frames were obtained for each chip. Any
  pixels with outlier fluxes beyond $4\sigma$ in at least {\em two} of
  the three frames, or beyond $3\sigma$ in all {\em three} of the
  frames were flagged as bad. This approach helps catch excessively
  variable pixels in addition to non-linear pixels, while still
  rejecting cosmic-ray hits. The flagging procedure was then repeated
  several times after boxcar smoothing of the original image along the
  readout direction with a selection of bin sizes from 2 to 20 pixels
  (this removes columns where individual pixels are not statistically
  bad, but are statistically bad when taken as an aggregate). Pixels
  lying in small gaps between bad pixels were then also iteratively
  flagged, with the aim of completely blocking out large regions of
  bad pixels while minimizing encroachment into good pixel
  regions. Any source detection with a flagged pixel within its
  photometry aperture is rejected and does not pass through the
  pipeline.

\item {\em Raw photometry}: Source detection and aperture photometry
  are performed using the IDL Astronomy User's
  Library 
  implementation of DAOPHOT \citep{IDLastro,DAOPHOT}, with minor
  modifications to return more significant figures in the output files
  and to handle the large number of sources in each CCD frame. Owing
  to the sometimes large offsets of the field on the detector from
  image to image, source detection is performed separately on every
  exposure obtained, with a $4\sigma$ detection threshold. A 4-pixel
  aperture radius provides a reasonable compromise between maximizing
  the flux enclosed for bright sources and minimizing sky-background
  noise for faint sources, while also being large enough to prevent
  significant photometric errors from imperfect centroiding of the
  apertures. This compares to median seeing of $2.5\arcsec$ at
  full-width-half-maximum (FWHM; where 1 pixel $\approx
  1\arcsec$). Inner and outer radii of 10 and 20 pixels, respectively,
  are chosen for the annulus used for background estimation around the
  aperture. The default limits for `sharpness' and
  `roundness' of the detected sources provide a first cut to reject
  cosmic rays and background galaxy detections.

\item {\em Initial zero point correction}: An initial zero-point
  correction is applied by matching detections against the USNO-B1.0
  catalog and correcting for the outlier-resistant\footnote{Throughout
    this text, where an outlier-resistant mean or standard deviation
    is referred to, these are found using the `resistant\_mean' and
    `robust\_sigma' functions, respectively, from the IDL Astronomy
    User's Library.} mean difference between raw and USNO-B magnitudes
  for each image from each chip. This places all photometry on an
  absolute scale, within the accuracy of the USNO-B catalog ($\sim
  0.3\vmag$). Sloan Digital Sky Survey (SDSS) photometry data would provide a
  more accurate correction, but were not available for the 25 Ori
  field.

\item{\em Creation of master source list}: After filtering out images
  with poor seeing or high background, a master source list for each
  chip is created by stepping through each of the 100 images with the
  highest numbers of source counts, in decreasing order of source
  counts. For each image, if a source is detected whose sky position
  cannot be matched with one already found (within a $2\arcsec$
  radius, the matching radius used throughout the pipeline), it is
  added to the master list. On every 20th image, sources with only one
  detection are pruned from the table, as are any sources with fewer
  than 10 detections at the end of the process. This ensures that
  cosmic rays, asteroids, and spurious detections are rejected from
  the table. Since we are primarily concerned with persistent stellar
  objects with relatively low-level variability, we accepted the
  consequent loss of short transient events that may appear from below
  the detection limit in return for the significant savings in data
  storage space in the final output data tables. This does not affect
  any of the sources that are suitable for planet transit searches.

\end{enumerate}

\subsection{Differential Photometry}\label{sec:diffphot}
After the initial raw photometry and preparation of the master
source lists, a single data table for each chip is prepared with
entries for each source in the lists at each epoch of observation. A
second pruning of the table is performed to remove sources if they
show very few total detections (fewer than 100) \emph{and} those detections
are sparsely distributed in time (the fraction of successful
detections within any 50 consecutive epochs is less than 0.5). This
helps remove sources which are due to badly behaved pixels or on the
extreme edge of the detection limit, while retaining genuine
transients that are only detected for a limited length of time. The
source list from each chip is then split into four equal-area
quadrants, each treated separately. This provides a first-order
accommodation for variations in photometry across the wide field of
view which may be caused by thin cloud, variations in point spread
function (PSF) due to the wide-field optics, effects of differential
airmass variation, and atmospheric refraction
\citep[see][]{EverettHowell}. On average $\approx 2600$ sources were
detected per chip quadrant for the 25~Ori field.

In a given quadrant, an outlier-resistant mean magnitude is then
calculated for each source across all epochs; this creates a reference
magnitude list against which to perform the differential
photometry. For each observation, the difference is calculated between
every source's measured magnitude and its corresponding reference
magnitude. Within each quadrant, the mode of the differences is
then calculated and subtracted from the measured magnitudes to provide
the differentially corrected magnitudes. The mode is calculated by
fitting a Gaussian curve to the peak of the histogram of magnitude
differences in the quadrant, and the peak of this Gaussian is taken to
represent the best-likelihood estimator of the true fine-scale
correction for the zero point of the frame. All stars in a given
quadrant are thus used in establishing the differential correction
applied for that quadrant.

Using the modal magnitude difference correction in this manner has the
advantage of being very robust to outliers and variable stars. It also
requires no selection of an ensemble set of stable reference stars,
provided that there is a sufficient number of stars in the field and that a
sufficient number of those stars are relatively stable. Even in the
event that only a small fraction of stars are stable, provided that the
total number of stars is large, the mode still represents a reasonable
estimate of the differential correction since there is no reason to
expect any correlation in intrinsic variability between the stars
included in the histogram. Additionally, the modal difference draws
information from the full ensemble of stars, providing a natural
`weighting' of the differences: stars that have higher photometric
uncertainties (i.e., lower fluxes) or that are more variable will
contribute a more spread-out Gaussian to the total distribution,
adding more to the wings of the histogram, though still peaking around the
true magnitude difference; those with lower uncertainties (high
fluxes) will contribute a tighter Gaussian to the total and contribute
more to the peak of the histogram.

To further improve the statistics, the differential photometry is then
repeated after identifying and omitting from the histogram sources
which still show an rms variation greater than $4\sigma$ above their
mean measurement error, or an rms variation greater than 2\% (or more
precisely, an rms greater than the quadrature sum of these two
values). This entire procedure is iterated until the total number of
sources rejected in this way changes by less than 3\%.

As a final step, once differential photometry is complete, the real
apparent magnitude is set by comparing the median magnitudes of each
differentially corrected time series against the list of stable
reference stars from the CIDA-QUEST survey, which in turn are calibrated
against the Landolt system (see section \ref{sec:ancillary}). The
median of the differences is used to provide a final single zero-point
correction for each entire chip.

After processing, the data are automatically checked and any
questionable sources, nights, or individual data points are flagged
according to a number of criteria. Flags are added for failed DAOPHOT
photometry or differential photometry; multiple source detections,
USNO-B reference sources, or master-list sources, within the FWHM;
high image background, very low zero-point, or high variation in
background across the image; exceptionally large FWHM; low number of
sources across the image; large variations with time in the measured
right ascension/declination for an individual source location;
saturation; high sigma or skew in individual background annuli for
sources; proximity to predicted optical ghosts; and indication of
crowding or contamination within the photometry aperture (assessed by
finding outlier values for the ratio of flux inside the aperture to that
in a thin annulus around the aperture compared to other sources in the
field).

%% file: dataquality.tex
\section{Data}\label{sec:dataquality}

The differential photometry pipeline yielded a total of 116,285
individual light curves (including some resulting from false multiple
detections on saturated stars). For an initial investigation of the
clearly variable stars, a subsample of all the light curves with an
outlier-resistant rms variation greater than three times the
outlier-resistant mean measurement-error for each light curve was
selected. This cut alone yielded $\approx 2060$ sources. These light
curves and their corresponding images were visually inspected to
select only those which clearly represented intrinsic photometric
variation, rather than crowded sources or image artifacts which had
escaped flagging, sources on unmasked bad pixels, or sources whose
variability was clearly due to known systematics (see section
\ref{sec:precisionandsensitivity}). This initial inspection yielded a
total of $\approx530$ sources, the difference in number from the first
cut being primarily due to the $\approx 4\mmag$ noise floor at the
bright end which was not accounted for, and to biasing toward light
curves that were not too sparsely sampled. (For comparison, selecting
only sources with $>1000$ of a possible $\approx 2400$ unflagged
detections, and rms greater than three times the quadrature sum of the
mean measurement-error and the noise floor, yields $\approx 700$ light
curves). Among the light curves selected from the initial inspection,
we identified visually 82 eclipsing binaries and 16 likely classical
T-Tauri star (CTTS) light curves. The results presented in this paper
are based on these last two groups of light curves.

\subsection{Precision and Sensitivity}\label{sec:precisionandsensitivity}
\begin{figure*}[tbp] 
\plotone{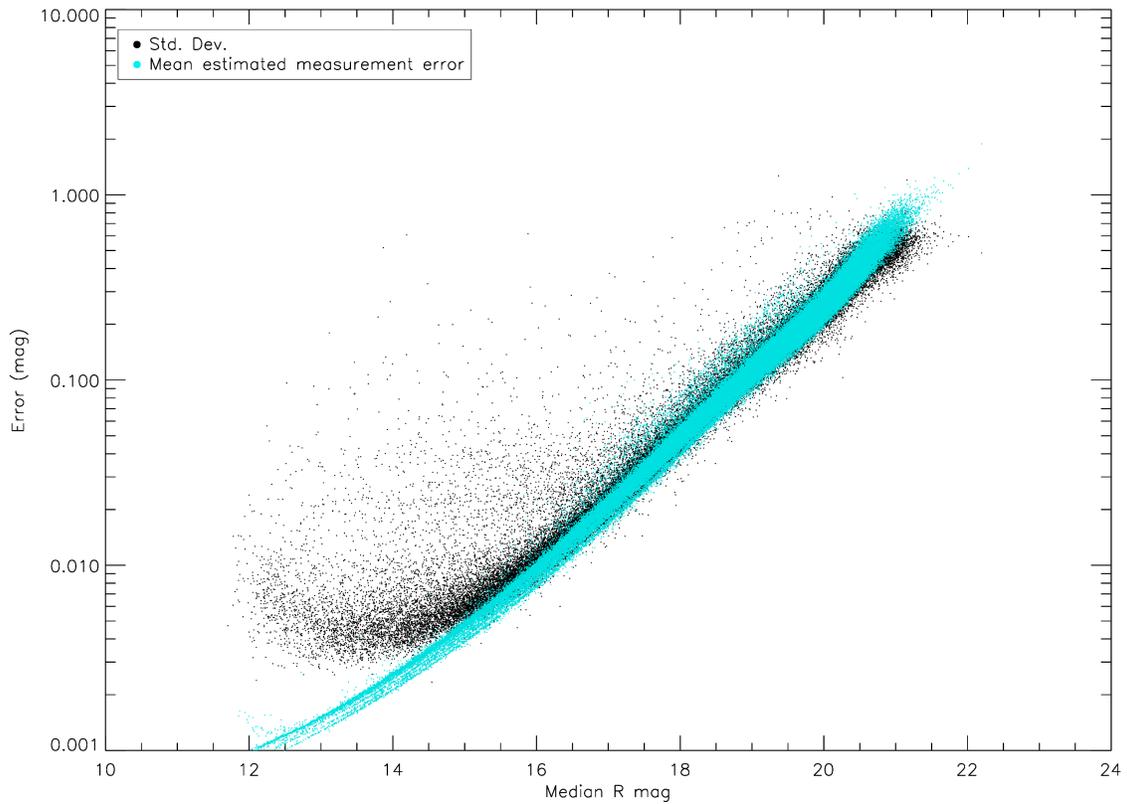}
\caption[RMS precision]{\label{fig:rmsvsmag} Standard deviation (black
  points) vs. median magnitude, compared to theoretical photometric
  errors (cyan), for all unflagged data from \emph{all} light curves
  obtained with more than 10 good data points. There is a slight
  overdense band around 0.02--$0.03\vmag$ attributable to chip 4,
  where a known systematic trend affects certain localized
  regions. See section \ref{sec:precisionandsensitivity}.}
\end{figure*}

Figure \ref{fig:rmsvsmag} shows a plot of the standard deviation of
unflagged data points versus $R$ magnitude for \emph{all} of the $\sim
100,000$ light curves obtained, on all 11 chips, over the full length
of the observations. The results are compared with the theoretical
errors including photon, background, and readout noise. Some chips are
better behaved than others, but the noise floor at around 4$\mmag$ is
evident. Approximately 8\% of all light curves with more than 10
detections have an rms $<1\%$. The banding in the theoretical errors
comes from the different chips, and is likely due to small differences
in quantum efficiency. Saturation begins to set in around $R\approx
13\vmag$, where the upturn in the rms appears. The faint-end limit
varies depending on the chip, but the turn-over in the number of
source detections with increasing faintness begins to set in around
$R\approx 19.5\vmag$, with few detections beyond $20.5\vmag$.

In addition to intrinsically variable stars, some of the scatter of
points above the noise floor at brighter magnitudes in figure
\ref{fig:rmsvsmag} is due to systematic trends that appear in small
regions of certain chips. After performing differential photometry, it
is not unusual to discover common systematic trends in some of the
resulting light curves that may be a function of position, color,
and/or airmass. An advantage of the `modal differencing'
technique employed in the differential photometry (section
\ref{sec:diffphot}) is that, provided
the fraction of sources in the ensemble exhibiting a given trend is
relatively small, the position of the peak of the difference histogram
should be insensitive to the trend. As a result, sources that
are not directly affected by the trend will also not
be compromised as a result of the differential photometry corrections.

The presence of trends appears to be fairly minimal in the PTF Orion
differential photometry, with the exception of one effect which is
attributed to known fogging of the detector during the period of the
Orion observations, due to slow deposition of oil contaminants from the
dry-air system used at the time to prevent condensation on the CCD
window \citep[see][]{Law2010}. This contributes a slow long-term trend
in certain regions of the detector over the length of the
observations, with the exception of the last night, just before which
the CCD window had been cleaned. After the cleaning the affected light
curves returned to their initial flux level. The trend causes
variation of up to $\approx0.2\vmag$ in the worst cases, but is
generally easy to recognize, and appears to be confined to relatively
small regions on the detector; only a small fraction of the light
curves are significantly affected at the level of variability with
which we are concerned here. Chip 4 suffered the most from the fogging
effect, which leads to a slight overdense band of sources from that
chip around 0.02--$0.03\vmag$ in figure \ref{fig:rmsvsmag}. At this
point light curves which clearly show the trend are rejected from our
current analysis for simplicity (four probable eclipsing binaries are
affected in this way). A number of sources on chip 4 alone also show
random noise on the few-percent level on the first three nights and the
last night of the run, which may be detector noise related. This
should be borne in mind with regard to results from that chip.

There are some suggestions of other small trends, possibly color
related, but no detrending of the data has been attempted at this
stage, and none of these effects are accounted for in figure
\ref{fig:rmsvsmag}. Nonetheless, the figure shows that the majority of
the data are well behaved down to the $4\mmag$ noise floor without any
detrending.

\subsection{Absolute Photometric Accuracy}\label{sec:absaccuracy}

The absolute accuracy of the magnitudes is dependent on the second
zero-point correction against the CIDA-QUEST reference stars (see
section \ref{sec:ancillary}). The outlier resistant standard deviation
in the magnitude difference between the CIDA-QUEST sources and the
least variable 50\% of the matched PTF Orion sources ranges from
0.025--0.044$\vmag$ depending on the chip. Combining this with the
overall 0.02--0.03$\vmag$ error of the CIDA-QUEST list (including
systematics), we therefore
conservatively estimate the zero points of the individual light curves
to be accurate to $\approx$3\%--5\% for chips 0--4 and 6--10.

The area covered by chips 5 and 11 in the Orion field had no coverage
in the reference star set, and therefore no direct secondary
zero-point correction could be determined. Instead, a correction was
made based on the weighted mean of the corrections from the other
chips, excluding only chip 4, owing to more significant fogging-trend
systematics on that chip which give rise to a distinctly bimodal,
non-Gaussian distribution in the zero-point histogram for that
chip. The mean of the secondary zero-point corrections for the
remaining chips was $0.159\vmag$, with a standard deviation of
$0.044\vmag$. Adding this whole-chip zero-point uncertainty in
quadrature with the typical uncertainty for the individual light
curves discussed above, and a $0.03\vmag$ error for the CIDA-QUEST list,
yields an absolute accuracy estimate of $0.069\vmag$ for the light
curves on chips 5 and 11.

Marginal color-dependent trends are seen in the magnitude residuals
against the CIDA-QUEST references (as a function of $R-I$), but they
are below the level of the random scatter in the residuals, implying
that errors due to mismatch between the PTF and CIDA-QUEST filters are
negligible at our accuracy level. Since the error estimates for the
zero points are based on the scatter in the residuals about the mean,
rather than the error in the mean itself, any systematic errors in
zero point should be included in the accuracy estimate.

%% file: binaries.tex
\section{Eclipsing Binary Systems}\label{sec:binaries}

Among the most easily identified of the light curves are the eclipsing
binaries. In this section we present those found in this initial
investigation. Section \ref{sec:EBdata} presents an overview of the
data; an analysis is presented in sections \ref{sec:youngbinaries},
where we identify candidate young binaries, and \ref{sec:lowmass},
where we identify some additional low-mass systems. In section
\ref{sec:binariesdiscussion} we summarize and discuss the specific
noteworthy sources.
\subsection{Eclipsing Binary Results}\label{sec:EBdata}
\begin{figure*}[tbp] \epsscale{0.85} 
    \centering
    \plotone{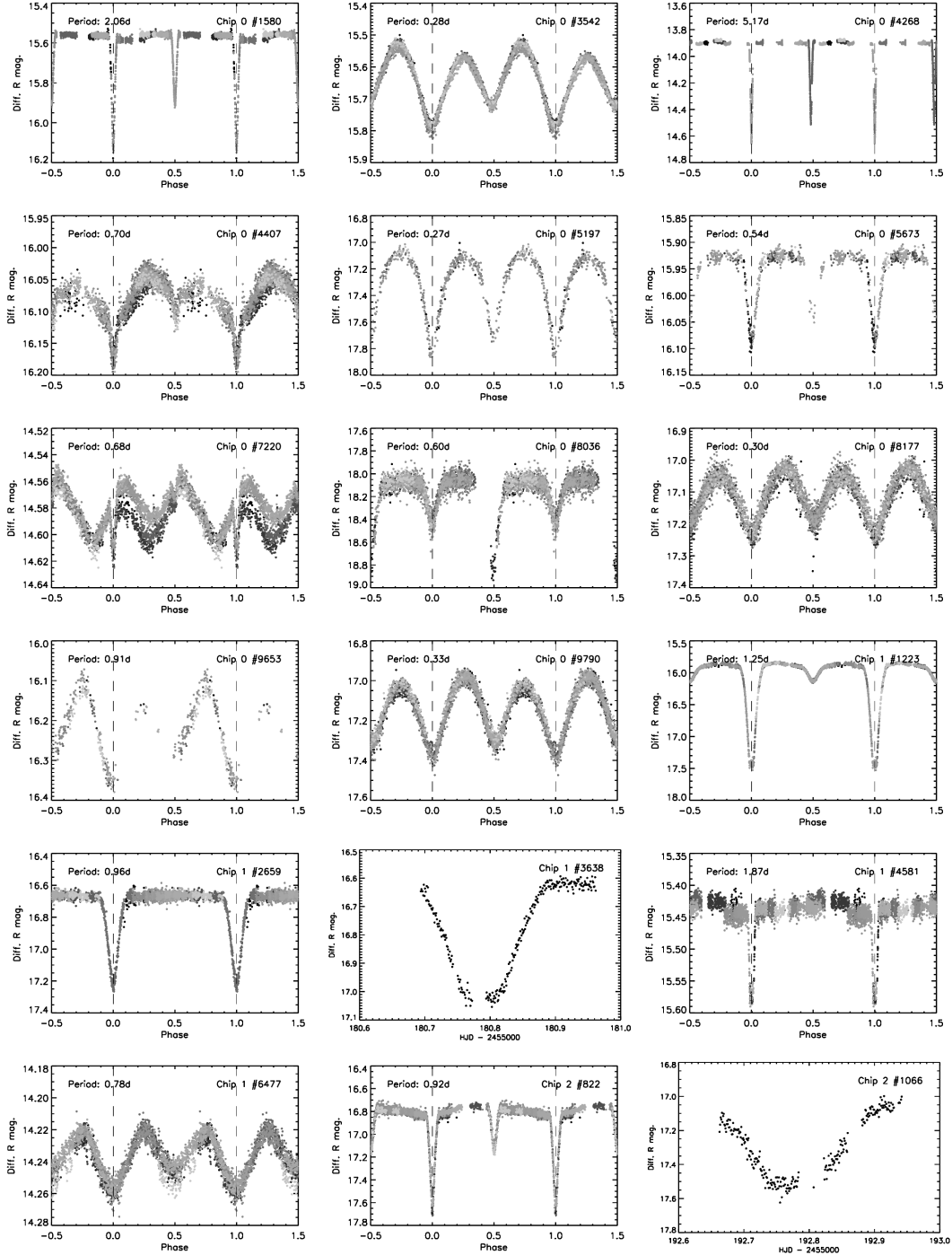}
    \caption{\label{fig:binaryLCs1}Folded light curves for the
      eclipsing binaries identified in the data (section
      \ref{sec:EBdata} and table \ref{tbl:binaries}). The gray scale
      represents time of observation to give a sense of any long-term
      variation, running from black at the beginning of the PTF Orion
      observing run to pale gray at the end. Error bars are omitted
      for clarity. In cases where insufficient coverage was obtained
      to determine a period, or where a regular period could not be
      found, data covering the best eclipse detection are instead
      plotted in black directly against heliocentric Julian date
      (complete light curves for these are shown in figures
      \ref{fig:unfoldedBinaries1}--\ref{fig:unfoldedBinaries2}). See
      table \ref{tbl:binaries} for the full precision in the measured
      periods. (Continued in figures
      \ref{fig:binaryLCs2}--\ref{fig:binaryLCs5}.)}
\end{figure*}

\begin{figure*}[tbp] \epsscale{0.95} 
    \plotone{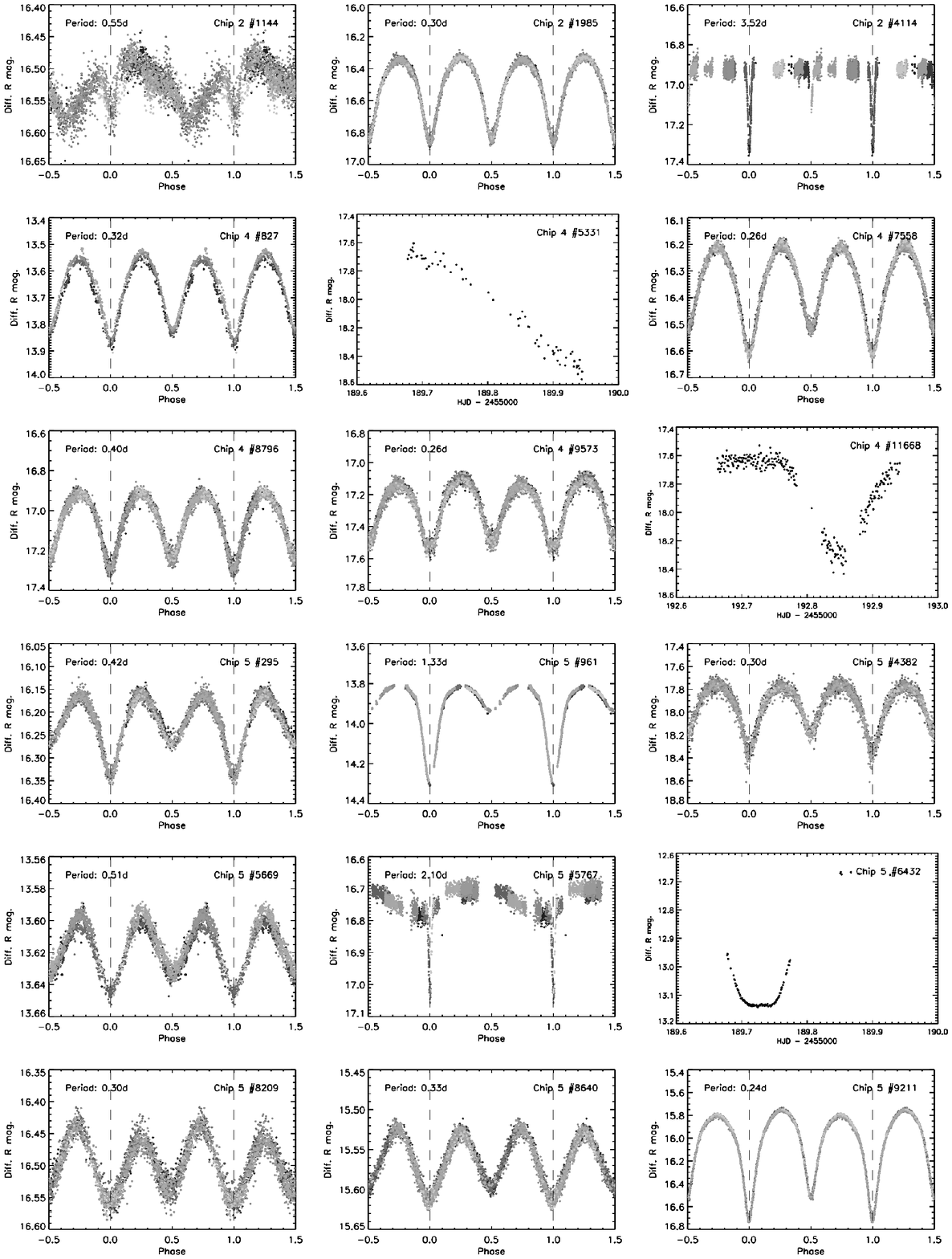}
    \caption{\label{fig:binaryLCs2}Folded binary system light curves - cont.}
\end{figure*}

\begin{figure*}[tbp] \epsscale{0.95} 
    \plotone{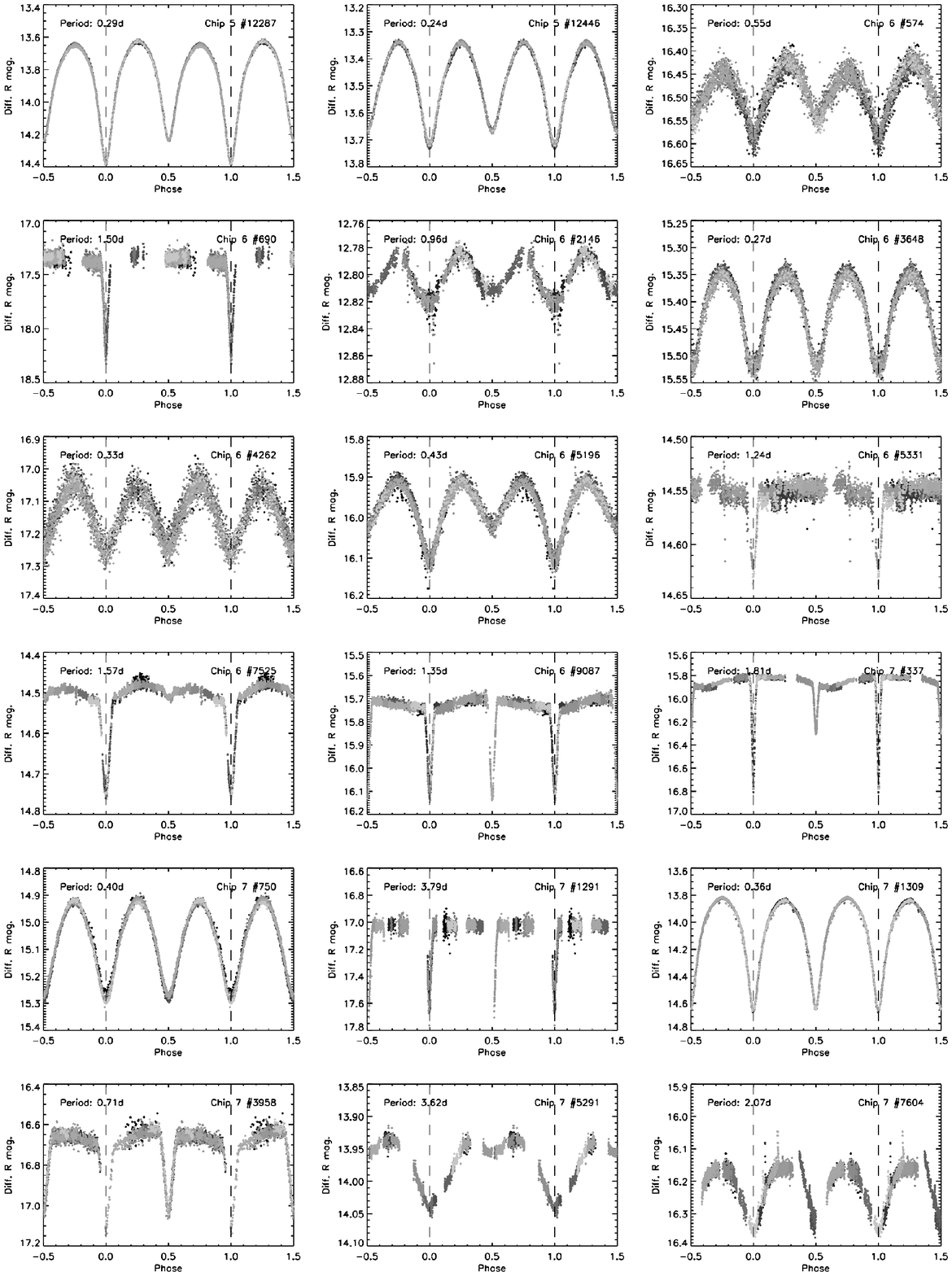}
    \caption{\label{fig:binaryLCs3}Folded binary system light curves - cont.}
\end{figure*}

\begin{figure*}[tbp] \epsscale{0.95} 
    \plotone{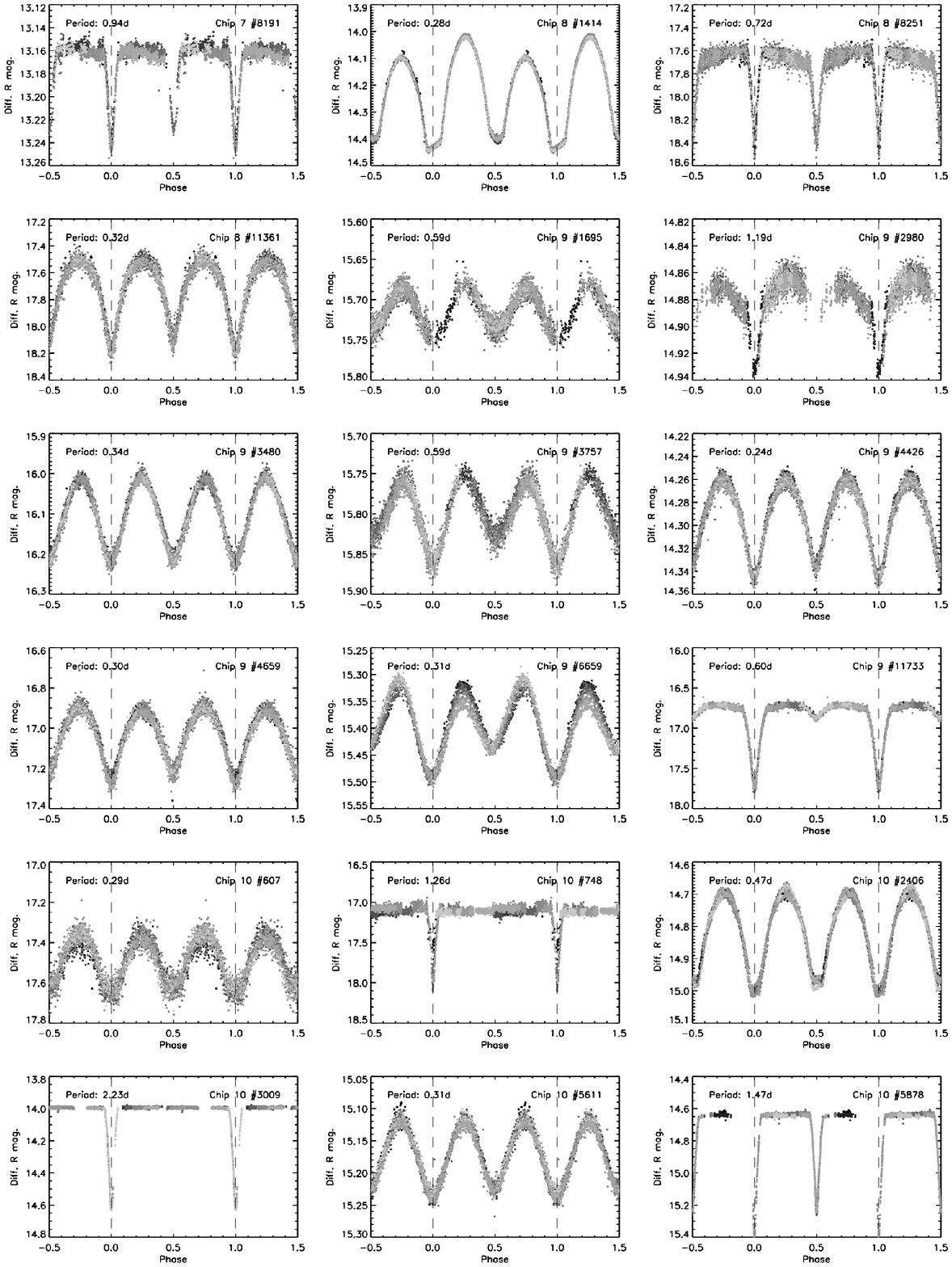}
    \caption{\label{fig:binaryLCs4}Folded binary system light curves - cont.}
\end{figure*}

\begin{figure*}[tbp] \epsscale{0.95} 
    \plotone{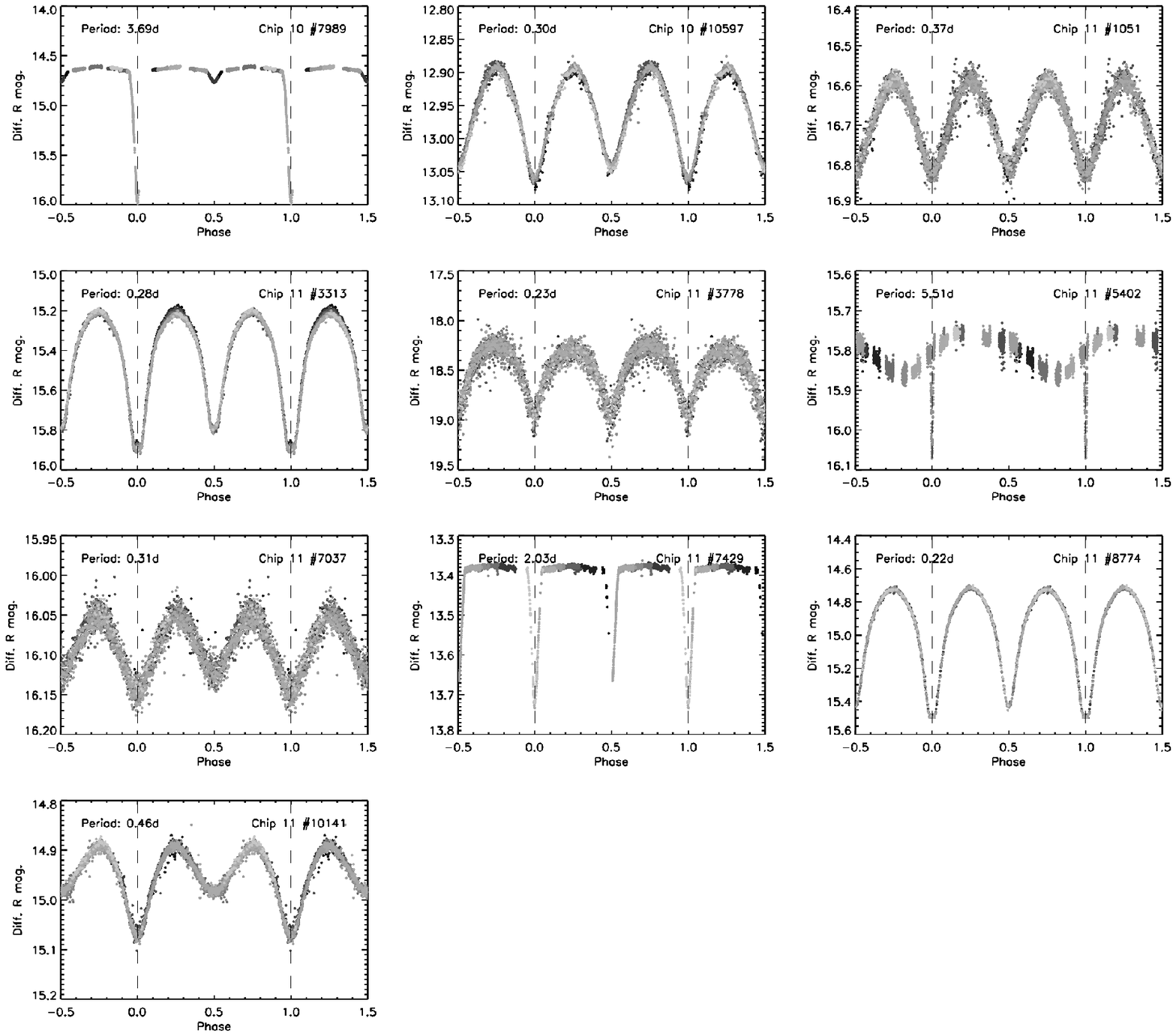}
    \caption{\label{fig:binaryLCs5}Folded binary system light curves - cont.}
\end{figure*}

\begin{figure*}[tbp] \epsscale{0.97}
\plotone{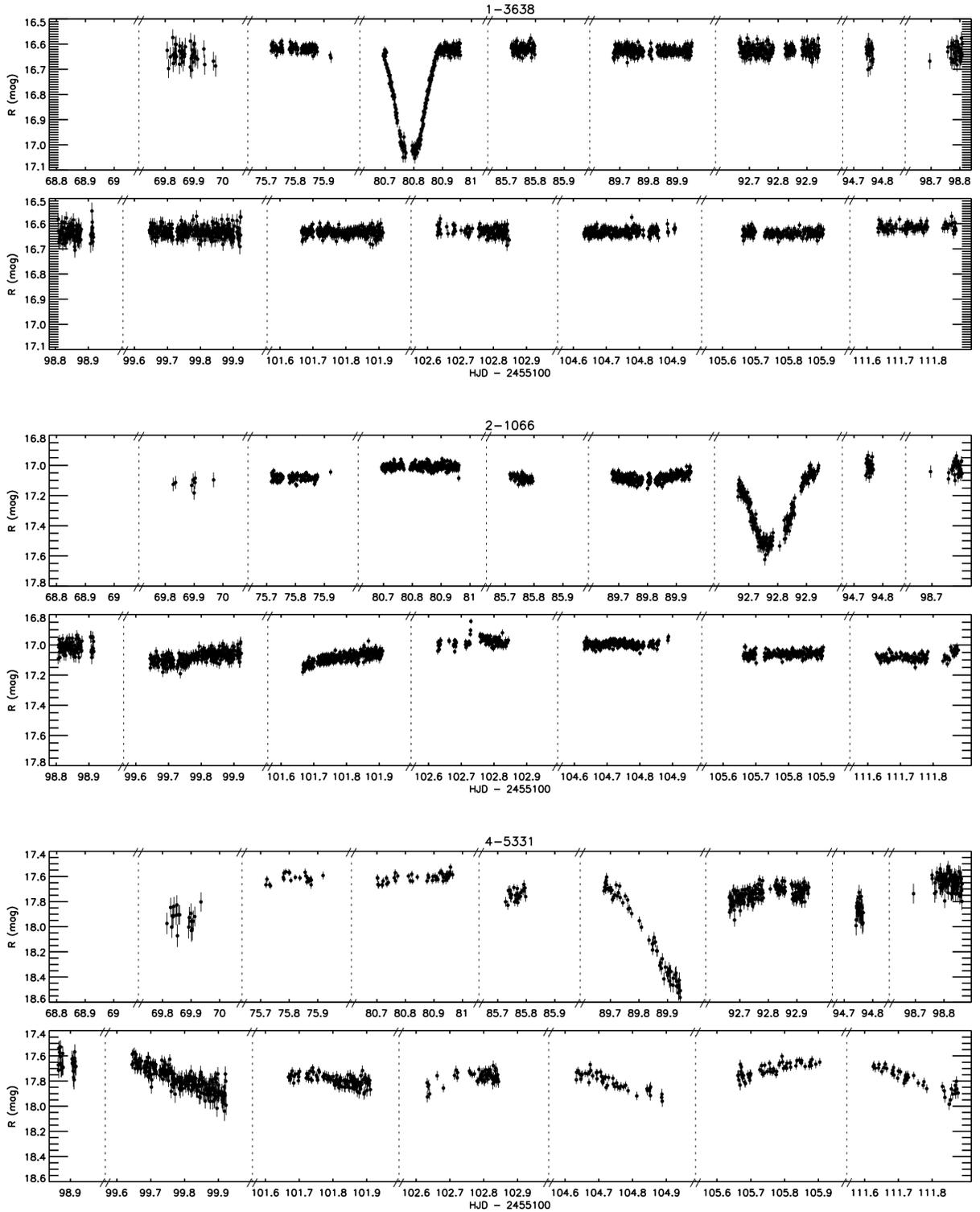}
\caption{\label{fig:unfoldedBinaries1}Full unfolded light curves for
  those sources for which a period could not be uniquely determined
  due to an insufficient number of eclipse detections or because
  the data did not exhibit a regular period. Continued in
  figure \ref{fig:unfoldedBinaries2}.}
\end{figure*}

\begin{figure*}[tbp] \epsscale{0.97}
\plotone{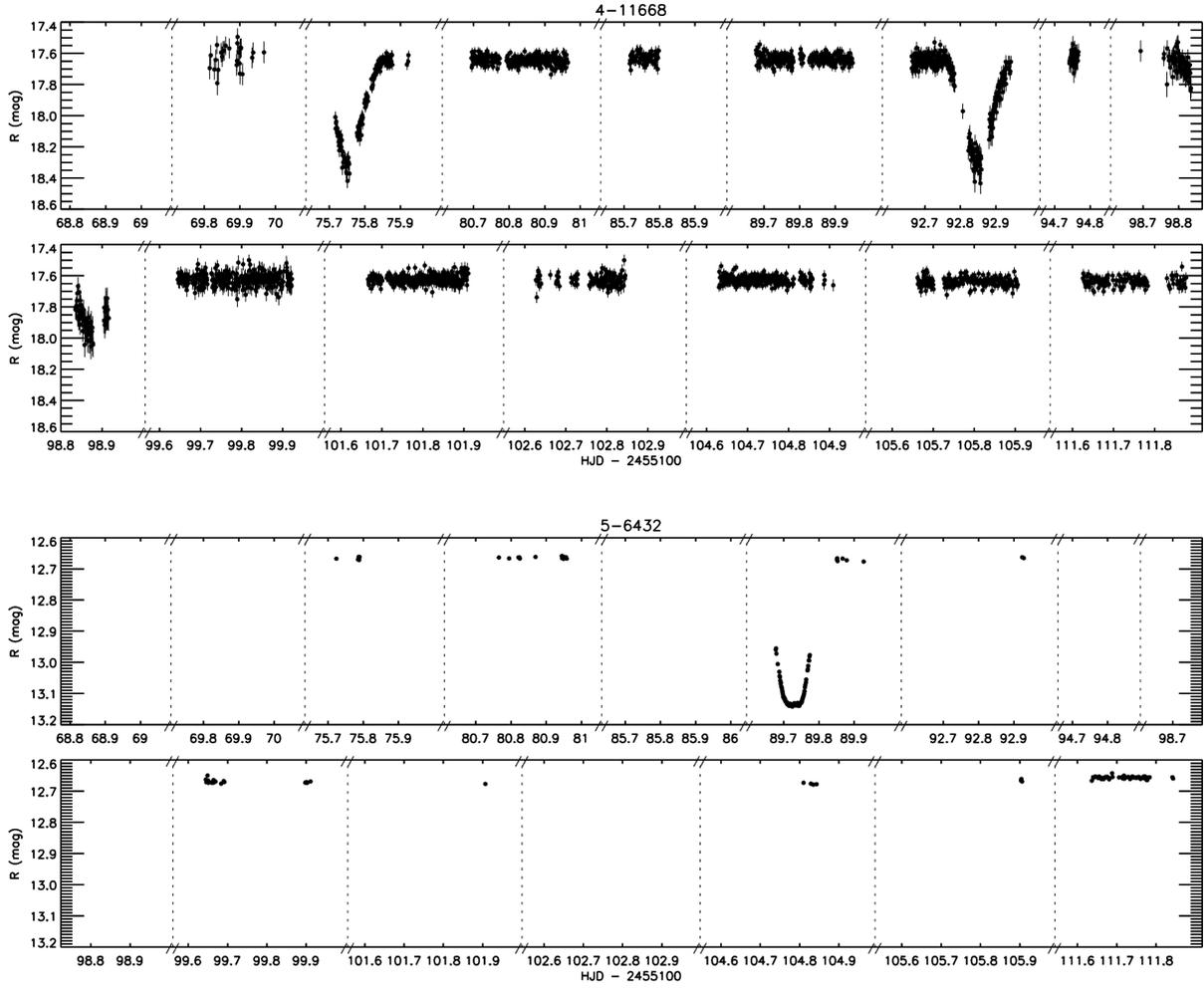}
\caption{\label{fig:unfoldedBinaries2}Unfolded light curves -
  continued. 5-6432 (bottom) shows sparse coverage because most of the data
  is saturated; the light curve drops below the saturation threshold
  during the eclipse, however.}
\end{figure*}

We identified a total of 82 clear eclipsing binaries in the PTF Orion
data, for which period-folded light curves are presented in figures
\ref{fig:binaryLCs1}--\ref{fig:binaryLCs5}. These were identified by
visual inspection of our short-list of strongly varying light curves,
most eclipsing binary light-curve types displaying distinctive
characteristic shapes. Light curves which were sinusoidal in
appearance were rejected due to the ambiguity in interpretation of
their nature. Among the binaries, three were found to have been
previously identified as candidate variable stars by
\citet{MOTESS-GNAT} in the MOTESS-GNAT survey (sources 0-9653,
5-12287, and 5-12446). In those cases where a period could not be
determined, either because too few eclipses were detected or because
the data did not show a regular period, the data covering only the
best captured eclipse are instead plotted against time to show the
eclipse shape. The complete light curves for the five such targets are
shown in addition in figures \ref{fig:unfoldedBinaries1} -
\ref{fig:unfoldedBinaries2}.

Table \ref{tbl:binaries} lists the basic properties of the eclipsing
binary systems: the assigned PTF Orion survey ID, $N$-$n$, consisting
of the chip number, $N$, on which the source was detected, followed by
a running sequence number, $n$; the mean measured J2000 coordinates
over the whole observing run (accurate to $\approx 1$\arcsec);
the corresponding 2MASS ID where a match was obtained \citep{2MASS};
the 2MASS $J$, $H$, and $K_s$ magnitudes; classification as `close'
(C) or `detached' (D) (see below); the period, $P$, in days; $T_0$,
the heliocentric Julian date for the epoch of the primary eclipse; the
estimated distance to the system; the median measured $R$ magnitude;
and $\Delta R$, the approximate peak-to-peak magnitude range of the
measured light curve. Specific sources discussed in the text are
broken out at the beginning of the table. The newly measured
properties in the table are obtained as described below.

\begin{description}
\item[Classification:] The 82 systems are categorized into one of two
  broad classifications: `close' (`C'), of which we find 45, and
  `detached' (`D'), of which we find 37. For the purposes of this
  paper, `close' systems are defined as those for which there is no
  apparent distinction between in- and out-of-eclipse regions in the
  light curve, i.e., where there is no clear discontinuity in the
  slope of the curve. These should represent, for the most part,
  contact and over-contact binary systems. Those for which such a
  distinction \emph{is} apparent are considered `detached', though
  this category can also include semi-detached and near-contact
  systems. In some cases semi-detached or near-contact systems may
  also fall into the `close' category instead, depending on the exact
  nature of the light curve. We have not attempted to further
  sub-classify the binary systems to avoid ambiguities that are better
  addressed with full light-curve modeling.

\item[Orbital parameters:] Orbital periods, $P$, are obtained by
  calculating the Plavchan periodogram of the light curve
  \citep{PlavchanPeriodogram}, using a stand-alone version of the
  NASA/IPAC/\linebreak[0]NExScI Star and Exoplanet Database (NStED) periodogram
  service.\footnote{\url{http://nsted.ipac.caltech.edu/periodogram/cgi-bin/Periodogram/nph-simpleupload}}
  The light curve is folded on the period corresponding to the highest
  peak in the periodogram, and visually inspected to confirm the
  period is reasonable. If not, the next highest peaks are
  investigated until a plausible folding is found. The error in the
  period is estimated from the width of a Gaussian fit to the
  corresponding peak. In cases where the shape of the peak was such
  that a good Gaussian fit could not be made, the approximate peak was
  estimated by hand, and an appropriately large error assigned.

  The time of mid-eclipse, $T_0$, for the systems is measured by
  fitting a symmetrical inverted trapezium to the deepest and
  best-sampled minimum of each light curve and calculating the
  center time of the fitted trapezium. This fit also yields the formal
  errors for the eclipse center time.

\item[Distance:] W~UMa systems display a relatively
  well-established empirical period-color-luminosity relationship
  resulting from their approximately uniform-temperature common
  envelope: Kepler's law constrains the orbital radius as a function
  of period, and the orbital radius in turn constrains the radiating
  surface area and, hence, the luminosity and absolute magnitude of the
  system \citep{Rucinski2004WUmaDistances}. They therefore can be used
  as distance estimators. \citet{Eker2009WUMaCalibration} provided an
  updated calibration of the relationship based on 2MASS
  ($J-H$ and $H-K_s$) colors; we use this relationship to estimate
  the distances for the `close' binaries in the table where the
  detections have 2MASS counterparts, on the assumption that the vast
  majority are W~UMa systems.

  \citet{Bilir2008DetachedCalibration} provided a similar empirical
  relationship between 2MASS colors and absolute magnitude for
  main-sequence detached binaries, analogous to the color-luminosity
  relationship for single main-sequence stars. We apply this
  relationship to the `detached' binaries in the table, on the
  assumption of dwarf star status.

  Errors in the derived distances are estimated by standard
  propagation of the 2MASS photometric errors through the respective
  empirical relationships. The errors in the coefficients in the
  relationships are neglected, but the reported standard deviations
  about the relationships ($0.26\vmag$ and $0.49\vmag$ for the W~UMa
  and detached-system calibrations respectively) are added in
  quadrature to account for astrophysical effects that are not
  included in the relationships. In both cases, the effects of
  reddening are also neglected: for our assumed value of $A_V$ at the
  distance of 25~Ori, using the extinction laws of
  \citet{Cardelli1989} gives reddening values of $E(J-H)=0.027\vmag$
  and $E(H-K)=0.022\vmag$, which in most cases are comparable to or
  less than the photometric measurement errors. Neglecting reddening
  does lead to a small systematic underestimation of the distances,
  generally by of order half the measurement error in most cases for
  our assumed mean $A_V$, and always less than the full size of the
  error, but it is simpler than attempting to realistically estimate
  the reddening for each individual source.

  We note that these distance relationships are likely to be
  inaccurate at PMS ages. Low mass stars
  ($<1.5\Msun$) younger than $10\Myr$ will still be on PMS tracks and
  can be expected to be brighter than predicted (albeit the close
  companions in our sample may complicate the evolutionary picture).
  The agreement between the list of young candidates from the
  color-magnitude diagram and the list from the distance estimates,
  however, provides some suggestion that the estimates are still
  meaningful in these cases (see sections \ref{sec:CMDselection} and
  \ref{sec:distances}). Some systems also represent extrapolations
  beyond the formal calibrated limits of the relationships provided by
  \citet{Eker2009WUMaCalibration} and
  \citet{Bilir2008DetachedCalibration}, particularly for the detached
  systems at apparent types later than mid-K. The distances provided
  in the table therefore, are only intended as a guideline and should
  be treated with appropriate caution.

\item[$R$ magnitudes:] A single representative value for the
  magnitude, $R_\mathrm{med}$, is given as a straight median of the measured
  differential $R$ magnitude for each light curve. We estimate this to
  be accurate on an absolute scale to $\approx 0.03$--$0.05\vmag$ for
  chips 0-4 and 6-10, and $\approx0.07\vmag$ for chips 5 and 11 (see
  section \ref{sec:absaccuracy}). The light-curve amplitude variation,
  $\Delta R$, is calculated as the difference between the 1st and
  99th percentiles of the measured $R$ magnitudes for each light
  curve, in order to reject possible unflagged outliers and allow for
  a certain amount of statistical deviation. This is similar to the
  range used by \citet{Basri2010}.

\end{description}

\begin{deluxetable}{ccccccccccccc}
\tablecolumns{13}
\tablewidth{0pt}  
\setlength{\tabcolsep}{0.035in} 
\tablecaption{\label{tbl:binaries}New Eclipsing Binaries from the PTF
Orion Data}
\tabletypesize{\scriptsize}
\rotate
\tablehead{ & & & & \multicolumn{3}{c}{2MASS} & & & & & & 
\\\cline{5-7}
\colhead{ID} & \colhead{R.A.} & \colhead{Decl.} & \colhead{2MASS/USNOB1.0} &
\colhead{$J$} & \colhead{$H$} & \colhead{$K_s$} & \colhead{Class.} &
\colhead{$P$} & \colhead{$T_0$} & \colhead{Dist.} &
\colhead{$R_\mathrm{med}$\tablenotemark{n}} & \colhead{$\Delta R$}
\\
 & \colhead{(deg)} & \colhead{(deg)} & & \colhead{(mag)} &
 \colhead{(mag)} & \colhead{(mag)} & & \colhead{(days)} &
 \colhead{(HJD-2455000)} &
 \colhead{(pc)} & \colhead{(mag)} & \colhead{(mag)}
}
\startdata
\input{binariestable}
\enddata

\tablecomments{A description of the columns is given in section
  \ref{sec:EBdata}. Corresponding light curves are shown in figures
  \ref{fig:binaryLCs1}--\ref{fig:unfoldedBinaries2}. Distance estimates are
  omitted in cases where 2MASS magnitudes reported are limits
  only. Where possible the matched 2MASS identifier is given
  based on the J2000 source coordinates; in cases where no 2MASS match
  was found, the corresponding USNO-B1.0 match running-number--based
  ID is given instead, with the format `NNNN-NNNNNNNN' (with no
  preceding `J'). The full 2MASS identifier has the format `2MASS
  JHHMMSSss+DDMMSSs,' where HH, MM, and SSss represent hours, minutes,
  and seconds of right ascension, respectively; and DD, MM, and SSs
  represent degrees, minutes, and seconds of declination,
  respectively. Lower case `s' represents digits to the right of the
  decimal point. Full PTF survey catalog IDs can be constructed
  based on these coordinates, using the format `PTF1
  JHHMMSS.ss+DDMMSS.s.' Since precision astrometry was not the primary
  goal of the PTF Orion project, it is preferable to use
  2MASS/USNO-B1.0 coordinates for this purpose rather than our PTF
  Orion measured coordinates.}
\tablenotetext{a}{Only one eclipse obtained - period indeterminate.}
\tablenotetext{b}{Partial coverage of only one (presumed) eclipse
  obtained - period indeterminate, $T_0$ not well constrained.}
\tablenotetext{c}{$T_0$ not well determined due to incomplete coverage
  of eclipse. Measurement error may be unreliable.}
\tablenotetext{d}{Only one (very good) eclipse observed; period based
  on out-of-eclipse variation.}
\tablenotetext{e}{Unusual light curve, possible semi-detached
  system. Listed distance is assuming a contact binary, but using a
  detached-system distance estimate yields a distance of $268\pm
  64\pc$ -- see section \ref{sec:youngclose}.}
\tablenotetext{f}{Very
  short period W UMa system - see section \ref{sec:otherclose}.}
\tablenotetext{g}{Used center of secondary eclipse for $T_0$, owing to
  poor coverage of primary.}
\tablenotetext{h}{Three clear eclipses,
  but unable to find coherent period - possible triple system?}
\tablenotetext{i}{Nearby second source in USNO-B; chance of slight
  contamination.}
\tablenotetext{j}{Secondary eclipse not evident, or
  primary and secondary eclipses indistinguishable -- possible factor
  of two ambiguity in $P$.}
\tablenotetext{k}{Only two eclipses
  obtained - period ambiguous.}
\tablenotetext{l}{Apparent pulsating
  binary -- short-period oscillations seen, with $\sim1/2\hr$
  period. See section \ref{sec:otherclose}.}
\tablenotetext{m}{Somewhat distorted light-curve shape -- may actually
  represent stellar pulsations rather than a binary system
  (D.~Bradstreet, priv. comm., 2010).}
\tablenotetext{n}{See section
  \ref{sec:absaccuracy} for a discussion of zero-point accuracy.}
\tablenotetext{o}{Identified as a candidate variable star by \citet{MOTESS-GNAT}}

\end{deluxetable}

\subsection{Identifying Young Binaries}\label{sec:youngbinaries}

We identify young binary candidates in our sample with a
color-magnitude selection, and then follow several approaches to check
for consistency, detailed below. We find nine candidates which are worthy
of followup (see beginning of table \ref{tbl:binaries}).

\subsubsection{Color-Magnitude Selection}\label{sec:CMDselection}

\begin{figure}[tbp] \epsscale{0.7} 
    \plotone{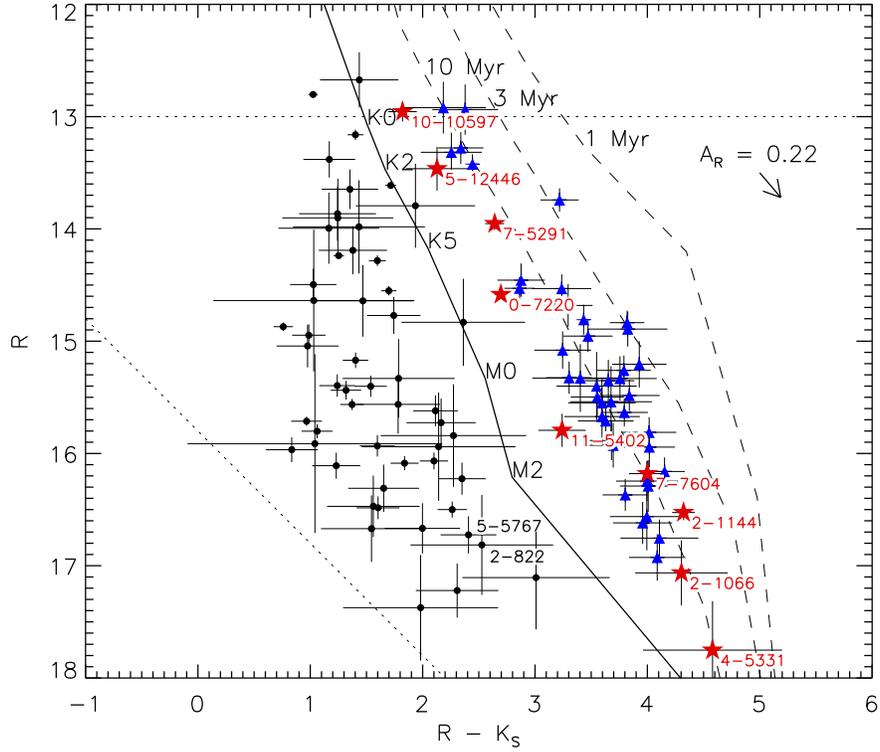}
    \caption{\label{fig:RvsR-K}Color-magnitude diagram for the binary
      systems identified in this paper (circles/star symbols), and for
      other sources previously identified as single WTTS sources by
      \citet{Briceno2007} which are found in the PTF Orion data
      (triangles). Star symbols highlight candidate young Orion systems,
      labeled with their PTF Orion IDs. The dotted lines mark rough
      completeness limits; the solid line marks the main sequence at
      $330\pc$, with the assumed mean reddening values for the 25 Ori
      association; the dashed lines represent PMS isochrones at 1, 3,
      and $10\Myr$ (see section \ref{sec:CMDselection}). Two further
      sources, 2-822 and 5-5767, are also labeled; these are discussed
      in section \ref{sec:otherlowmass}.}
\end{figure}

In figure \ref{fig:RvsR-K} we plot the complete set of 82 binary
systems in a color-magnitude diagram, using our median measured $R$
values and the single-epoch 2MASS $K_s$ magnitudes (excepting those
with poor $K_s$ measurements, i.e., those with 2MASS quality flags
worse than `C' or error bars $>0.1\vmag$ - see section
\ref{sec:ancillary}). Isochrones are over-plotted which are based on
the \citet{SiessModels} models. These are tuned to fit the Pleiades
main sequence at $100\Myr$
\citep{GuieuIC2118,Stauffer2007,Jeffries2007}, and then converted to
$R$ versus $R-K_s$ using polynomial fits to young-star colors
\citep{KenyonHartmann1995}. The blue triangles mark the locations of
previously identified (single) WTTS stars identified by
\citet{Briceno2007} which are also found in our data, and in the 2MASS
catalog. They lie along the $10\Myr$ isochrone, in good agreement with
the $10\Myr$ age determined by \citet{Briceno2007} also using Siess et
al.\ isochrones, and supporting our placement of the isochrones in the
diagram.

The uncertainties of magnitudes and colors are dominated by the
amplitude of the physical photometric variability. As a result the error bars in figure
\ref{fig:RvsR-K} are assigned to represent the overall variability of
the light curves (half the peak-to-peak amplitude). In the absence of
multi-epoch 2MASS measurements, we assume the variability in $K_s$ to
be the same as that in the PTF $R$ measurements for the purposes of
the error bars, and since the $R$ and $K_s$ are not contemporaneous,
we assume the errors to be independent. We treat the WTTS sources in
the same way to allow for their variability as well.

Those systems lying below the marked main sequence are taken to be
unassociated with the 25~Ori/OB1a association. Those falling above the
main sequence (the PMS region for the assumed 25~Ori
distance and reddening) are highlighted with star symbols, and
identified as candidate young systems. This criterion yields nine
systems: 0-7220, 2-1066, 2-1144, 4-5331, 5-12446, 7-5291, 7-7604,
10-10597, and 11-5402 (see table \ref{tbl:binaries}). For comparison,
we ran a Besan\c{c}on Galactic model simulation for the PTF Orion
field using the default model
parameters\footnote{\url{http://model.obs-besancon.fr/}}
\citep{BesanconModel}. This allows us to estimate how many field stars
might be expected to lie in the PMS region. Scaling the
fraction of sources found to lie above the $330\pc$ main sequence in
the simulation to the size of our binary sample yields $\approx 2$--5
expected field sources, depending on the exact location of the main
sequence cut.

There is an ambiguity with regard to binary systems on the
color--magnitude diagram in that the mass ratios of the systems are
unknown. For systems with a small mass ratio, one star dominates the
luminosity, and hence the system's location on the diagram would fall
as expected for a single star of the same type and color (diverging
evolutionary effects due to the binarity notwithstanding). However, for
mass ratios near unity, the two stars contribute similar fluxes,
appearing up to $\approx 0.75\vmag$ brighter in total than a single
star of the same color. There is a possibility, therefore, that some
binaries could be shifted systematically upward in the diagram
compared to single stars, and that location near the OB1a/25-Ori group
sequence could simply be a result of field binaries suffering from
this effect. These nine systems should therefore only be considered a
list of good candidate young systems from the sample, for which
further spectroscopic follow-up is needed for proper confirmation. At
the same time, we note that shifting downward by $0.75\vmag$ any of
those candidates on the low edge of the cluster sequence in the diagram
(assuming that $R-K$ remains the same) still places them near the main
sequence for the distance of the association, and does not rule out
Orion membership. It is possible, for example, that binarity has
affected their evolution compared to that of single stars,
particularly given the close separations implied by the relatively
short orbital periods of our sample. In the following sections, we
investigate other consistency tests for the proposed young ages
of the candidates.

\subsubsection{Distances}\label{sec:distances}

\begin{figure}[tbp] \epsscale{0.7} 
    \plotone{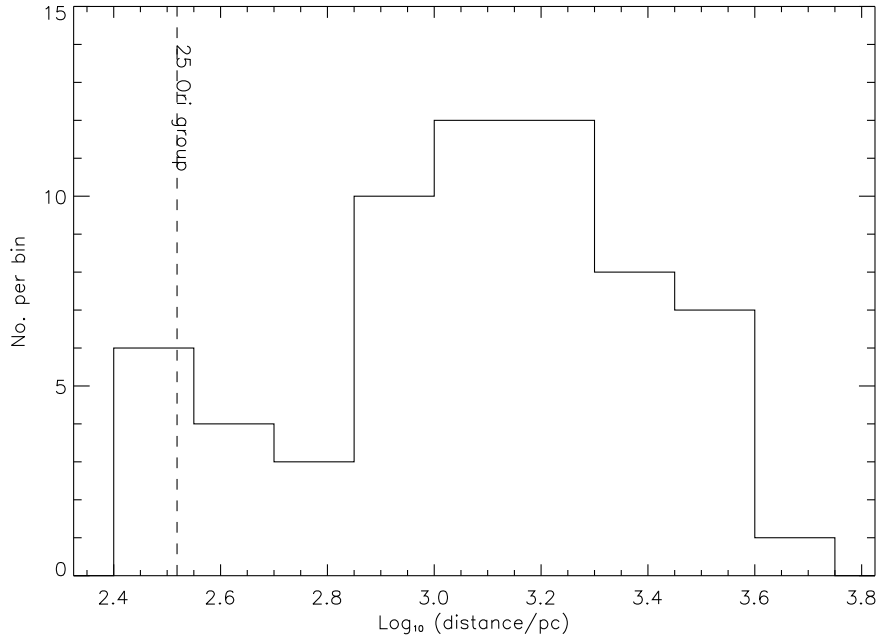}
    \caption{\label{fig:distancehistogram}Histogram of estimated
      distances to the binary systems (section \ref{sec:distances}). A
      peak is seen around $330\pc$, the estimated distance of the 25
      Ori group and Orion OB1a ($\log_{10}$ distance = 2.52, marked with a
      dashed line). Bin size is 0.15 $\log_{10}$ (distance/pc).}
\end{figure}

Figure \ref{fig:distancehistogram} shows a histogram of the log of the
derived distances for the binaries in table \ref{tbl:binaries}
(excluding those with 2MASS quality flags worse than `C' -- see
section \ref{sec:ancillary}). The number of binaries per logarithmic
bin increases with distance, peaking at around
log(distance/pc) = 3.1 and then tailing off. This we attribute to the
background field distribution of binaries in the galaxy in combination
with the sensitivity limit of the sample and the logarithmic bin
sizes. In addition we see a distinct peak at around $330\pc$, the
distance to the OB1a and 25 Ori associations adopted by
\citet{Briceno2005,Briceno2007}. We take systems with derived
distances falling at the 25 Ori/OB1a distance within a factor of 1.5
times their respective errors, and with errors below $150\pc$, to be
good candidates for membership in the association. This criterion
yields exactly the same list of candidates as the color-magnitude
diagram selection, consistent with their proposed young nature, with
the exception only of source 7-5291. This source presents some
peculiarities, however, and there are indications that the initial
distance estimate is probably incorrect. An estimate based on the
detached-system calibration is likely more appropriate than the
presumed W~UMa distance calibration, and satisfies the above
criteria. We therefore do not rule it out as a young candidate (see
section \ref{sec:youngclose} for more discussion).

The same mass-ratio ambiguity as discussed in the previous section can
also affect the distance estimates, although probably to a lesser
extent since the distance calibrations used are based on broad samples
of binary systems, and therefore should represent the relation for a
`typical' mass ratio. \citet{Bilir2008DetachedCalibration} report a
standard deviation of $0.49\vmag$ about the relation for detached
systems, pointing out that this could be largely explained by the
mass-ratio ambiguity, which should lead on average to an error less
than $0.4\vmag$; \citet{Eker2009WUMaCalibration} report a standard
deviation of $0.26\vmag$ for W~UMa systems. In the worst possible case
where a mass ratio of unity leads to a $\approx 0.75\vmag$
underestimate of the distance modulus, this would cause a distance
underestimate of $\approx 40$\%. Even at this extreme, however, an
estimated distance at the $330\pc$ mark still places the true
value near the more distant OB1b association \citep[$\approx
440\pc$;][]{Briceno2005,Brown1999}, and such a source would remain of
potential interest.

\subsubsection{Stellar Variability}\label{sec:variability}

\begin{figure}[tbp] \epsscale{0.8} 
    \plotone{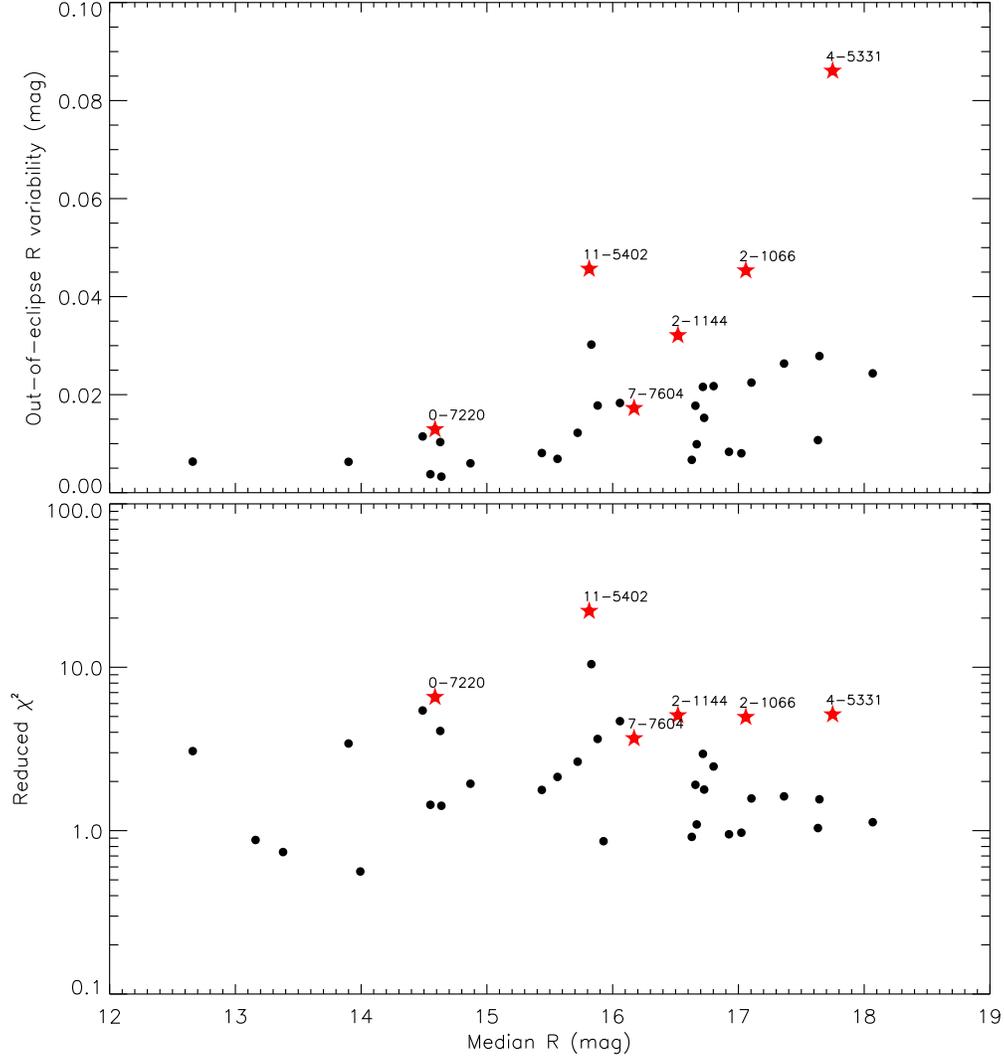}
    \caption{\label{fig:vindex}Intrinsic stellar variability estimates
      for all the binary systems classified as detached (section
      \ref{sec:variability}). Star symbols highlight those which are
      the candidate young systems selected from the color-magnitude
      diagram (figure \ref{fig:RvsR-K}). Variability (top panel) is
      calculated as the quadrature difference between the
      out-of-eclipse rms and the median measurement-error for each
      light curve (with an additional estimated $0.004\vmag$ noise floor
      added in quadrature); sources where the rms is below the
      median measurement error are omitted. The reduced $\chi^2$
      against the weighted mean $R$ magnitude is shown for the same
      data (bottom panel), including those sources with rms smaller than
      the measurement error. System 5-961 is omitted from both panels since it shows
      significant ellipsoidal variation (figure
      \ref{fig:binaryLCs2}).}
\end{figure}

Intrinsic stellar variability can also be a useful indicator for young
stars, which tend to be more spotted and active
\citep[e.g.,][]{Briceno2001, Lamm2004}. In figure \ref{fig:vindex} we
plot an estimate of the intrinsic variability of the detached binary
systems in our sample for which a meaningful estimate could be
calculated. All `close' systems are omitted, as is 5-961, a likely
semi-detached system, since they show continuous strong ellipsoidal
variation which confuses the intrinsic variability measurement. To
quantify the intrinsic stellar variability, the out-of-eclipse rms $R$
magnitude is first calculated as the outlier-resistant standard
deviation for a given light curve between orbital phases 0.1--0.4 and
0.6--0.9, effectively cutting out the eclipses in the detached
binaries (assuming circular orbits, which is reasonable given the
short orbital periods in our sample).  To account for the apparent
noise floor in figure \ref{fig:rmsvsmag} a constant $0.004\vmag$ is
added in quadrature to the measurement errors. Variability is then
calculated by subtracting the median measurement error for each light
curve (in quadrature) in order to leave only intrinsic
variability. This quantity provides an estimate of the stellar
activity, though with the caveat that low-level ellipsoidal variations
may also be included in the metric. The value of the reduced $\chi^2$
compared to a constant $R$ magnitude at the weighted mean for each
source is also plotted, to provide an indication of the statistical
significance of the measure.

The effects of photon error having been removed from the variability diagram, the
general increasing trend with fainter magnitudes is attributed to the
tendency toward increased activity at later spectral types, since more
of these are seen at the faint end of the sample. The same candidate
young binary systems (section \ref{sec:CMDselection}) are again marked
with star symbols (excluding the three `close' systems, 7-5291,
5-12446, and 10-10597). They are intrinsically more variable on
average than the other detached systems in the sample, consistent with
their suspected young ages. (The elevated source below 11-5402 is an
additional system that shows marked variation probably caused by hot
or cold spots (7-337 -- see figure \ref{fig:binaryLCs3}).)

\subsubsection{Proper Motions}\label{sec:propermotions}

\begin{figure*}[tbp] \epsscale{0.85} 
    \plotone{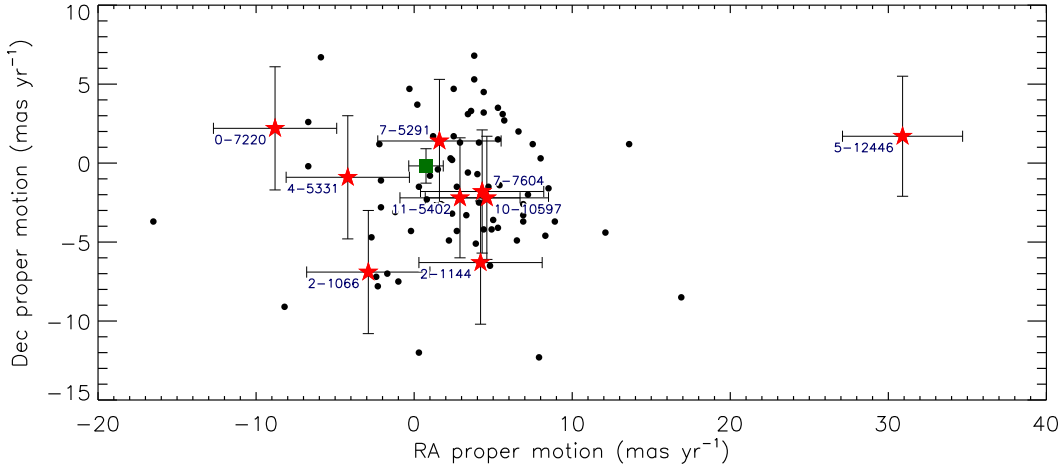}
    \caption{\label{fig:propermotions}PPMXL proper motions for all the
      identified eclipsing binary star systems. Candidate young systems are
      marked with star symbols, and have error bars indicated. Error
      bars for all other systems are omitted for clarity, but are
      similar in size. The square symbol marks the proper motion of
      the 25 Ori cluster measured by \citet{Kharchenko2005}, with
      error bars representing the dispersion derived from RV
      measurements by \citet{Briceno2007}.  (See section
      \ref{sec:propermotions}.)}
\end{figure*}

Proper motions provide a further consistency check on the candidate
young binary systems, although for the 25~Ori cluster and the broader
Orion region, the proper motion is near zero, making it hard to
distinguish between 25~Ori association members, Ori OB1 members, and
more distant background sources
\citep[see][]{deZeeuw1999,Kharchenko2005}. Figure
\ref{fig:propermotions} shows the proper motions of each of the binary
systems, taken from the recently released PPMXL catalog of positions
and proper motions \citep{PPMXL}. \citet{Kharchenko2005} reported a
proper motion for the 25~Ori cluster (identified as ASCC 16 in their
catalog) of $0.75\pm 0.22\masyr$ in right ascension and $-0.18\pm
0.29\masyr$ in declination based on the brighter stars in the
cluster. \citet{Briceno2007} report a $\pm1.7\kms$ width to the core
of the peak of their distribution of radial velocity (RV) measurements for
the cluster, which corresponds to a $\pm1.09\masyr$ dispersion in
proper motions at $330\pc$ (indicated in the figure), assuming a
three-dimensionally isotropic distribution of velocities.

The nine young candidate systems are again highlighted with red star
symbols. Their proper motions are largely consistent with 25~Ori/OB1a
membership, with the exception of 5-12446, with a proper motion of
$30.0\pm3.8\masyr$ in right ascension despite its estimated distance
of $317\pm41\pc$, which is extremely close to that of the 25~Ori/OB1a
association. This suggests it is probably an interloper rather
than a true member (see section \ref{sec:youngclose}).

\subsubsection{Infrared Analysis}\label{sec:binariesspitzer}

\begin{figure*}[tbp] \epsscale{0.85}
    \plotone{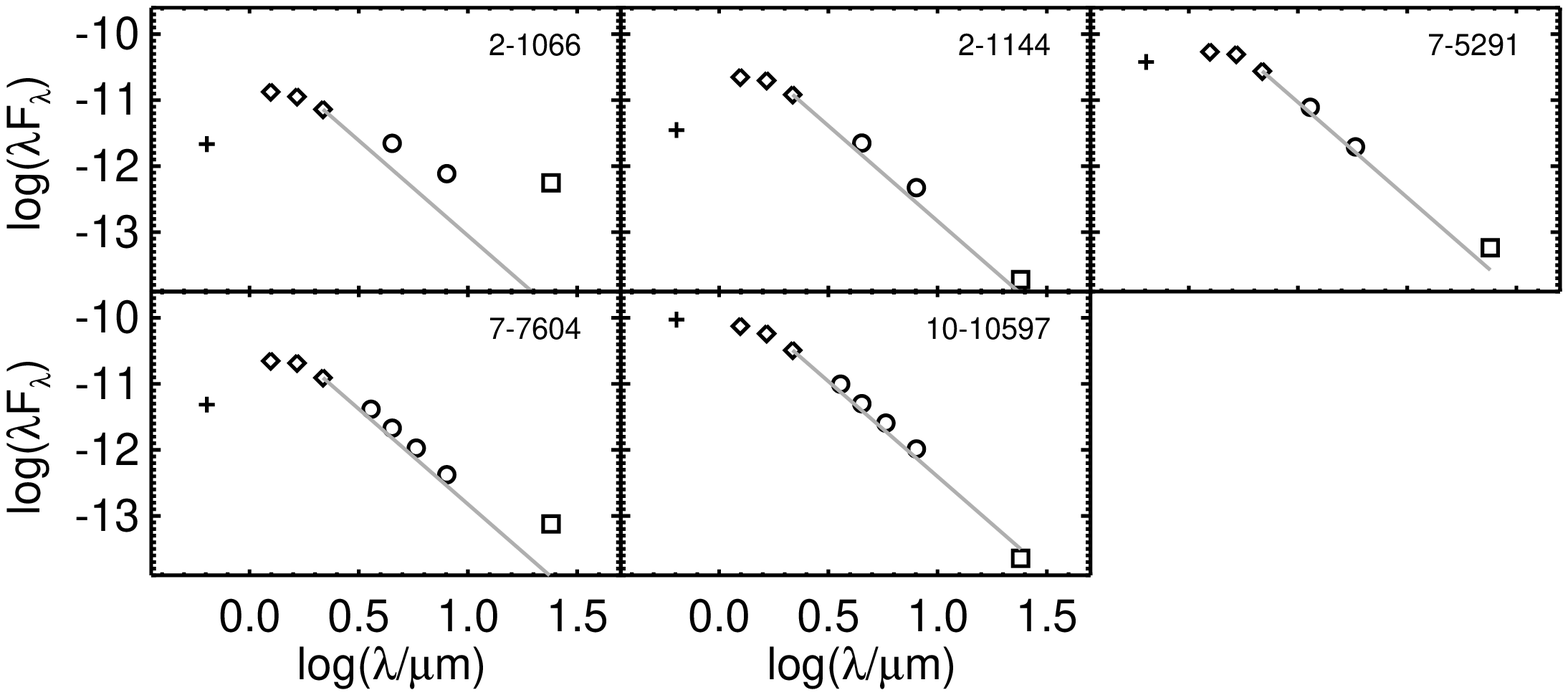}
    \caption{\label{fig:binariesspitzer}SEDs for the candidate young
      binary systems for which IR {\em Spitzer} data could be found (section
      \ref{sec:binariesspitzer} and table
      \ref{tbl:binariesspitzer}). The SEDs are all in log $\lambda
      F_{\lambda}$ in cgs units (erg s$^{-1}$ cm$^{-2}$), against log
      $\lambda$ in microns. Crosses are median PTF $R$, diamonds are
      2MASS ($JHK_s$, 1.25, 1.65, and $2.17\um$), circles are IRAC
      (3.6, 4.5, 5.8, $8\um$), and squares are MIPS ($24\um$).  A gray line
      is extended from $K_s$ to guide the eye, representing an
      extrapolated Rayleigh-Jeans slope. If the object is a star,
      there is no excess at $K_s$, no significant reddening, and the
      star is earlier than mid-M, this would be the approximate
      location of the photosphere. 2-1066 shows a clear possible IR
      excess; a very weak excess appears to be present at $24\um$ for
      7-7604.}
\end{figure*}

Finally, we also searched for any existing {\em Spitzer} IRAC/MIPS data, to
check for any infrared excess that might be indicative of T-Tauri-like
disks in the nine candidate young systems. Data were found for 2-1066,
2-1144, 7-5291, 7-7604, and 10-10597. These are listed in table
\ref{tbl:binariesspitzer}, and SEDs are shown in figure
\ref{fig:binariesspitzer}, where the gray lines represent an
extrapolation of the Rayleigh-Jeans slope from the $K_s$ measurements
to guide the eye with regard to the expected photospheric flux. A
color difference between any two points on such a gradient will have a
value of zero, and anything lying above the slope is suggestive of an
IR excess. As discussed in section \ref{sec:ancillary}, comparing to
$K_s$ is not optimal, but [3.6] is in many cases not available. 2-1066
seems to show a clear excess, with $K_s-[24] = 4.75\pm 0.06\vmag$. Though
it lacks a [3.6] measurement, the two IRAC bands which are measured
([4.5] and [8]) also lie above the extrapolated photosphere as can be
seen in the figure, supporting this conclusion (though it is also
possible that the excess may be caused by confusion from the secondary
in the system). The remainder show no detected excess, or at most, the
possibility of a marginal very weak excess. These results are not
surprising: by design, the age of the region in the field is such that
for the most part the disks have largely dissipated.

\input{pmsbinariestable}

\subsection{Identifying Low-mass Binaries}\label{sec:lowmass}

\begin{figure*}[tbp] \epsscale{0.7} 
    \plotone{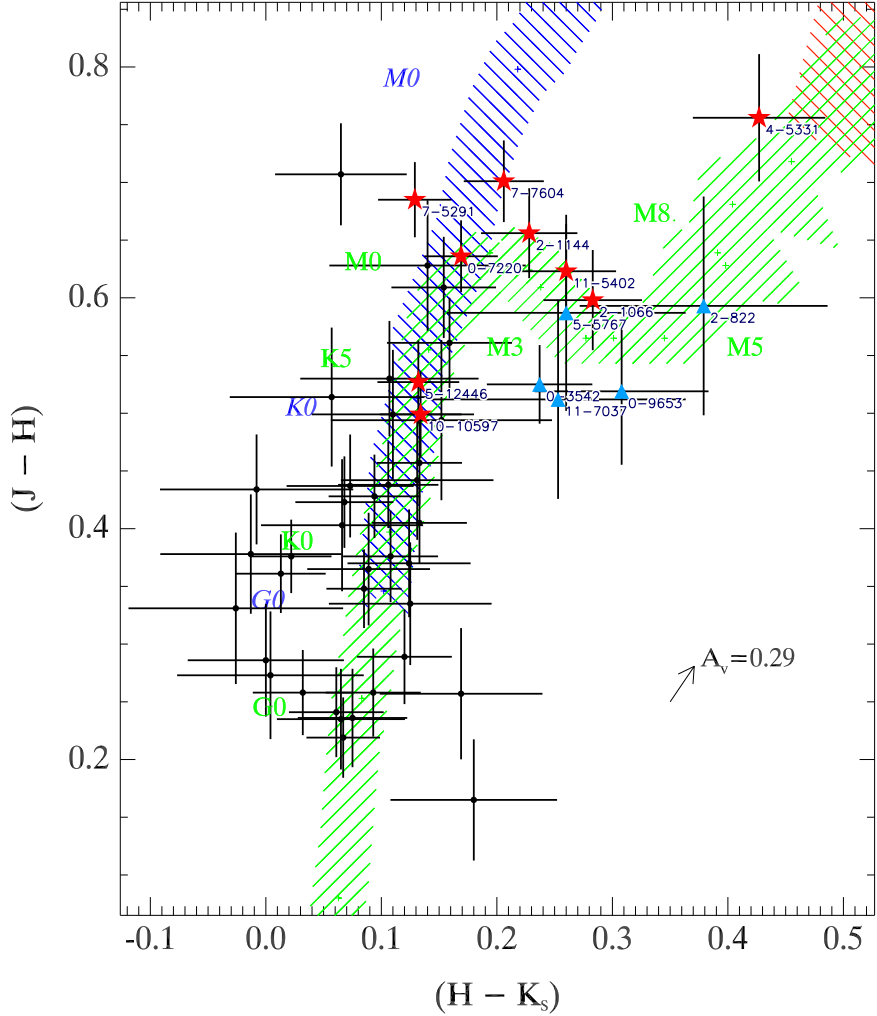}
    \caption{\label{fig:jhk}Color-color diagram for the new binaries
      based on 2MASS colors (section \ref{sec:lowmass}). The green
      hatched region marks the main sequence; the blue region marks
      the giant branch; and the red marks the beginning of the L dwarf
      regime. The assumed mean reddening vector for the field is
      indicated. The candidate young binary systems are marked with a
      star symbol as in previous figures. In addition, five other
      potential low mass binary systems are marked with triangles
      on the basis of their location in the diagram.}
\end{figure*}

In figure \ref{fig:jhk} we show a $JHK_s$ color-color diagram plotting
all of the 82 eclipsing binary systems for which good quality
photometry are found in the 2MASS point source catalog. Also indicated
are the regions encompassed by the main sequence and the giant
branch. The nine young binary candidates are highlighted with star
symbols as before.

From the figure we can see that, with the exception of the two contact
binaries, 5-12446 and 10-10597, all of the previously
discussed candidate young systems also have 2MASS colors that --
assuming no major reddening -- are consistent with M dwarf
primaries. These systems are therefore consistent with being both
young \emph{and} low mass.

In addition to the candidate young systems already discussed, we also
mark on the diagram the five other systems which fit in or near the
M-dwarf region of the plot, but are not necessarily young or
associated with Orion: 0-3542, 0-9653, 2-822, 5-5767, and 11-7037, of
which three (0-3542, 0-9653, and 11-7037) are classified as `close'
systems. These five systems are also broken out at the beginning of
table \ref{tbl:binaries}, and are discussed in sections
\ref{sec:otherlowmass} and \ref{sec:otherclose}.

Dereddening any of these systems by the mean assumed reddening,
indicated on the plot, would make little difference to their
status. For the young-system candidates, it only brings the sources
closer to the main sequence. However, it remains possible that some
sources are either obscured by tight localized regions of interstellar
dust, or intrinsically reddened objects. The small IR excesses in the
{\em Spitzer} data for 2-1066 (section \ref{sec:binariesspitzer}), and
possibly 7-7604, may be indicative of intrinsic reddening. It should 
be noted that the possibility of unresolved source confusion remains,
which could also lead to a distortion of the measured colors.


\subsection{Discussion}\label{sec:binariesdiscussion}

In the previous sections (\ref{sec:youngbinaries} and
\ref{sec:lowmass}) we discussed the approaches employed for
identifying particular sources of interest in the eclipsing binary
sample. We identified nine PMS candidate systems
(0-7220, 2-1066, 2-1144, 4-5331, 5-12446, 7-5291, 7-7604, 10-10597,
and 11-5402), of which two (5-12446 and 10-10597) appear to be contact
binary systems, one is also categorized as `close' but is somewhat
anomalous (7-5291), and the remainder are likely low-mass
systems. \citep[Contact binaries are not normally found as late as
M-type -- see, e.g.,][]{Rucinski2004WUmaDistances, Bilir2005}. We also
identified three notable very red contact-binaries (0-3542, 0-9653,
and 11-7037). In addition, two more systems (not previously discussed)
are particularly interesting: 11-8774, a contact binary with an
exceptionally short period, and 9-2980, a detached system which
appears to exhibit stellar pulsations. We summarize the results here
and discuss some of the specific systems. Since there is significant
overlap between the results from the analysis above, we break down
the specific sources identified into the four categories below, which
are reflected at the beginning of table \ref{tbl:binaries}.

\subsubsection{Young Low-mass Binary Candidates}\label{sec:younglowmass}
Six of the nine young candidate systems we found fell also into the
category of candidate low mass (M0 or later) systems: 0-7220, 2-1066,
2-1144, 4-5331, 7-7604, and 11-5402. These specific systems are
discussed in more detail below. The remaining three are close
binaries, discussed in section \ref{sec:youngclose} (though 7-5291 may
also qualify as low mass).

\begin{description}
\item[0-7220:] Presents an unusual light curve. The regular variation
  outside the primary and (small, $\sim1\%$) secondary eclipses is
  comparable with the depth of the primary eclipse itself (see figure
  \ref{fig:binaryLCs1}), and its phase is not consistent with
  ellipsoidal variation. It may be attributable to star spots: either
  one spot system with the star rotating at twice the orbital period,
  or more likely, two spot systems, separated by $\approx180\degr$ in
  longitude, rotating synchronously with the 0.680 day orbit. This
  would also be consistent with the long-term variation of the
  out-of-eclipse modulation during the length of the observations,
  which can be seen from the gray scale in the light curve plot
  (figure \ref{fig:binaryLCs1}).

\item[2-1066:] Although coverage for this object was good, only one
  eclipse was observed (figure \ref{fig:binaryLCs1}, bottom right), so
  the period is indeterminate. The eclipse is flat-bottomed, making it
  a noteworthy source for further observations. Out-of-eclipse
  variation is on the level of $\approx4\%$. It is also notable for its
  significant apparent IR excess, with $K_s - [24] = 4.75\pm0.06\vmag$
  (section \ref{sec:binariesspitzer} and figure
  \ref{fig:binariesspitzer}), which is suggestive of a disk (although it may
  also be attributable to the secondary companion).

\item[2-1144:] Displays a similar light curve to 0-7220, though a secondary
  eclipse is not clearly evident in the data, and the out-of-eclipse
  variation peaks only once per orbit. Again, this is not consistent
  with ellipsoidal variation, and is likely attributable to star
  spot/hot spot variation with the stellar rotation period synchronous
  with the 0.554 day orbital period.

\item[4-5331:] Only part of one eclipse was captured for this object
  (figure \ref{fig:binaryLCs2}), and therefore the period is not
  constrained, although coverage was good. The rest of the light curve
  is variable with an amplitude of $\sim 0.15\vmag$, with no clear period
  (figure \ref{fig:unfoldedBinaries1}). It is particularly noticeable
  for its very red colors, located as late as M9 in figure
  \ref{fig:jhk}; it is likely significantly reddened since if it had
  such a low mass it would appear much fainter at the distance of
  Orion. No coverage was found in the {\em Spitzer} data, but this reddening
  could be intrinsic, perhaps due to a disk. Though it seems likely
  that a stellar eclipse is the best explanation, it remains possible
  that the sudden drop in brightness is not in fact an eclipse from a
  binary companion, but that 4-5331 is a TTS, perhaps a
  `dipper'-like CTTS similar to 9-6886, shown in figure
  \ref{fig:cttslc4} (section \ref{sec:ctts}).

\item[7-7604:] Shows a small but significant level of flaring
  consistent with its proposed young age and a low mass -- hence the
  apparent discontinuity in the folded light curve (figure
  \ref{fig:binaryLCs3}). The object shows a possible very weak IR
  excess in the {\em Spitzer} data at $K_s$ only, with $K_s-[24] = 2.00\pm
  0.11\vmag$ and $[3.6]-[24] = 1.79\pm 0.13\vmag$ (table
  \ref{tbl:binariesspitzer} and figure
  \ref{fig:binariesspitzer}). 7-7604 lies closer to the giant branch
  than the main sequence in the $JHK_s$ color-color diagram (figure
  \ref{fig:jhk}), but if it is intrinsically reddened, this would
  bring its true color closer to the $\approx$M0 main sequence
  region. The apparent broadness of the eclipses suggests quite a
  closely separated system, with significant tidal
  distortion. Similarly to 7-5291 (see below, section
  \ref{sec:youngclose}), its short 2.07 day period argues against
  it being a giant system, since a giant would be larger than the
  implied orbital separation.

\item[11-5402:] Shows clear, roughly sinusoidal out-of-eclipse
  variation (figure \ref{fig:binaryLCs5}), again most likely due to
  star spots given the phase of the variation relative to the
  eclipse. Only one eclipse was observed for the object. The eclipse
  has good coverage despite being narrow, and is shown in more detail
  in figure \ref{fig:11-5402eclipse}, but the period of the binary is
  therefore not unambiguously determined. We take the period of the
  out-of-eclipse variations as the best guess of the orbital period,
  on the assumption of star spots co-rotating with the orbital period
  as for 0-7220 and 2-1144 (though the period here is longer, at
  5.5 days).

  \begin{figure}[tbp] \epsscale{0.6}
    \plotone{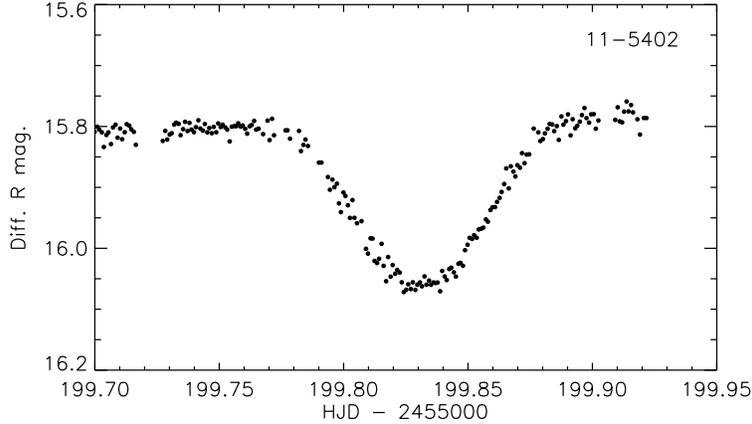}
    \caption{\label{fig:11-5402eclipse} Stretched view of the light curve for
      11-5402, showing the single eclipse observed for the object. The
      period for this system is estimated from the periodic out-of-eclipse
      variation (see section \ref{sec:younglowmass}). Error bars are
      omitted for clarity.}
  \end{figure}

\end{description}

\subsubsection{Other Low-mass Binary Candidates}\label{sec:otherlowmass}
We identified two other detached binary systems as low-mass candidates
on the basis of the 2MASS $JHK_s$ color-color selection (figure \ref{fig:jhk},
section \ref{sec:lowmass}). Given their distance estimates, and the
fact that they do not appear in the PMS region of the
color-magnitude diagram (figure \ref{fig:RvsR-K}), they are most
likely background systems unconnected with the Orion association.

\begin{description}
\item[2-822:] Shows a straightforward grazing-transit detached light
  curve, with out-of-eclipse variation suggesting longitudinal
  temperature variation on one of the components, synchronous with the
  0.9204 days orbital period (figure \ref{fig:binaryLCs1}). The
  $JHK_s$ colors suggest a spectral type of $\approx$M5V (though see
  the caveat below). Dereddening by the assumed mean reddening vector
  makes little difference to this estimation; dereddening by a greater
  amount would move the system off the main sequence, unless it is a
  late F-dwarf system, which would require a rather extreme reddening
  ($A_V\approx 5\vmag$).

\item[5-5767:] Only two grazing eclipses were detected for this
  detached system, leaving a `factors-of-two' ambiguity in the period,
  which is resolved by the periodic out-of-eclipse variation, if we
  assume that it is synchronous with the orbital period. Folding on
  the resulting 2.105 day period suggests that both are primary (or
  both secondary) eclipses. From the phase coverage (figure
  \ref{fig:binaryLCs2}) we see that the other eclipse was missed,
  assuming it would have fallen at phase 0.5. The out-of-eclipse
  variation appears to be caused by star spots. The $JHK_s$ colors
  suggest a spectral type of $\approx$M3V, unless it is a heavily
  reddened G-type system, with $A_V\approx 2.5\vmag$. Again, see the
  caveat below, however.

\end{description}

In both cases, we note that given the sizes of the their respective
2MASS measurement errors, the spectral type estimates are quite
uncertain. The same two systems are labeled in the color-magnitude
diagram in figure
\ref{fig:RvsR-K}. At face value, their $R-K_s$ values suggest somewhat
earlier spectral types, around late K to M0 depending on
reddening. However, the difference in epochs for the $R$ and $K_s$
observations leads also to large errors in $R-K_s$, as reflected in the
error bars; given the errors in both the color-magnitude and the
color-color diagrams, the measurements are not inconsistent. The true
spectral types likely lie somewhere between the two values, around
early M type. Followup spectroscopy is needed to provide a better estimate.

\subsubsection{Young `Close' Binary Candidates}\label{sec:youngclose}

The majority of the `close'-type binary systems in our sample appear
most likely to be contact binaries from the shape of their light
curves. Two W~UMa systems are identified in our binary sample as
candidate young 25 Ori/OB1a-association members. It remains possible
that their colors are distorted by unresolved third sources within the
PSF, which could confuse their apparent young status, but if their
ages are in indeed in the $7-10\Myr$ range, then there would be
significant constraints on the mechanisms by which they could have
formed, and they may merit further investigation. A third candidate
young `close' system is less likely to be a contact binary, but shows
some unusual properties.

\begin{description}
\item[5-12446:] Shows a straightforward W~UMa-type light curve (figure
  \ref{fig:binaryLCs3}), with period 0.23663 days. This system is an
  outlier in proper-motion space, with a proper motion of
  $30.0\pm3.8\masyr$ in right ascension (figure
  \ref{fig:propermotions}). It is probably an interloper; given
  its location somewhat below the cluster sequence in figure
  \ref{fig:RvsR-K}, it could also be an unrelated system with a mass
  ratio near unity that has been shifted upward in the color magnitude
  diagram due to the resulting luminosity enhancement, as discussed in
  section \ref{sec:CMDselection}. The expected location of such
  contact-systems at young ages with respect to single-star isochrones
  is uncertain, however, and the system's estimated distance
  ($317\pm41\pc$) is extremely close to that of the 25 Ori cluster.
  Even if this is an underestimate due to a near-unity mass ratio
  (section \ref{sec:distances}), it may still lie in the range
  of the broader Orion association. The system was previously
  identified as a candidate variable star by \citet{MOTESS-GNAT}.

\item[7-5291:] This object presents a peculiar light curve. Classified
  as a `close' type system under our definition, it has the appearance
  of a contact binary system, but with an unusual form. Assuming the
  two minima are indeed eclipses, then the width of the eclipses alone
  would suggest anything from a contact to a semi-detached
  system. However, the maxima and minima are rather sharp, and the
  primary ingress and egress are somewhat asymmetric. The period of
  the system, 3.62 day, is much longer than that expected for a W
  UMa system, and appears as an outlier in figure
  \ref{fig:PeriodHistogram}. The 2MASS $JHK_s$ colors are consistent with
  a mid- to late-K-type giant or possibly a late K or M0 dwarf primary
  (see figure \ref{fig:jhk}), but the period does not allow for either
  of these: a dwarf system would not even be significantly tidally
  distorted at such an orbital separation ($\approx 10\Rsun$), and
  therefore could not be near contact; a giant would be larger than
  the orbital separation. This suggests a subgiant-sized primary.

  One possibility is that distorted shape of the light curve results
  from a semi-detached (accreting) system with a disk. A semi-detached
  system would also be more consistent with the marked difference
  between the two eclipse depths, indicative of a relatively large
  temperature difference. If we assume that the system cannot be a
  contact binary owing to its period, then using the alternative
  (detached) distance calibration puts it at a distance of $268\pm
  64\pc$, which is $\approx 1\sigma$ from the 25~Ori/OB1a distance. A
  $10\Myr$-old inflated young low-mass primary, however, still would
  not be large enough. An alternative explanation for the light curve
  may be periodic occultation by a warped disk, in an analogue to AA
  Tau \citep{Bouvier2003}, similar to the objects reported by
  \citet{Morales2011}. The available (single-epoch) {\em Spitzer} photometry
  of this object (figure \ref{fig:binariesspitzer}, section
  \ref{sec:binariesspitzer}), however, shows at most only a very
  marginal IR excess at $24\um$, with $K_s-[24] = 0.84\pm0.12$ and
  $[3.6]-[24]= 0.73\pm0.13\vmag$ (table \ref{tbl:binariesspitzer}). The
  estimated spectral type is on the border between K and M, however,
  and source confusion at the relatively low spatial resolution of the
  MIPS data is a potential concern. Additional followup is needed to
  better constrain the spectral type and $A_V$ and better understand
  this object.

\item[10-10597:] Again shows a straightforward W~UMa-type light curve (figure
  \ref{fig:binaryLCs5}), with a period of 0.29953 days. {\em Spitzer} data
  were found for this object showing no IR excess; this is
  unsurprising, as we might expect any disk to be disrupted in such a
  system. Like 5-12446, this system also lies somewhat below the
  cluster sequence in figure \ref{fig:RvsR-K}, with a similar
  distance estimate ($348\pm45\pc$), so the same comments with regard
  to possible near-unity mass ratio may apply.

\end{description}

\subsubsection{Other Binaries of Interest}\label{sec:otherclose}

Among the remaining binaries, we identified a further five which are not
necessarily young or associated with Orion, but which are noteworthy
for other reasons.

\begin{description}

\item[0-3542, 0-9653, and 11-7037:] In the $JHK_s$ color-color diagram
  (figure \ref{fig:jhk}), while all the other low-mass (M0V and later)
  systems are classified as `detached' and lie on or very close to the
  main sequence, these three alone lie somewhat below the M-dwarf
  branch, and show W~UMa-like light curves. If they are genuine
  contact binaries and not near-contact or semi-detached systems,
  their location on the diagram suggests that they are either
  anomalously low-mass systems \citep[W~UMa systems are not normally
  found as late as M-type -- see,
  e.g.][]{Rucinski2004WUmaDistances,Bilir2005}, or probably more
  likely, particularly reddened W~UMa systems ($A_V\approx 2$--$3\vmag$)
  of earlier spectral type. In either case they may be interesting for
  further investigation. Given their unusual colors, we note that they
  lie outside the formal validity range of the
  \citet{Eker2009WUMaCalibration} calibrations, and their respective
  distance estimates may be unreliable. 0-9653 was previously
  identified as a candidate variable star by \citet{MOTESS-GNAT}.

\item[9-2980:] Shows small but pronounced $\sim 0.01\vmag$, $\approx
  45\min$ modulations superimposed on the eclipsing binary light
  curve. Stellar pulsations seem the most likely cause. The 2MASS
  colors of the source imply a spectral type of $\sim$G0, and the
  significant out-of-eclipse variations suggest both spot modulation
  and ellipsoidal variation indicative of a tidally distorted
  system. More observations are needed to sample the secondary
  eclipse.

\item[11-8774:] Figure \ref{fig:PeriodHistogram} shows a histogram of
  the periods of all the binaries from table \ref{tbl:binaries},
  highlighting the distribution of the `close' type systems. W UMa
  systems have a known sharp cutoff in the period distribution at the
  short-period end below around 0.22 days, just below a peak in the
  distribution \citep[see, e.g.,][]{Rucinski2007ShortPeriodWUMas}. The
  cause of this cutoff remains largely unclear. We have not attempted
  to correct for the effects of the observing windowing function or
  magnitude completeness effects,
  but it is still clear that a significant number of our targets lie
  near the short-period end of this distribution, which is not
  currently well sampled \citep{Rucinski2007ShortPeriodWUMas}. In
  particular, the shortest-period source, 11-8774, has a period of
  $0.2156509 \pm 0.0000071$ days, putting it among the most rapidly
  orbiting W UMa systems known.
  11-8774 also displays clear flat-bottomed primary eclipses, meaning
  the eclipses are total rather than grazing (figure
  \ref{fig:11-8774eclipse}). This makes the object a particularly good
  target for constraining parameters via light curve modeling. The
  derived distance of 11-8774, $450\pm 60\pc$, may be too distant for
  the 25 Ori cluster and the larger Orion OB1a association ($\approx
  330\pc$), but is consistent with that of the more distant
  \citep[$\approx 440\pc$;][]{Briceno2005,Brown1999} -- and younger --
  OB1b association. It could perhaps be an outlier of the OB1b
  association, although at such a young age, it would be far from the
  main-sequence, and the distance estimate would likely be
  systematically wrong. If, however, the system does lie within the
  general distance of the broader Orion association, it could
  constitute a young W UMa system. As such it would be useful in
  understanding and constraining the formation and evolution of such
  systems.

  \begin{figure}[tbp] \epsscale{0.7}
    \plotone{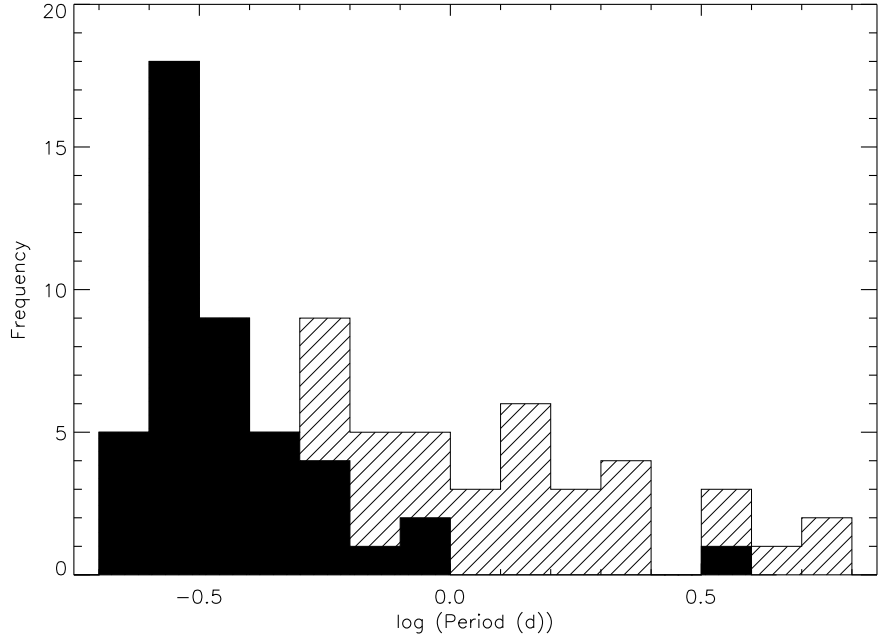}
    \caption{\label{fig:PeriodHistogram}Orbital period histogram for
      all eclipsing binaries for which a period estimate could be
      found (section \ref{sec:otherclose}). The solid regions
      indicate the fraction which are `close' (probably mostly
      contact binary) sources; the hatched regions indicate the
      remaining detached sources. At the shortest end of the
      distribution is 11-8774, with among the shortest W~UMa periods
      known (section \ref{sec:otherclose}). The outlying `close'
      source with a long period at just above log(period) = 0.5 is due
      to object 7-5291, which presents a somewhat unusual light curve
      (see section \ref{sec:youngclose}).}
  \end{figure}

  \begin{figure}[tbp] \epsscale{0.6}
    \plotone{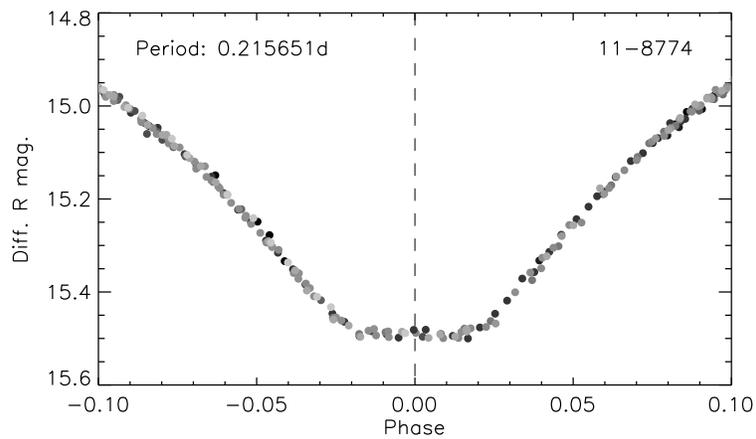}
    \caption{\label{fig:11-8774eclipse} Stretched view of the folded light
      curve for the very short period W~UMa system, 11-8774, centered
      on the primary eclipse (see section \ref{sec:otherclose}.). The
      flat bottom of the eclipse can clearly be seen. The gray-scale
      is again used to represents time of observation, running from
      dark gray at the beginning of the PTF Orion run to pale gray at
      the end of the run. The secondary eclipse (not shown) is also
      flat-bottomed. Error bars are omitted for clarity.}
  \end{figure}

\end{description}

%% file: binariestable.tex
%

\\
\sidehead{Pre-main-sequence Low-mass Binary Candidates}
0-7220                      & 80.26648 & 2.67426 & J05210397+0240275 & $12.69\pm 0.02$ & $12.06\pm 0.02$ & $11.89\pm 0.02$ & D & $0.679631\pm 0.001025$ & $205.83808\pm 0.00025$ & $349 \pm 83   $ & 14.58 & 0.06\\
2-1066\tablenotemark{a}     & 81.49801 & 1.96661 & J05255953+0157593 & $13.64\pm 0.03$ & $13.04\pm 0.03$ & $12.76\pm 0.03$ & D & \nodata                & $192.77787\pm 0.00050$ & $428 \pm 109  $ & 17.06 & 0.58\\
2-1144                      & 81.45159 & 1.97448 & J05254838+0158276 & $13.09\pm 0.03$ & $12.43\pm 0.03$ & $12.20\pm 0.03$ & D & $0.554431\pm 0.001839$ & $201.70360\pm 0.00075$ & $337 \pm 83   $ & 16.53 & 0.14\\
4-5331\tablenotemark{b}     & 82.40622 & 2.35695 & J05293750+0221251 & $14.36\pm 0.04$ & $13.60\pm 0.04$ & $13.17\pm 0.04$ & D & \nodata                & $189.94145\pm 1.00000$ & $270 \pm 72   $ & 17.75 & 0.87\\
7-7604\tablenotemark{c}     & 80.77483 & 1.64755 & J05230596+0138511 & $13.09\pm 0.03$ & $12.39\pm 0.02$ & $12.18\pm 0.03$ & D & $2.071092\pm 0.009377$ & $211.69810\pm 0.00158$ & $322 \pm 78   $ & 16.18 & 0.23\\
11-5402\tablenotemark{d}    & 83.18037 & 1.31577 & J05324332+0118566 & $13.44\pm 0.03$ & $12.81\pm 0.03$ & $12.55\pm 0.03$ & D & $5.505740\pm 0.355271$ & $199.83101\pm 0.00032$ & $391 \pm 98   $ & 15.79 & 0.29\\
                                                                                                                                                                                                
\sidehead{Other Low-mass Candidates}                                                                                                                                                            
2-822                       & 81.59643 & 1.93894 & J05262316+0156196 & $15.26\pm 0.07$ & $14.67\pm 0.07$ & $14.29\pm 0.08$ & D & $0.920431\pm 0.000166$ & $180.83174\pm 0.00012$ & $694 \pm 245  $ & 16.82 & 0.89\\
5-5767                      & 83.10952 & 2.38606 & J05322632+0223097 & $15.16\pm 0.06$ & $14.58\pm 0.06$ & $14.32\pm 0.08$ & D & $2.104862\pm 0.002494$ & $211.71483\pm 0.00044$ & $946 \pm 326  $ & 16.73 & 0.33\\
                                                                                                                                                                                                
\sidehead{Pre-main-sequence Close-Binary Candidates}                                                                                                                                            
5-12446\tablenotemark{o}    & 83.34683 & 2.97482 & J05332318+0258281 & $11.99\pm 0.03$ & $11.47\pm 0.02$ & $11.34\pm 0.03$ & C & $0.236633\pm 0.000017$ & $204.65544\pm 0.00003$ & $317 \pm 41   $ & 13.46 & 0.39\\
7-5291\tablenotemark{e}     & 80.80171 & 1.35431 & J05231240+0121151 & $12.13\pm 0.02$ & $11.44\pm 0.02$ & $11.31\pm 0.02$ & C & $3.623108\pm 0.023858$ & $192.86419\pm 0.00265$ & $1442\pm 183  $ & 13.95 & 0.11\\
10-10597                    & 82.53469 & 1.81344 & J05300832+0148489 & $11.77\pm 0.03$ & $11.27\pm 0.03$ & $11.13\pm 0.02$ & C & $0.299531\pm 0.000050$ & $192.72915\pm 0.00008$ & $348 \pm 45   $ & 12.96 & 0.18\\
                                                                                                                                                                                                
\sidehead{Other Binaries of Interest}                                                                                                                                                           
0-3542                      & 80.00204 & 2.25436 & J05200049+0215159 & $14.27\pm 0.03$ & $13.74\pm 0.02$ & $13.51\pm 0.04$ & C & $0.284425\pm 0.000077$ & $199.77639\pm 0.00023$ & $934 \pm 122  $ & 15.62 & 0.28\\
0-9653\tablenotemark{o}     & 80.29957 & 2.93869 & J05211191+0256189 & $14.70\pm 0.05$ & $14.18\pm 0.04$ & $13.88\pm 0.06$ & C & $0.910619\pm 0.001313$ & $202.84000\pm 0.10000$ & $2273\pm 339  $ & 16.23 & 0.29\\
9-2980\tablenotemark{l}     & 81.86439 & 0.98081 & J05272748+0058506 & $14.39\pm 0.04$ & $14.12\pm 0.04$ & $14.11\pm 0.07$ & D & $1.190701\pm 0.003662$ & $175.79416\pm 0.00062$ & $2923\pm 900  $ & 14.87 & 0.07\\
11-7037                     & 82.87870 & 1.50007 & J05313088+0130004 & $15.01\pm 0.05$ & $14.50\pm 0.07$ & $14.25\pm 0.09$ & C & $0.306948\pm 0.000228$ & $199.74559\pm 0.00116$ & $1395\pm 235  $ & 16.09 & 0.13\\
11-8774\tablenotemark{f}    & 83.32166 & 1.68053 & J05331720+0140498 & $13.23\pm 0.03$ & $12.62\pm 0.03$ & $12.47\pm 0.03$ & C & $0.215651\pm 0.000007$ & $180.94621\pm 0.00004$ & $450 \pm 60   $ & 14.83 & 0.78\\
                                                                                                                                                                                                
\sidehead{Other Binaries}                                                                                                                                                                       
0-1580                      & 79.94125 & 2.02664 & J05194592+0201361 & $14.35\pm 0.04$ & $13.91\pm 0.04$ & $13.78\pm 0.05$ & D & $2.061758\pm 0.001051$ & $198.80653\pm 0.00012$ & $1334\pm 376  $ & 15.56 & 0.52\\
0-4268                      & 80.46153 & 2.33903 & J05215079+0220205 & $13.03\pm 0.03$ & $12.67\pm 0.02$ & $12.66\pm 0.03$ & D & $5.167944\pm 0.176421$ & $202.79638\pm 0.00003$ & $1233\pm 307  $ & 13.90 & 0.70\\
0-4407                      & 80.17316 & 2.35519 & J05204154+0221186 & $14.60\pm 0.03$ & $14.07\pm 0.04$ & $13.97\pm 0.07$ & D & $0.701540\pm 0.001087$ & $189.75110\pm 0.00039$ & $1296\pm 389  $ & 16.07 & 0.15\\
0-5197                      & 80.45244 & 2.44664 & J05214859+0226480 & $16.20\pm 0.13$ & $15.42\pm 0.11$ & $15.60\pm 0.25$ & C & $0.273236\pm 0.000064$ & $198.80522\pm 0.00047$ & $2006\pm 627  $ & 17.24 & 0.80\\
0-5673                      & 80.06399 & 2.50293 & J05201536+0230104 & $14.91\pm 0.05$ & $14.39\pm 0.04$ & $14.34\pm 0.08$ & D & $0.543151\pm 0.000185$ & $175.83572\pm 0.00037$ & $1786\pm 602  $ & 15.93 & 0.19\\
0-8036\tablenotemark{g}     & 80.03662 & 2.76290 & 0927-0078105      & \nodata         & \nodata         & \nodata         & D & $0.597359\pm 0.000604$ & $204.70458\pm 0.00048$ & \nodata         & 18.09 & 0.75\\
0-8177                      & 80.28231 & 2.77769 & J05210774+0246397 & $15.98\pm 0.08$ & $15.82\pm 0.14$ & $>15.21$        & C & $0.304414\pm 0.000214$ & $204.74716\pm 0.00047$ & \nodata         & 17.11 & 0.25\\
0-9790                      & 80.42457 & 2.95301 & J05214195+0257109 & $15.96\pm 0.10$ & $15.77\pm 0.14$ & $15.49\pm 0.23$ & C & $0.326311\pm 0.000147$ & $199.72763\pm 0.00036$ & $3842\pm 1112 $ & 17.12 & 0.43\\
1-1223                      & 80.94578 & 2.00642 & J05234700+0200225 & $15.26\pm 0.05$ & $14.80\pm 0.07$ & $14.87\pm 0.13$ & D & $1.252181\pm 0.000334$ & $192.80870\pm 0.00012$ & $3488\pm 1528 $ & 15.91 & 1.59\\
1-2659                      & 80.80171 & 2.18552 & J05231240+0211079 & $15.99\pm 0.11$ & $15.90\pm 0.19$ & $15.12\pm 0.16$ & D & $0.960161\pm 0.002785$ & $189.89204\pm 0.00025$ & $1045\pm 585  $ & 16.67 & 0.59\\
1-3638\tablenotemark{a}     & 81.08221 & 2.31406 & J05241975+0218507 & $15.68\pm 0.07$ & $15.46\pm 0.12$ & $15.43\pm 0.20$ & D & \nodata                & $180.78801\pm 0.00028$ & $5650\pm 3628 $ & 16.63 & 0.43\\
1-4581\tablenotemark{j}     & 80.66826 & 2.43163 & J05224038+0225541 & $14.48\pm 0.03$ & $14.11\pm 0.04$ & $14.12\pm 0.07$ & D & $1.870680\pm 0.002107$ & $175.74000\pm 0.10000$ & $2489\pm 760  $ & 15.44 & 0.17\\
1-6477                      & 80.54039 & 2.66092 & J05220972+0239392 & $13.30\pm 0.03$ & $13.06\pm 0.03$ & $12.98\pm 0.03$ & C & $0.779850\pm 0.001626$ & $202.77152\pm 0.00119$ & $2159\pm 284  $ & 14.24 & 0.05\\
2-1985                      & 81.48699 & 2.06813 & J05255688+0204050 & $15.27\pm 0.05$ & $15.12\pm 0.10$ & $14.91\pm 0.14$ & C & $0.303446\pm 0.000047$ & $202.78324\pm 0.00015$ & $3083\pm 630  $ & 16.47 & 0.55\\
2-4114                      & 81.23841 & 2.30490 & J05245720+0218173 & $16.02\pm 0.09$ & $15.88\pm 0.18$ & $>15.29 $       & D & $3.519536\pm 0.005422$ & $189.81694\pm 0.00034$ & \nodata         & 16.93 & 0.45\\
4-827                       & 82.72646 & 1.93051 & J05305436+0155497 & $12.78\pm 0.03$ & $12.40\pm 0.03$ & $12.29\pm 0.03$ & C & $0.322230\pm 0.000100$ & $199.75955\pm 0.00006$ & $738 \pm 96   $ & 13.65 & 0.35\\
4-7558                      & 82.67281 & 2.56394 & J05304150+0233503 & $15.19\pm 0.06$ & $14.83\pm 0.07$ & $14.66\pm 0.10$ & C & $0.258346\pm 0.000040$ & $204.70188\pm 0.00013$ & $1904\pm 337  $ & 16.31 & 0.42\\
4-8796                      & 82.58078 & 2.67704 & J05301942+0240378 & $16.14\pm 0.12$ & $16.13\pm 0.23$ & $>15.51 $       & C & $0.397136\pm 0.000170$ & $198.83439\pm 0.00056$ & \nodata         & 17.04 & 0.41\\
4-9573                      & 82.76732 & 2.74706 & J05310413+0244486 & $16.01\pm 0.11$ & $15.55\pm 0.16$ & $14.91\pm 0.14$ & C & $0.264623\pm 0.000081$ & $201.84733\pm 0.00021$ & $1596\pm 424  $ & 17.22 & 0.48\\
4-11668\tablenotemark{h}    & 82.35476 & 2.93170 & J05292516+0255538 & $16.21\pm 0.12$ & $15.76\pm 0.15$ & $>15.63 $       & D & \nodata                & $175.75243\pm 0.00069$ & \nodata         & 17.64 & 0.75\\
5-295                       & 83.34093 & 1.88633 & J05332181+0153105 & $15.68\pm 0.09$ & $>15.32$        & $>15.66$        & C & $0.422437\pm 0.000242$ & $201.85629\pm 0.00035$ & \nodata         & 16.20 & 0.20\\
5-961                       & 82.91029 & 1.95100 & J05313843+0157027 & $12.91\pm 0.02$ & $12.65\pm 0.03$ & $12.62\pm 0.03$ & D & $1.331246\pm 0.000334$ & $169.97516\pm 0.00027$ & $1417\pm 353  $ & 13.87 & 0.48\\
5-4382                      & 83.05060 & 2.26285 & 0922-0084153      & \nodata         & \nodata         & \nodata         & C & $0.303254\pm 0.000128$ & $205.85421\pm 0.00060$ & \nodata         & 17.88 & 0.68\\
5-5669                      & 83.21871 & 2.37652 & J05325252+0222357 & $12.42\pm 0.02$ & $11.99\pm 0.03$ & $11.90\pm 0.03$ & C & $0.510305\pm 0.000534$ & $189.76562\pm 0.00054$ & $772 \pm 99   $ & 13.61 & 0.05\\
5-6432\tablenotemark{a}     & 82.90660 & 2.44429 & J05313758+0226397 & $11.67\pm 0.03$ & $11.32\pm 0.02$ & $11.24\pm 0.02$ & D & \nodata                & $189.72730\pm 0.00004$ & $554 \pm 134  $ & 12.67 & 0.49\\
5-8209                      & 83.10664 & 2.60591 & J05322562+0236214 & $14.88\pm 0.05$ & $14.39\pm 0.05$ & $14.24\pm 0.08$ & C & $0.295733\pm 0.000223$ & $189.87958\pm 0.00071$ & $1441\pm 228  $ & 16.50 & 0.14\\
5-8640\tablenotemark{i}     & 83.26033 & 2.64550 & J05330249+0238437 & $14.50\pm 0.04$ & $14.17\pm 0.05$ & $14.19\pm 0.08$ & C & $0.329931\pm 0.000178$ & $189.89227\pm 0.00017$ & $2001\pm 307  $ & 15.56 & 0.10\\
5-9211                      & 83.39732 & 2.69212 & J05333535+0241314 & $14.41\pm 0.04$ & $13.91\pm 0.04$ & $13.80\pm 0.06$ & C & $0.243963\pm 0.000012$ & $199.86100\pm 0.00009$ & $1052\pm 151  $ & 15.94 & 0.96\\
5-12287\tablenotemark{o}    & 83.34737 & 2.96147 & J05332334+0257401 & $12.45\pm 0.03$ & $11.99\pm 0.03$ & $11.86\pm 0.03$ & C & $0.287096\pm 0.000012$ & $204.75612\pm 0.00002$ & $500 \pm 64   $ & 13.79 & 0.75\\
6-574                       & 80.02016 & 0.77828 & J05200487+0046426 & $15.50\pm 0.05$ & $15.01\pm 0.07$ & $14.88\pm 0.13$ & C & $0.549560\pm 0.000737$ & $180.79434\pm 0.00120$ & $2893\pm 539  $ & 16.48 & 0.20\\
6-690\tablenotemark{j}      & 79.97470 & 0.79530 & J05195397+0047438 & $16.03\pm 0.10$ & $15.34\pm 0.09$ & $15.39\pm 0.18$ & D & $1.499817\pm 0.000954$ & $180.91794\pm 0.00043$ & $2672\pm 1611 $ & 17.37 & 0.94\\
6-2146\tablenotemark{c,}\tablenotemark{m}   & 79.97923 & 0.98111 & J05195505+0058522 & $12.06\pm 0.03$ & $11.85\pm 0.02$ & $11.78\pm 0.02$ & C & $0.964920\pm 0.002042$ & $199.85344\pm 0.00165$ & $1454\pm 186  $ & 12.80 & 0.05\\
6-3648                      & 80.38690 & 1.16913 & J05213285+0110085 & $14.58\pm 0.03$ & $14.15\pm 0.04$ & $14.16\pm 0.07$ & C & $0.270660\pm 0.000069$ & $201.76570\pm 0.00024$ & $1507\pm 218  $ & 15.40 & 0.20\\
6-4262                      & 80.51065 & 1.24251 & J05220252+0114330 & $16.10\pm 0.10$ & $15.91\pm 0.15$ & $15.46\pm 0.20$ & C & $0.328103\pm 0.000301$ & $180.90403\pm 0.00088$ & $3579\pm 1002 $ & 17.13 & 0.28\\
6-5196                      & 80.46700 & 1.35628 & J05215210+0121224 & $15.33\pm 0.05$ & $15.04\pm 0.09$ & $15.13\pm 0.17$ & C & $0.425248\pm 0.000221$ & $201.82021\pm 0.00044$ & $3887\pm 832  $ & 15.97 & 0.22\\
6-5331\tablenotemark{k}     & 80.29305 & 1.37240 & J05211032+0122202 & $13.40\pm 0.03$ & $12.96\pm 0.03$ & $12.85\pm 0.03$ & D & $1.242543\pm 0.003772$ & $201.69228\pm 0.00063$ & $930 \pm 233  $ & 14.55 & 0.08\\
6-7525                      & 80.29326 & 1.63339 & J05211040+0138003 & $13.77\pm 0.03$ & $13.53\pm 0.04$ & $13.47\pm 0.04$ & D & $1.569462\pm 0.000698$ & $189.87010\pm 0.00016$ & $2023\pm 530  $ & 14.50 & 0.28\\
6-9087                      & 80.50852 & 1.82028 & J05220211+0149141 & $14.28\pm 0.03$ & $13.72\pm 0.03$ & $13.56\pm 0.05$ & D & $1.351641\pm 0.000382$ & $180.74149\pm 0.00018$ & $894 \pm 242  $ & 15.73 & 0.43\\
7-337                       & 80.77330 & 0.74753 & J05230561+0044515 & $14.34\pm 0.03$ & $13.63\pm 0.04$ & $13.57\pm 0.04$ & D & $1.811381\pm 0.000233$ & $180.90122\pm 0.00009$ & $844 \pm 224  $ & 15.84 & 0.91\\
7-750                       & 80.95319 & 0.80217 & J05234877+0048078 & $14.41\pm 0.03$ & $14.25\pm 0.04$ & $14.07\pm 0.06$ & C & $0.402915\pm 0.000064$ & $199.66972\pm 0.00022$ & $2467\pm 351  $ & 15.04 & 0.38\\
7-1291                      & 80.72298 & 0.87163 & J05225353+0052179 & $15.99\pm 0.09$ & $>15.53$        & $>15.26$        & D & $3.790587\pm 0.023490$ & $180.75767\pm 0.00068$ & \nodata         & 17.03 & 0.65\\
7-1309                      & 80.92900 & 0.87349 & J05234296+0052244 & $12.96\pm 0.03$ & $12.67\pm 0.03$ & $12.55\pm 0.03$ & C & $0.363835\pm 0.000015$ & $185.84963\pm 0.00006$ & $1000\pm 131  $ & 13.98 & 0.83\\
7-3958                      & 80.92392 & 1.19610 & J05234174+0111454 & $15.45\pm 0.07$ & $15.07\pm 0.09$ & $14.67\pm 0.11$ & D & $0.706710\pm 0.000315$ & $205.73030\pm 0.00040$ & $1193\pm 497  $ & 16.67 & 0.45\\
7-8191                      & 81.06074 & 1.71670 & J05241461+0143004 & $12.16\pm 0.02$ & $11.78\pm 0.02$ & $11.76\pm 0.03$ & D & $0.937275\pm 0.000397$ & $199.87813\pm 0.00017$ & $776 \pm 188  $ & 13.16 & 0.09\\
8-1414                      & 81.67115 & 0.83737 & J05264107+0050147 & $13.30\pm 0.03$ & $12.88\pm 0.03$ & $12.81\pm 0.03$ & C & $0.281382\pm 0.000020$ & $199.83044\pm 0.00007$ & $820 \pm 107  $ & 14.19 & 0.43\\
8-8251                      & 81.17790 & 1.53097 & J05244268+0131514 & $15.99\pm 0.12$ & $15.80\pm 0.17$ & $>14.98$        & D & $0.718476\pm 0.000247$ & $211.64936\pm 0.00042$ & \nodata         & 17.66 & 0.89\\
8-11361                     & 81.57440 & 1.82366 & 0918-0060816      & \nodata         & \nodata         & \nodata         & C & $0.318375\pm 0.000106$ & $180.84402\pm 0.00037$ & \nodata         & 17.65 & 0.75\\
9-1695                      & 82.19286 & 0.85685 & J05284629+0051248 & $15.15\pm 0.04$ & $14.96\pm 0.07$ & $14.75\pm 0.12$ & C & $0.594296\pm 0.001069$ & $175.76614\pm 0.00116$ & $4098\pm 747  $ & 15.71 & 0.08\\
9-3480                      & 81.96793 & 1.02633 & J05275232+0101345 & $15.14\pm 0.05$ & $14.83\pm 0.08$ & $14.88\pm 0.14$ & C & $0.342832\pm 0.000113$ & $204.68358\pm 0.00041$ & $2921\pm 557  $ & 16.11 & 0.23\\
9-3757\tablenotemark{i}     & 82.15440 & 1.05077 & J05283707+0103024 & $15.11\pm 0.05$ & $14.83\pm 0.07$ & $14.74\pm 0.11$ & C & $0.585259\pm 0.000765$ & $201.80116\pm 0.00075$ & $3775\pm 655  $ & 15.80 & 0.13\\
9-4426                      & 81.81013 & 1.11631 & J05271446+0106585 & $13.22\pm 0.03$ & $12.82\pm 0.02$ & $12.69\pm 0.03$ & C & $0.239152\pm 0.000056$ & $201.86540\pm 0.00024$ & $698 \pm 91   $ & 14.28 & 0.10\\
9-4659                      & 81.86969 & 1.14085 & J05272875+0108267 & $15.92\pm 0.09$ & $15.50\pm 0.13$ & $15.49\pm 0.22$ & C & $0.297846\pm 0.000110$ & $180.84125\pm 0.00034$ & $3024\pm 833  $ & 17.00 & 0.42\\
9-6659                      & 82.18806 & 1.33594 & J05284514+0120089 & $14.33\pm 0.03$ & $13.93\pm 0.05$ & $13.86\pm 0.05$ & C & $0.310999\pm 0.000134$ & $204.68321\pm 0.00031$ & $1458\pm 207  $ & 15.40 & 0.19\\
9-11733                     & 82.00983 & 1.80036 & J05280235+0148015 & $15.93\pm 0.10$ & $15.80\pm 0.18$ & $>16.30$        & D & $0.602753\pm 0.000100$ & $192.72042\pm 0.00007$ & \nodata         & 16.75 & 1.06\\
10-607                      & 82.46420 & 0.77599 & J05295140+0046337 & $16.29\pm 0.14$ & $16.12\pm 0.19$ & $>15.84$        & C & $0.289616\pm 0.000200$ & $189.72307\pm 0.00075$ & \nodata         & 17.48 & 0.38\\
10-748\tablenotemark{j}     & 82.52387 & 0.79020 & J05300573+0047249 & $14.87\pm 0.04$ & $14.24\pm 0.04$ & $14.10\pm 0.07$ & D & $1.261511\pm 0.001564$ & $192.67356\pm 0.00200$ & $1051\pm 337  $ & 17.11 & 0.92\\
10-2406                     & 82.78685 & 0.96683 & J05310883+0058008 & $13.52\pm 0.03$ & $13.15\pm 0.03$ & $13.03\pm 0.04$ & C & $0.468237\pm 0.000143$ & $199.83276\pm 0.00016$ & $1313\pm 178  $ & 14.77 & 0.33\\
10-3009\tablenotemark{j}    & 82.49377 & 1.03404 & J05295852+0102022 & $13.13\pm 0.03$ & $12.89\pm 0.03$ & $12.83\pm 0.03$ & D & $2.234000\pm 0.006000$ & $202.82948\pm 0.00011$ & $1504\pm 370  $ & 13.99 & 0.63\\
10-5611                     & 82.57626 & 1.31356 & J05301832+0118484 & $14.23\pm 0.04$ & $13.89\pm 0.04$ & $13.77\pm 0.06$ & C & $0.308967\pm 0.000117$ & $180.87753\pm 0.00034$ & $1484\pm 211  $ & 15.17 & 0.14\\
10-5878                     & 82.81800 & 1.34175 & J05311632+0120299 & $13.63\pm 0.03$ & $13.26\pm 0.04$ & $13.17\pm 0.04$ & D & $1.466659\pm 0.000361$ & $199.66223\pm 0.00012$ & $1294\pm 339  $ & 14.64 & 0.64\\
10-7989                     & 82.71804 & 1.55714 & J05305236+0133253 & $14.03\pm 0.04$ & $13.78\pm 0.04$ & $13.61\pm 0.06$ & D & $3.690487\pm 0.008470$ & $204.88043\pm 0.00031$ & $1611\pm 458  $ & 14.64 & 1.26\\
11-1051\tablenotemark{j}    & 82.94454 & 0.82398 & J05314668+0049260 & $15.91\pm 0.10$ & $15.62\pm 0.12$ & $15.57\pm 0.23$ & C & $0.372703\pm 0.000221$ & $201.79040\pm 0.00050$ & $4185\pm 1216 $ & 16.68 & 0.27\\
11-3313                     & 83.03217 & 1.07005 & J05320774+0104117 & $14.06\pm 0.03$ & $13.62\pm 0.03$ & $13.55\pm 0.04$ & C & $0.281535\pm 0.000019$ & $180.87347\pm 0.00006$ & $1128\pm 152  $ & 15.33 & 0.70\\
11-3778                     & 83.06084 & 1.12772 & 0911-0062767      & \nodata         & \nodata         & \nodata         & C & $0.227754\pm 0.000108$ & $180.90388\pm 0.00060$ & \nodata         & 18.40 & 0.91\\
11-7429                     & 83.03414 & 1.54127 & J05320820+0132287 & $12.56\pm 0.03$ & $12.30\pm 0.03$ & $12.21\pm 0.03$ & D & $2.034245\pm 0.002301$ & $211.73886\pm 0.00010$ & $1014\pm 252  $ & 13.38 & 0.32\\
11-10141                    & 83.42733 & 1.82439 & J05334253+0149276 & $14.25\pm 0.04$ & $13.96\pm 0.03$ & $13.96\pm 0.06$ & C & $0.456071\pm 0.000184$ & $201.79630\pm 0.00022$ & $2319\pm 327  $ & 14.95 & 0.19\\
                                                                                                                                                                                                

%% file: pmsbinariestable.tex
\begin{deluxetable}{lrrrrrrrrrr}
\tablecolumns{11}
\tablewidth{0pt}
\tablecaption{\label{tbl:binariesspitzer}Existing {\em Spitzer} Data
  for Candidate Young Binaries}
\tabletypesize{\scriptsize}
\rotate
\tablehead{
ID & \colhead{$J$} & \colhead{$H$} & \colhead{$K_s$} & \colhead{[3.6]} & \colhead{[4.5]} & \colhead{[5.8]} & \colhead{[8]} & \colhead{[24]} & \colhead{$K_s-[24]$} & \colhead{$[3.6]-[24]$}
}

\startdata

2-1066   & 13.64$\pm$0.03 & 13.04$\pm$0.03 & 12.76$\pm$0.03 & \nodata        & 11.84$\pm$0.06 & \nodata        & 11.24$\pm$0.06 & 8.01 $\pm$0.05  &  4.75$\pm$0.06  &  \nodata       \\
2-1144   & 13.09$\pm$0.03 & 12.43$\pm$0.03 & 12.20$\pm$0.03 & \nodata        & 11.82$\pm$0.06 & \nodata        & 11.77$\pm$0.06 & 11.68$\pm$0.34  &  0.52$\pm$0.34  &  \nodata       \\
7-5291   & 12.13$\pm$0.02 & 11.44$\pm$0.02 & 11.31$\pm$0.02 & 11.20$\pm$0.06 & \nodata        & 11.21$\pm$0.06 & \nodata        & 10.47$\pm$0.12  &  0.84$\pm$0.12  &  0.73$\pm$0.13 \\
7-7604   & 13.09$\pm$0.03 & 12.39$\pm$0.02 & 12.18$\pm$0.03 & 11.97$\pm$0.06 & 11.87$\pm$0.06 & 11.87$\pm$0.06 & 11.89$\pm$0.06 & 10.18$\pm$0.11  &  2.00$\pm$0.11  &  1.79$\pm$0.13 \\
10-10597 & 11.77$\pm$0.03 & 11.27$\pm$0.03 & 11.14$\pm$0.02 & 10.93$\pm$0.06 & 10.94$\pm$0.06 & 10.92$\pm$0.06 & 10.92$\pm$0.06 & 11.48$\pm$0.20  & -0.34$\pm$0.20  & -0.55$\pm$0.21\\

\enddata

\tablecomments{All candidate young binaries are listed for which
  {\em Spitzer} IRAC/MIPS data could be found (section
  \ref{sec:binariesspitzer}). 2MASS $J,H$, and $K_s$ measurements are
  repeated from table \ref{tbl:binaries} for comparison. All values
  are in magnitudes.}

\end{deluxetable}

%% file: ctts.tex
\clearpage
\section{CTTS Sources}\label{sec:ctts}

\begin{figure*}[tbp] \epsscale{1.0} 
    \plotone{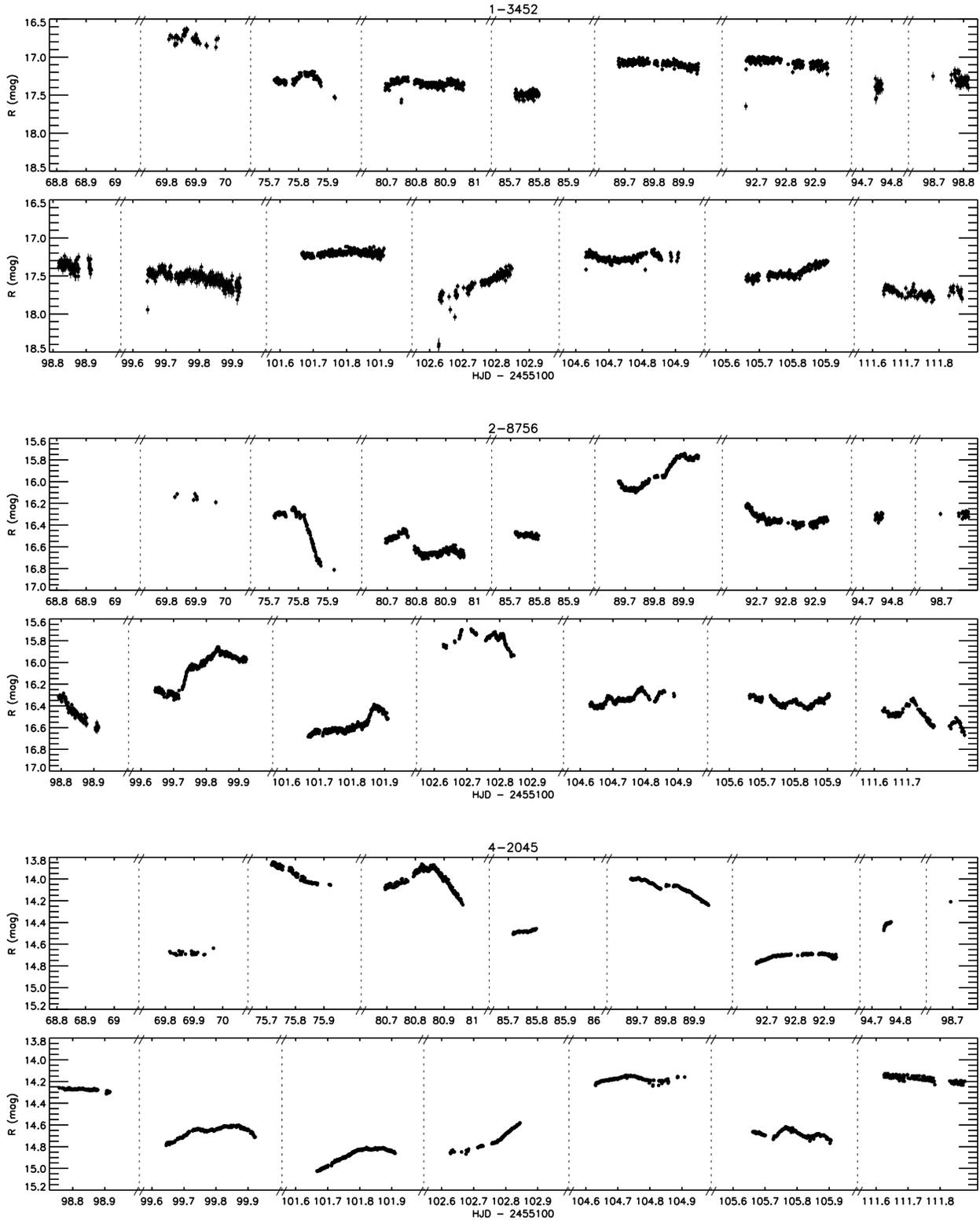}
    \caption{\label{fig:cttslc0}Light curves for the candidate CTTS
      (see section \ref{sec:ctts}).  Continued in figures
      \ref{fig:cttslc1}--\ref{fig:cttslc5}.  }
\end{figure*}

\begin{figure*}[tbp] \epsscale{1.0} 
    \plotone{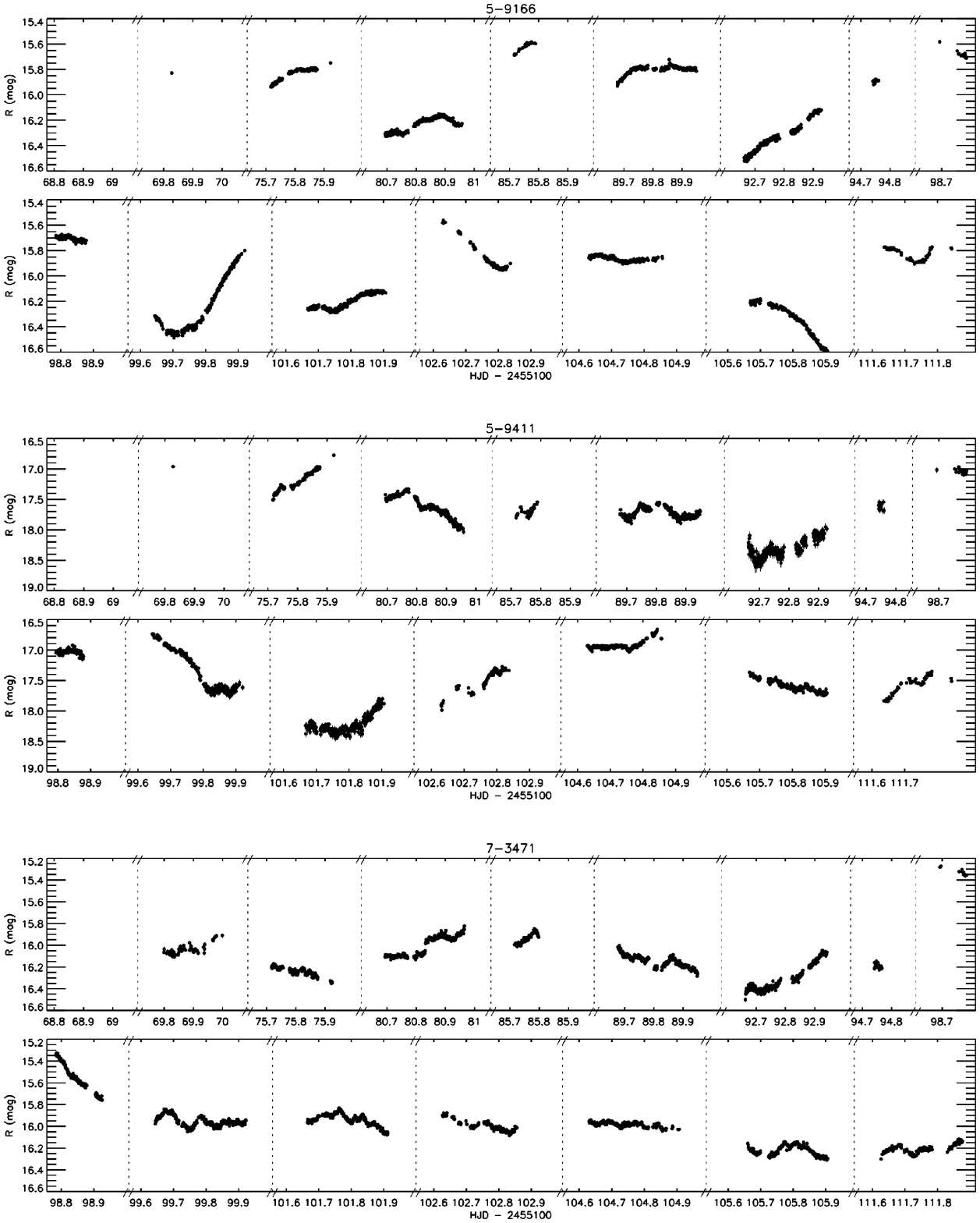}
    \caption{\label{fig:cttslc1}Light curves for the candidate CTTS - cont.}
\end{figure*}

\begin{figure*}[tbp] \epsscale{1.0} 
    \plotone{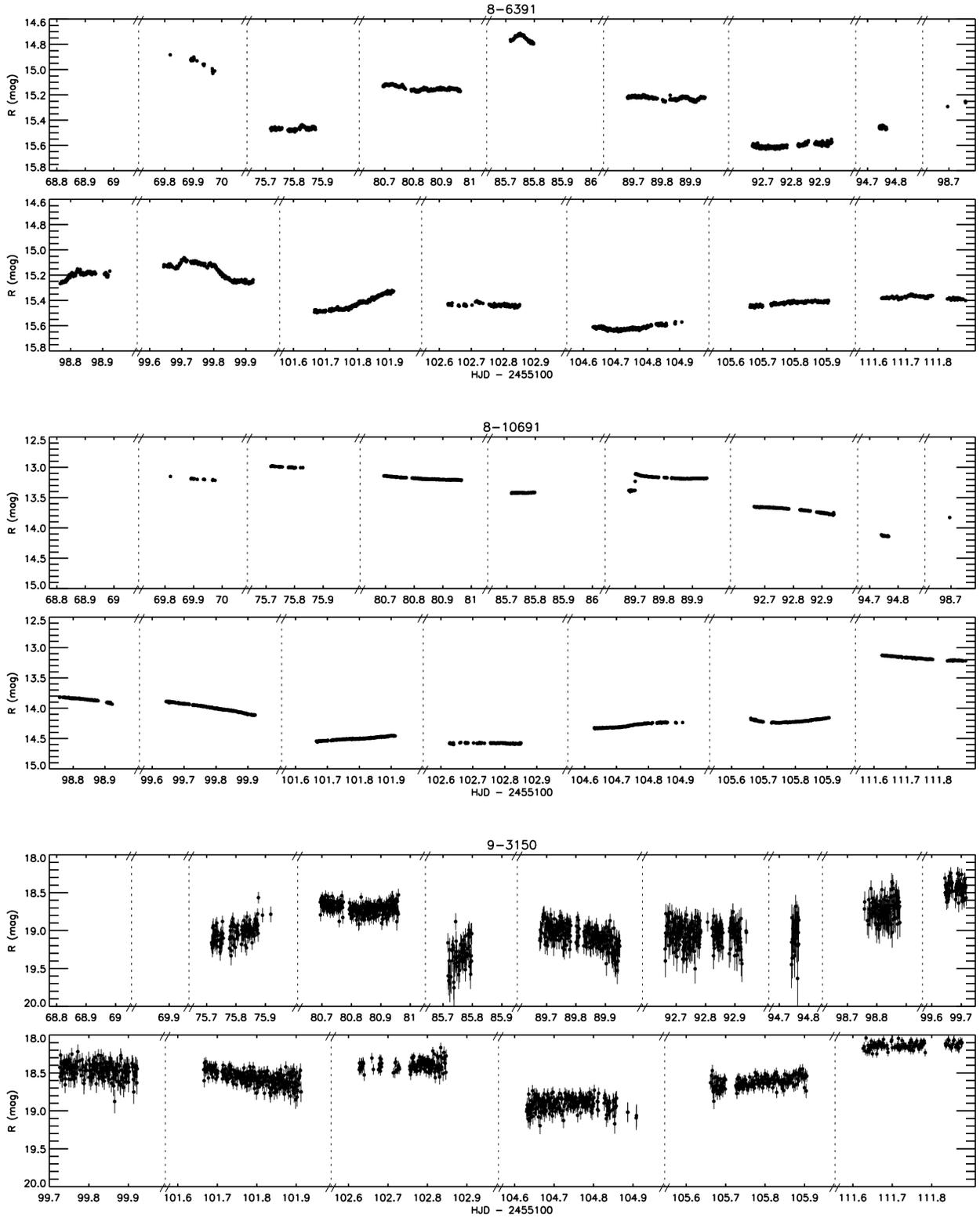}
    \caption{\label{fig:cttslc2}Light curves for the candidate CTTS -
      cont. 8-10691 (center panel) is a known CTTS \citep{Briceno2005}.}
\end{figure*}

\begin{figure*}[tbp] \epsscale{1.0} 
    \plotone{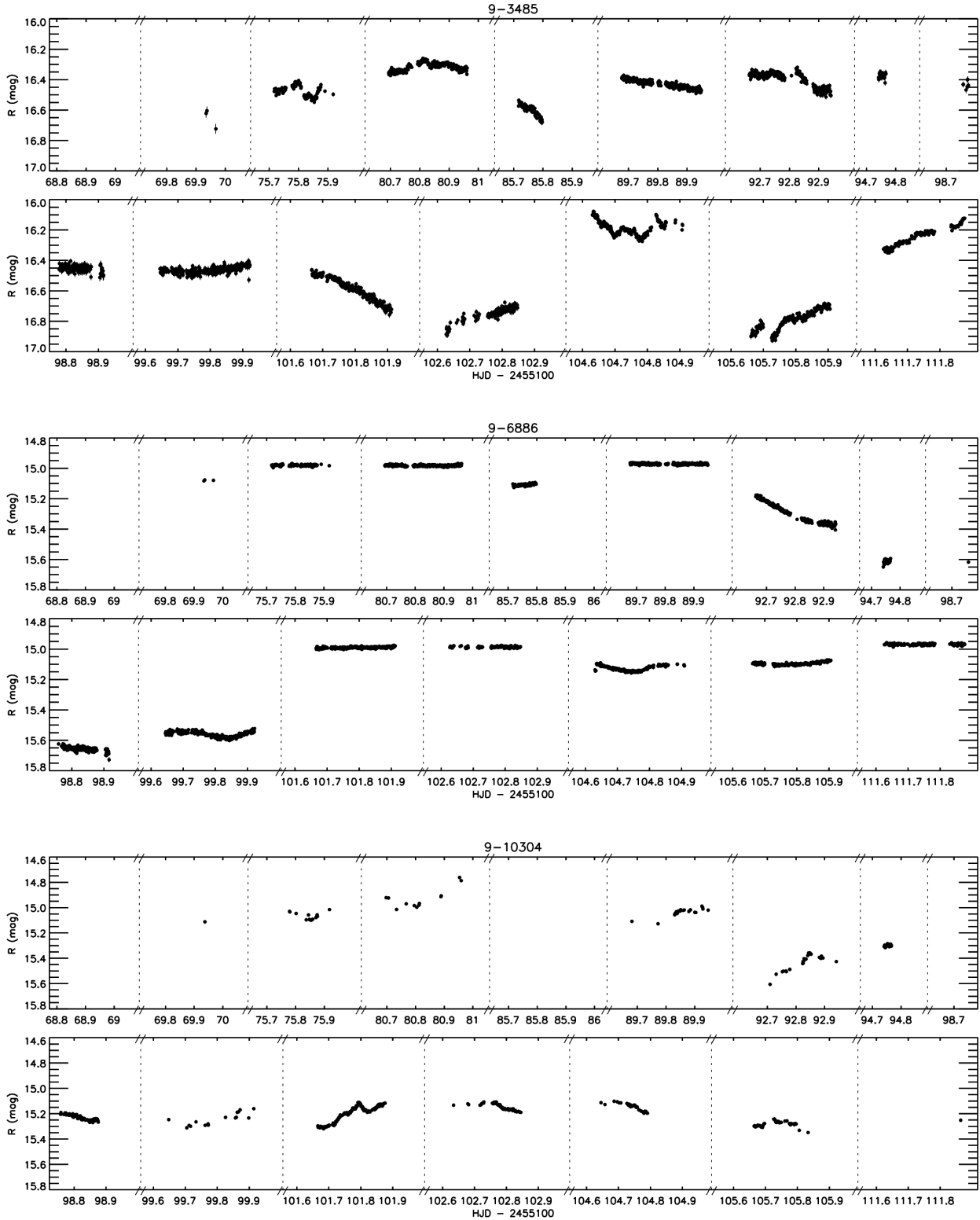}
    \caption{\label{fig:cttslc3}Light curves for the candidate CTTS -
      cont. 9-3485 (top panel) was previously reported as a candidate weak-lined
      TTS by \citet{McGehee2006}. 9-6886 (center panel) resembles the
      `dipper' sources discussed by \citet{Morales2011}.}
\end{figure*}

\begin{figure*}[tbp] \epsscale{1.0} 
    \plotone{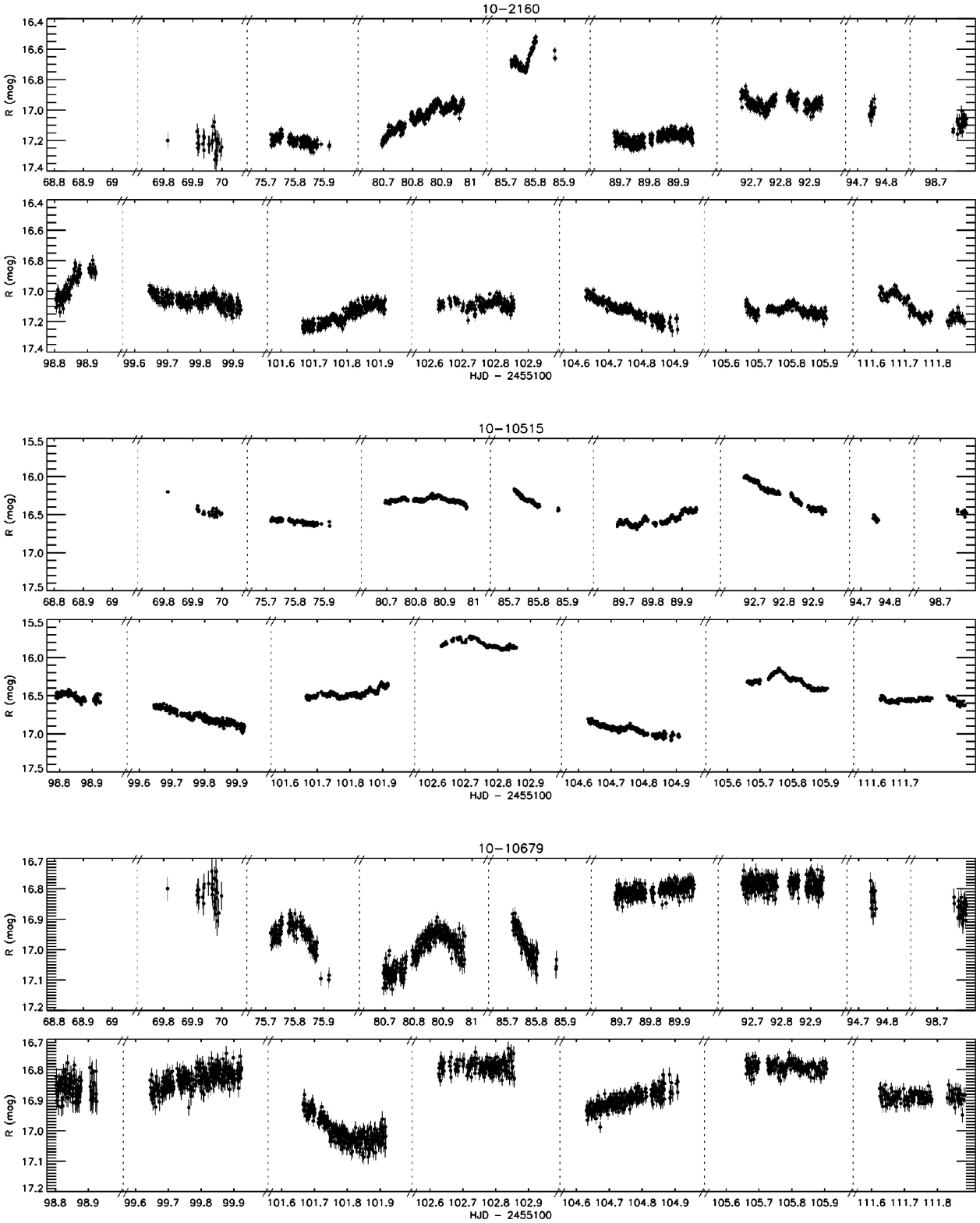}
    \caption{\label{fig:cttslc4}Light curves for the candidate CTTS -
      cont.}
\end{figure*}

\begin{figure*}[tbp] \epsscale{1.0} 
    \plotone{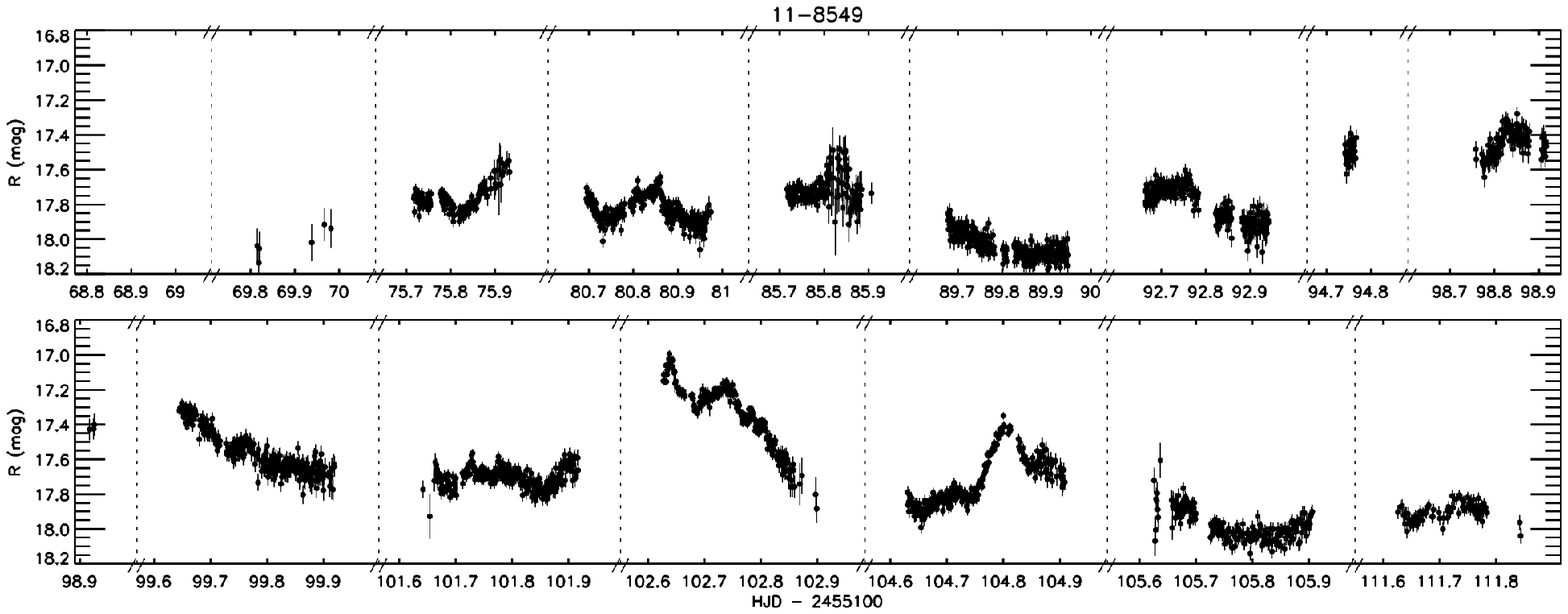}
    \caption{\label{fig:cttslc5}Light curves for the candidate CTTS - cont.}
\end{figure*}

Another class of light curve found in our data is that of CTTSs. These young systems exhibit particularly large
and irregular brightness modulations, thought to be related to
photospheric spots, short-lived accretion hot-spots, and/or obscuration
due to circumstellar dust \citep{Herbst1994}. On the basis of
this variability, we identified visually 16 candidate CTTS
light-curves which showed irregular variability on scales of $\sim
1\vmag$ or more, listed in table \ref{tbl:CTTS}. These include 14
candidates which are new to our knowledge, along with one known CTTS,
CVSO 35 ({\bf 8-10691}), previously reported by
\citet{Briceno2005,Briceno2007}, and one previously reported as a
candidate WTTS, SDSS J052700.12+010136.8
({\bf 9-3485}) \citep{McGehee2006}. The light curves for all 16 of the
sources are shown in figures \ref{fig:cttslc0}-\ref{fig:cttslc5}.

\begin{figure}[tbp] \epsscale{0.8} 
    \plotone{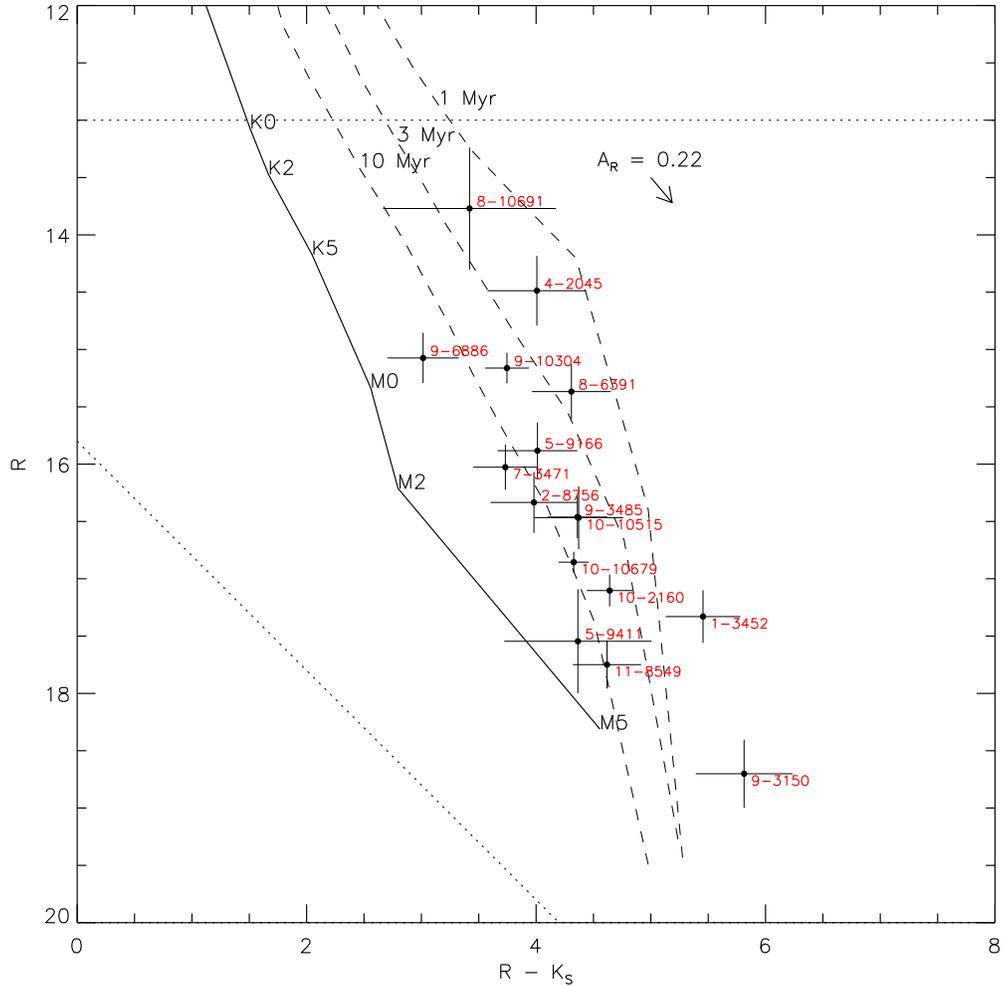}
    \caption{\label{fig:cttsRvsRK}Color-magnitude diagram as for
      figure \ref{fig:RvsR-K}, for the CTTS identified
      in the PTF Orion data matched against the 2MASS catalog (see
      section \ref{sec:ctts}). Errors in
      $R$ represent the standard deviation in the clipped (1st-99th
      percentile) standard deviation of the light curves to account
      for variability; errors in $K_s$ are taken to be the same in the
      absence of other information, added in quadrature with the
      intrinsic measurement errors in $K_s$.}
\end{figure}

Figure \ref{fig:cttsRvsRK} shows
the location of the candidate CTTS sources on the same
color-magnitude diagram as previously shown for the binaries (figure
\ref{fig:RvsR-K}). All clearly lie in the PMS
region for the distance and mean reddening of the 25~Ori region.

\begin{figure*}[tbp] \epsscale{0.75} 
    \plotone{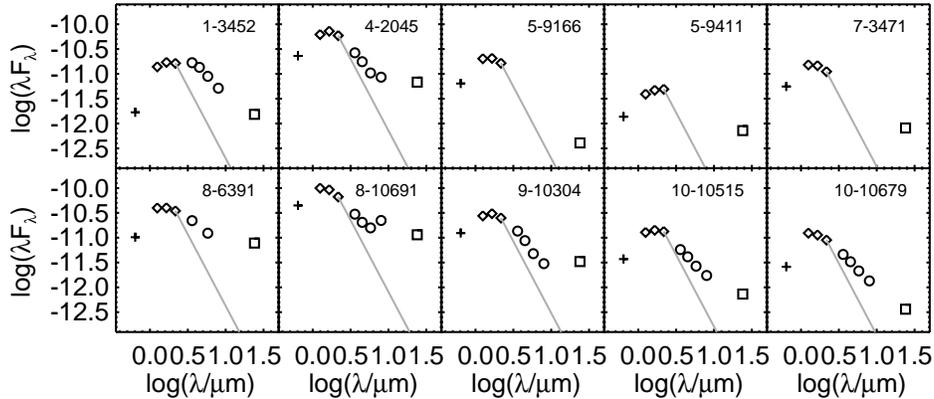}
    \caption{\label{fig:cttsSEDS}SEDs for the 10 CTTS-classified
      sources for which {\em Spitzer} archive data were available (nine new
      candidates and the known CTTS, 8-10691 \citep[CVSO
      35][]{Briceno2005,Briceno2007}).  Symbols are as for figure
      \ref{fig:binariesspitzer}, and the gray line again represents an
      extrapolated photosphere. All appear to have clear excesses, as
      expected for such sources.}
  \end{figure*}

Of the 16 CTTS sources, 10 were found with coverage in the {\em Spitzer}
archive data (including the one known CTTS), and all show IR excesses
suggestive of disk presence. The SEDs of these sources are shown in
figure \ref{fig:cttsSEDS}.
All the objects appear to have clear excesses, and for those where
multi-band {\em Spitzer} data are available, the excess is seen in more than
one band, lending credence to the identification of a disk excess. For
the three where there are only $24\um$ detections, the apparent $24\um$
excess is very large ($K_s-[24]=3.52\pm0.06, 5.45\pm0.06$, and
$4.69\pm0.06$, for sources 5-9166, 5-9411, and 7-3471, respectively),
and seems to be real, although the uncertainty is larger than the
formal error on the points because of the uncertainty in $A_V$ and
spectral type.

{\bf 1-3452}: Seems to show a slight discontinuity between $JHK_s$ and
the {\em Spitzer} IRAC bands (figure \ref{fig:cttsSEDS}). This may
either be a
result of the contribution from the black-body function of the
disk, or due to intrinsic variability between the epochs of 2MASS and
{\em Spitzer} observations.

{\bf 9-6886}: Displays a light curve which appears flat for a
substantial portion of the time, but with downward deviations of up to
almost a magnitude (figure \ref{fig:cttslc3}). This is reminiscent of
`dipper' objects discussed by \citet{Morales2011}, for which AA~Tau is
a prototype \citep{Bouvier2003}. (These objects were previously alluded to in
section \ref{sec:youngclose} with regard to object 7-5291.)

The new sources are spatially distributed fairly evenly across the PTF
field of view, with no obvious clustering. This suggests that they are
likely associated with the broader Orion OB1a association rather than
with the 25 Ori group itself, which is at approximately the same
distance.

\begin{deluxetable}{lrrlrrrrrrrrrr}
\tablecolumns{14}
\tablewidth{0pt}
\setlength{\tabcolsep}{0.06in} 
\tablecaption{\label{tbl:CTTS}Candidate CTTS: Sources with
  Irregular Light Curves and $\Delta R \sim 1.0\vmag$ or
  More.}
\tabletypesize{\scriptsize}
\rotate
\tablehead{
\colhead{ID} & \colhead{$\alpha$} & \colhead{$\delta$} &
\colhead{2MASS} & \colhead{PM $\alpha$} & \colhead{PM $\delta$} &
\colhead{$R_{\mathrm{med}}$\tablenotemark{d}} & \colhead{$\Delta R$} &
\colhead{$\sigma_{R}$} &
\colhead{$J$} & \colhead{$H$} & \colhead{$K_s$} \\

 & \colhead{(deg)} & \colhead{(deg)} & & \colhead{(mas $\mathrm{yr}^{-1}$)} &
 \colhead{(mas $\mathrm{yr}^{-1}$)} & \colhead{(mag)} & \colhead{(mag)} & \colhead{(mag)} &
 \colhead{(mag)} & \colhead{(mag)} & \colhead{(mag)} \\

\cline{5-12}

 & & & & \colhead{[3.6]} & \colhead{[4.5]} & \colhead{[5.8]} & \colhead{[8]} &
\colhead{[24]} & \colhead{$K_s-[24]$} & \colhead{$[3.6]-[24]$} & \\

 & & & & \colhead{(mag)} & \colhead{(mag)} & \colhead{(mag)} & \colhead{(mag)}
& \colhead{(mag)} & \colhead{(mag)} & \colhead{(mag)} & }

\startdata
\input{ctts_table}
\enddata

\tablecomments{See section \ref{sec:ctts}. 2MASS quality flag is `A'
  in all three bands for all sources (see section
  \ref{sec:ancillary}). A description of how to construct general PTF
  catalog identifiers is given in table \ref{tbl:binaries}.} 
\tablenotetext{a}{Known CTTS CVSO 35 \citep{Briceno2005,Briceno2007}}
\tablenotetext{b}{Reported as a candidate WTTS by \citet{McGehee2006}
  (SDSS J052700.12+010136.8).}
\tablenotetext{c}{Resembles the `dipper' objects discussed by
  \citet{Morales2011}.}
\tablenotetext{d}{See section \ref{sec:absaccuracy} for a discussion of
  zero-point accuracy.}
\end{deluxetable}


\clearpage

%% file: ctts_table.tex


1-3452 & 81.07590 & 2.28890 & 05241824+0217200 & 7.0$\pm$3.9 & 2.7$\pm$3.9 & 17.33 & 1.08 & 0.23 & 13.60$\pm$0.02 & 12.59$\pm$0.02 & 11.88$\pm$0.03 \\
 & & & &  10.36$\pm$0.05 & 9.86$\pm$0.05 & 9.56$\pm$0.05 & 9.17$\pm$0.05 & 6.91$\pm$0.05 & 4.96$\pm$0.05 & 3.45$\pm$0.07 & \\
\\
2-8756 & 81.42163 & 2.81835 & 05254119+0249059 & 0.3$\pm$3.9 & -7.7$\pm$3.9 & 16.33 & 1.09 & 0.27 & 13.59$\pm$0.03 & 12.86$\pm$0.03 & 12.35$\pm$0.03 \\
 & & & &  \nodata & \nodata & \nodata & \nodata & \nodata & \nodata & \nodata & \\
\\
4-2045 & 82.63182 & 2.05147 & 05303164+0203051 & 3.8$\pm$4.0 & -0.3$\pm$4.0 & 14.49 & 1.13 & 0.30 & 11.97$\pm$0.02 & 11.02$\pm$0.02 & 10.48$\pm$0.02 \\
 & & & &  9.86$\pm$0.05 & 9.58$\pm$0.05 & 9.39$\pm$0.05 & 8.62$\pm$0.05 & 5.29$\pm$0.05 & 5.19$\pm$0.05 & 4.57$\pm$0.07 & \\
\\
5-9166 & 82.87327 & 2.68835 & 05312959+0241183 & -0.8$\pm$4.0 & 0.2$\pm$4.0 & 15.88 & 0.92 & 0.24 & 13.19$\pm$0.03 & 12.39$\pm$0.03 & 11.87$\pm$0.03 \\
 & & & &  \nodata & \nodata & \nodata & \nodata & 8.35$\pm$0.05 & 3.52$\pm$0.06 & \nodata & \\
\\
5-9411 & 82.97474 & 2.70859 & 05315396+0242310 & 6.1$\pm$3.8 & -4.4$\pm$3.8 & 17.55 & 1.72 & 0.45 & 14.97$\pm$0.05 & 13.99$\pm$0.04 & 13.18$\pm$0.04 \\
 & & & &  \nodata & \nodata & \nodata & \nodata & 7.73$\pm$0.05 & 5.45$\pm$0.06 & \nodata & \\
\\
7-3471 & 80.97543 & 1.13702 & 05235410+0108128 & 0.5$\pm$3.9 & -10.9$\pm$3.9 & 16.03 & 1.14 & 0.20 & 13.50$\pm$0.03 & 12.76$\pm$0.03 & 12.29$\pm$0.03 \\
 & & & &  \nodata & \nodata & \nodata & \nodata & 7.60$\pm$0.05 & 4.69$\pm$0.06 & \nodata & \\
\\
8-6391 & 81.26950 & 1.34760 & 05250468+0120509 & -0.4$\pm$3.9 & -1.6$\pm$3.9 & 15.37 & 0.90 & 0.24 & 12.46$\pm$0.03 & 11.67$\pm$0.02 & 11.06$\pm$0.02 \\
 & & & &  10.06$\pm$0.05 & \nodata & 9.21$\pm$0.05 & \nodata & 5.15$\pm$0.04 & 5.91$\pm$0.05 & 4.92$\pm$0.07 & \\
\\
8-10691\tablenotemark{a} & 81.44118 & 1.76381 & 05254589+0145500 & 6.1$\pm$3.9 & -1.7$\pm$3.9 & 13.77 & 1.60 & 0.53 & 11.46$\pm$0.03 & 10.76$\pm$0.03 & 10.35$\pm$0.02 \\
 & & & &  9.74$\pm$0.05 & 9.42$\pm$0.05 & 8.94$\pm$0.05 & 7.58$\pm$0.05 & 4.73$\pm$0.04 & 5.62$\pm$0.05 & 5.01$\pm$0.07 & \\
\\
9-3150 & 82.23882 & 0.99533 & 05285732+0059431 & 5.6$\pm$16.3 & -31.4$\pm$16.3 & 18.70 & 1.38 & 0.30 & 13.90$\pm$0.03 & 13.26$\pm$0.03 & 12.89$\pm$0.03 \\
 & & & &  \nodata & \nodata & \nodata & \nodata & \nodata & \nodata & \nodata & \\
\\
9-3485\tablenotemark{b} & 81.75044 & 1.02696 & 05270012+0101368 & 2.3$\pm$3.9 & -3.2$\pm$3.9 & 16.46 & 0.78 & 0.18 & 13.18$\pm$0.02 & 12.47$\pm$0.02 & 12.10$\pm$0.03 \\
 & & & &  \nodata & \nodata & \nodata & \nodata & \nodata & \nodata & \nodata & \\
\\
9-6886\tablenotemark{c} & 82.08428 & 1.35453 & 05282024+0121159 & 3.5$\pm$3.9 & -0.4$\pm$3.9 & 15.07 & 0.71 & 0.22 & 13.05$\pm$0.02 & 12.31$\pm$0.02 & 12.06$\pm$0.02 \\
 & & & &  \nodata & \nodata & \nodata & \nodata & \nodata & \nodata & \nodata & \\
\\
9-10304 & 81.73075 & 1.67283 & 05265537+0140224 & -16.0$\pm$3.9 & 2.9$\pm$3.9 & 15.16 & 0.82 & 0.13 & 12.85$\pm$0.03 & 11.96$\pm$0.03 & 11.42$\pm$0.02 \\
 & & & &  10.59$\pm$0.05 & 10.34$\pm$0.06 & 10.24$\pm$0.06 & 9.76$\pm$0.06 & 6.08$\pm$0.05 & 5.34$\pm$0.05 & 4.51$\pm$0.07 & \\
\\
10-2160 & 82.72717 & 0.94125 & 05305452+0056286 & 5.4$\pm$4.0 & -1.5$\pm$4.0 & 17.10 & 0.76 & 0.14 & 13.34$\pm$0.03 & 12.80$\pm$0.03 & 12.46$\pm$0.03 \\
 & & & &  \nodata & \nodata & \nodata & \nodata & \nodata & \nodata & \nodata & \\
\\
10-10515 & 82.55167 & 1.80581 & 05301240+0148214 & 3.9$\pm$3.9 & -1.7$\pm$3.9 & 16.47 & 1.29 & 0.27 & 13.69$\pm$0.03 & 12.79$\pm$0.02 & 12.10$\pm$0.02 \\
 & & & &  11.52$\pm$0.06 & 11.16$\pm$0.06 & 10.87$\pm$0.06 & 10.35$\pm$0.06 & 7.71$\pm$0.05 & 4.38$\pm$0.05 & 3.81$\pm$0.07 & \\
\\
10-10679 & 82.41315 & 1.82079 & 05293913+0149156 & -3.8$\pm$3.9 & 5.0$\pm$3.9 & 16.86 & 0.38 & 0.09 & 13.72$\pm$0.02 & 13.03$\pm$0.03 & 12.53$\pm$0.02 \\
 & & & &  11.77$\pm$0.06 & 11.41$\pm$0.06 & 11.11$\pm$0.06 & 10.63$\pm$0.06 & 8.48$\pm$0.05 & 4.05$\pm$0.06 & 3.29$\pm$0.07 & \\
\\
11-8549 & 82.99254 & 1.66050 & 05315820+0139383 & 4.1$\pm$3.8 & -1.6$\pm$3.8 & 17.75 & 0.96 & 0.21 & 14.40$\pm$0.04 & 13.72$\pm$0.04 & 13.13$\pm$0.04 \\
 & & & &  \nodata & \nodata & \nodata & \nodata & \nodata & \nodata & \nodata & \\



%% file: conclusions.tex
\section{Summary}\label{sec:summary}

We have presented some of the initial results from the PTF Orion
survey based on the most strongly variable light curves, relating to
eclipsing binary systems and PMS stars, two of the main
science objectives of the survey. We have identified 82 eclipsing binary
systems, of which 37 are detached, and 45 are `close' (mostly contact
binaries). Of the close systems, 11-8774 is among the shortest-period
W-UMa type systems known, lying in an as-yet poorly sampled region of
W-UMa parameter space. Furthermore, the distance estimate for this
system puts it within the range of the Orion association, implying it may
be very young: it may therefore be particularly interesting for
constraining evolutionary models of contact binaries.

Among the eclipsing binaries, we find a total of nine candidate
PMS systems, with implied ages of $\approx 7$--$10\Myr$
if we assume Orion 25 Ori and/or Orion OB1a association
membership. Their positions in the color-magnitude diagram, estimated
distances (with the exception of one uncertain estimate),
out-of-eclipse variability, and proper motions (with the exception of
one), are all consistent with their proposed youth and association
membership, making them good candidate PMS binaries.
Such systems are potentially valuable for constraining evolutionary
models of both binary systems and individual stars. Furthermore, two
of these systems are potentially young contact binaries. If
these are genuinely as young as 7--$10\Myr$, this could have
significant implications regarding the timescales of mechanisms by
which they could have formed.

A number of the binary systems are also candidate low-mass systems,
apparently with M-dwarf primaries; this is also a regime that remains
poorly sampled to date. With the exception of the two contact systems
(which are not expected at such late spectral types), most, if not
all, of the above mentioned candidate young systems appear to have
low-mass primaries based on their 2MASS colors. We also find an
additional two detached systems whose $JHK_s$ colors are suggestive of
low mass, but with distance estimates more consistent with being
background stars; and three contact (or near-contact) systems which appear
to be unusually reddened.

Some of the systems show somewhat unusual light curves. One in
particular, 7-5291, has the appearance of a distorted contact- or
near-contact binary curve, but has a much longer period than is found
for contact binaries, and its position on the color-magnitude diagram
is consistent with a young age at the distance of Orion. One possible
explanation is that it is a semi-detached system with an inflated
PMS primary, perhaps with a disk -- although the
available {\em Spitzer} data do not seem to show a clear excess. Assuming
it is not a contact binary system gives it a distance estimate nearer
the 25 Ori/Orion OB1a association. Alternatively, the light curve
might be explained by a star being occulted by a warped disk, in an
analogue of AA Tau \citep{Bouvier2003} and the `dipper' systems
identified by \citet{Morales2011}. Further data and analysis are
needed for this object.

One of the detached systems, 9-2980 appears to
exhibit stellar pulsations superimposed on the eclipsing binary light
curve. The pronounced ellipsoidal variation suggests a fairly closely
separated system. Further data are needed for coverage of the
secondary eclipse, but the pulsations could potentially allow for
valuable asteroseismological measurements of the system to complement
those that can be obtained from standard light curve fitting
and RV measurements.

Finally, we have identified 16 candidate CTTS systems, 14 of which are
new to our knowledge, 1 of which is already known \citep[8-10691,
CVSO 35,][] {Briceno2005,Briceno2007}, and 1 of which has
previously been reported as a candidate WTTS \citep[9-3485,
J052700.12+010136.8;][]{McGehee2006}. One of the new CTTS, 9-6886, has
the appearance of another `dipper' source.

Absolute photometric accuracy in the data set is estimated to be
around $3\%$--$5\%$ ($\approx 7\%$ for chips 5 and 11). The noise floor in
the differential photometry sets in at a precision of around $4\mmag$
for a single exposure, without any attempt at detrending. The
differential photometry algorithm developed appears to be relatively
robust against systematic trends in all but the worst cases, although,
owing to known instrument issues at the time of these observations,
detrending the data may reveal more light curves that could be of
interest, particularly on chip 4.

The data reported here represent an initial cut of only the most
strongly variable stars, but have yielded a wealth of
interesting data. Further followup data on the same field taken with
PTF at the time of writing will enable confirmation of interesting
sources, and help to distinguish persistent sources of variability
(binaries, planets, etc.) from more transient sources (star spots,
stellar activity, etc.). A finer search of the PTF Orion data at a
lower variability threshold will explore a much larger sample of
sources. Much remains to be mined in the data.

%% file: acknowledgments.tex
The authors thank Cesar Brice\~no for providing the
reference stars for zero-point calibration, and for his advice
regarding field-selection; and Tim Lister, Rachel Street, Andrej
Pr\v{s}a, David Bradstreet, Ed Guinan, and Adam Krauss, for valuable
discussion and input. S.B.C.~wishes to acknowledge generous support
from Gary and Cynthia Bengier, the Richard and Rhoda Goldman Fund,
National Aeronautics and Space Administration (NASA)/\textit{Swift}
grant NNX10AI21G, NASA/\textit{Fermi} grant NNX1OA057G, and National
Science Foundation (NSF) grant AST--0908886. Observations obtained
with the Samuel Oschin Telescope at the Palomar Observatory as part of
the Palomar Transient Factory project, a scientific collaboration
between the California Institute of Technology, Columbia University,
Las Cumbres Observatory, the Lawrence Berkeley National Laboratory,
the National Energy Research Scientific Computing Center, the
University of Oxford, and the Weizmann Institute of Science. This
publication makes use of data products from the Two Micron All Sky
Survey, which is a joint project of the University of Massachusetts
and the Infrared Processing and Analysis Center/California Institute
of Technology, funded by the National Aeronautics and Space
Administration and the National Science Foundation. This work is based
in part on archival data obtained with the {\em Spitzer Space Telescope},
which is operated by the Jet Propulsion Laboratory, California
Institute of Technology under a contract with NASA. This research has
made use of the NASA/IPAC/NExScI Star and Exoplanet Database, which is
operated by the Jet Propulsion Laboratory, California Institute of
Technology, under contract with the National Aeronautics and Space
Administration. This research has made use of the NASA/IPAC Infrared
Science Archive, which is operated by the Jet Propulsion Laboratory,
California Institute of Technology, under contract with the National
Aeronautics and Space Administration. This research has made use of
the VizieR catalogue access tool, CDS, Strasbourg, France. Support for
this work was provided by an award issued by JPL/Caltech.

{\it Facilities:} \facility{PO:1.2m (PTF)}, \facility{Spitzer}

%% file: PTFOrionOverview.bbl
\begin{thebibliography}{71}
\expandafter\ifx\csname natexlab\endcsname\relax\def\natexlab#1{#1}\fi

\bibitem[{{Aigrain} {et~al.}(2007){Aigrain}, {Hodgkin}, {Irwin}, {Hebb},
  {Irwin}, {Favata}, {Moraux}, \& {Pont}}]{Monitor}
{Aigrain}, S., {Hodgkin}, S., {Irwin}, J., {Hebb}, L., {Irwin}, M., {Favata},
  F., {Moraux}, E., \& {Pont}, F. 2007, \mnras, 375, 29

\bibitem[{{Armitage}(2009)}]{ArmitagePlanetFormation}
{Armitage}, P.~J. 2009, in AIP Conf. Ser., Vol. 1192, XIII Special Courses at
  the National Observatory of Rio de Janeiro, ed. {F.~Roig, D.~Lopes, R.~de La
  Reza, \& V.~Ortega} (Melville, NY: AIP), 3--42

\bibitem[{{Basri} {et~al.}(2010){Basri}, {Walkowicz}, {Batalha}, {Gilliland},
  {Jenkins}, {Borucki}, {Koch}, {Caldwell}, {Dupree}, {Latham}, {Meibom},
  {Howell}, \& {Brown}}]{Basri2010}
{Basri}, G., {Walkowicz}, L.~M., {Batalha}, N., {Gilliland}, R.~L., {Jenkins},
  J., {Borucki}, W.~J., {Koch}, D., {Caldwell}, D., {Dupree}, A.~K., {Latham},
  D.~W., {Meibom}, S., {Howell}, S., \& {Brown}, T. 2010, \apjl, 713, L155

\bibitem[{{Bertin}(2006)}]{SCAMP}
{Bertin}, E. 2006, in Astronomical Society of the Pacific Conference Series,
  Vol. 351, Astronomical Data Analysis Software and Systems XV, ed.
  {C.~Gabriel, C.~Arviset, D.~Ponz, \& S.~Enrique}, 112

\bibitem[{{Bertin} \& {Arnouts}(1996)}]{sextractor}
{Bertin}, E. \& {Arnouts}, S. 1996, \aaps, 117, 393

\bibitem[{{Bilir} {et~al.}(2008){Bilir}, {Ak}, {Soydugan}, {Soydugan}, {Yaz},
  {Filiz Ak}, {Eker}, {Demircan}, \& {Helvaci}}]{Bilir2008DetachedCalibration}
{Bilir}, S., {Ak}, T., {Soydugan}, E., {Soydugan}, F., {Yaz}, E., {Filiz Ak},
  N., {Eker}, Z., {Demircan}, O., \& {Helvaci}, M. 2008, Astronomische
  Nachrichten, 329, 835

\bibitem[{{Bilir} {et~al.}(2005){Bilir}, {Karata{\c s}}, {Demircan}, \&
  {Eker}}]{Bilir2005}
{Bilir}, S., {Karata{\c s}}, Y., {Demircan}, O., \& {Eker}, Z. 2005, \mnras,
  357, 497

\bibitem[{{Bouvier} {et~al.}(2003){Bouvier}, {Grankin}, {Alencar}, {Dougados},
  {Fern{\'a}ndez}, {Basri}, {Batalha}, {Guenther}, {Ibrahimov}, {Magakian},
  {Melnikov}, {Petrov}, {Rud}, \& {Zapatero Osorio}}]{Bouvier2003}
{Bouvier}, J., {Grankin}, K.~N., {Alencar}, S.~H.~P., {Dougados}, C.,
  {Fern{\'a}ndez}, M., {Basri}, G., {Batalha}, C., {Guenther}, E., {Ibrahimov},
  M.~A., {Magakian}, T.~Y., {Melnikov}, S.~Y., {Petrov}, P.~P., {Rud}, M.~V.,
  \& {Zapatero Osorio}, M.~R. 2003, \aap, 409, 169

\bibitem[{{Brice{\~n}o} {et~al.}(2005){Brice{\~n}o}, {Calvet}, {Hern{\'a}ndez},
  {Vivas}, {Hartmann}, {Downes}, \& {Berlind}}]{Briceno2005}
{Brice{\~n}o}, C., {Calvet}, N., {Hern{\'a}ndez}, J., {Vivas}, A.~K.,
  {Hartmann}, L., {Downes}, J.~J., \& {Berlind}, P. 2005, \aj, 129, 907

\bibitem[{{Brice{\~n}o} {et~al.}(2007){Brice{\~n}o}, {Hartmann},
  {Hern{\'a}ndez}, {Calvet}, {Vivas}, {Furesz}, \&
  {Szentgyorgyi}}]{Briceno2007}
{Brice{\~n}o}, C., {Hartmann}, L., {Hern{\'a}ndez}, J., {Calvet}, N., {Vivas},
  A.~K., {Furesz}, G., \& {Szentgyorgyi}, A. 2007, \apj, 661, 1119

\bibitem[{{Brice{\~n}o} {et~al.}(2001){Brice{\~n}o}, {Vivas}, {Calvet},
  {Hartmann}, {Pacheco}, {Herrera}, {Romero}, {Berlind}, {S{\'a}nchez},
  {Snyder}, \& {Andrews}}]{Briceno2001}
{Brice{\~n}o}, C., {Vivas}, A.~K., {Calvet}, N., {Hartmann}, L., {Pacheco}, R.,
  {Herrera}, D., {Romero}, L., {Berlind}, P., {S{\'a}nchez}, G., {Snyder},
  J.~A., \& {Andrews}, P. 2001, Science, 291, 93

\bibitem[{{Brown} {et~al.}(1999){Brown}, {Walter}, \& A.}]{Brown1999}
{Brown}, A. G.~A., {Walter}, F.~M., \& A., B. 1999, in Astronomical Society of
  the Pacific Conference Series, Vol. 351, The Orion Complex Revisited, ed.
  {M.~J.~McCaughrean \& A.~Burkert}, unpublished, arXiv:astro-ph/9802054v2

\bibitem[{{Cardelli} {et~al.}(1989){Cardelli}, {Clayton}, \&
  {Mathis}}]{Cardelli1989}
{Cardelli}, J.~A., {Clayton}, G.~C., \& {Mathis}, J.~S. 1989, \apj, 345, 245

\bibitem[{{Cargile} {et~al.}(2008){Cargile}, {Stassun}, \&
  {Mathieu}}]{Cargile2008}
{Cargile}, P.~A., {Stassun}, K.~G., \& {Mathieu}, R.~D. 2008, \apj, 674, 329

\bibitem[{{Covino} {et~al.}(2000){Covino}, {Catalano}, {Frasca}, {Marilli},
  {Fern{\'a}ndez}, {Alcal{\'a}}, {Melo}, {Paladino}, {Sterzik}, \&
  {Stelzer}}]{Covino2000}
{Covino}, E., {Catalano}, S., {Frasca}, A., {Marilli}, E., {Fern{\'a}ndez}, M.,
  {Alcal{\'a}}, J.~M., {Melo}, C., {Paladino}, R., {Sterzik}, M.~F., \&
  {Stelzer}, B. 2000, \aap, 361, L49

\bibitem[{{Covino} {et~al.}(2004){Covino}, {Frasca}, {Alcal{\'a}}, {Paladino},
  \& {Sterzik}}]{Covino2004}
{Covino}, E., {Frasca}, A., {Alcal{\'a}}, J.~M., {Paladino}, R., \& {Sterzik},
  M.~F. 2004, \aap, 427, 637

\bibitem[{{de Zeeuw} {et~al.}(1999){de Zeeuw}, {Hoogerwerf}, {de Bruijne},
  {Brown}, \& {Blaauw}}]{deZeeuw1999}
{de Zeeuw}, P.~T., {Hoogerwerf}, R., {de Bruijne}, J.~H.~J., {Brown}, A.~G.~A.,
  \& {Blaauw}, A. 1999, \aj, 117, 354

\bibitem[{{Eggleton} \& {Kisseleva-Eggleton}(2006)}]{Eggleton2006}
{Eggleton}, P.~P. \& {Kisseleva-Eggleton}, L. 2006, \apss, 304, 75

\bibitem[{{Eker} {et~al.}(2009){Eker}, {Bilir}, {Yaz}, {Demircan}, \&
  {Helvaci}}]{Eker2009WUMaCalibration}
{Eker}, Z., {Bilir}, S., {Yaz}, E., {Demircan}, O., \& {Helvaci}, M. 2009,
  Astronomische Nachrichten, 330, 68

\bibitem[{{Eker} {et~al.}(2006){Eker}, {Demircan}, {Bilir}, \& {Karata{\c
  s}}}]{Eker2006}
{Eker}, Z., {Demircan}, O., {Bilir}, S., \& {Karata{\c s}}, Y. 2006, \mnras,
  373, 1483

\bibitem[{{Everett} \& {Howell}(2001)}]{EverettHowell}
{Everett}, M.~E. \& {Howell}, S.~B. 2001, \pasp, 113, 1428

\bibitem[{{Fazio} {et~al.}(2004){Fazio}, {Hora}, {Allen}, {Ashby}, {Barmby},
  {Deutsch}, {Huang}, {Kleiner}, {Marengo}, {Megeath}, {Melnick}, {Pahre},
  {Patten}, {Polizotti}, {Smith}, {Taylor}, {Wang}, {Willner}, {Hoffmann},
  {Pipher}, {Forrest}, {McMurty}, {McCreight}, {McKelvey}, {McMurray}, {Koch},
  {Moseley}, {Arendt}, {Mentzell}, {Marx}, {Losch}, {Mayman}, {Eichhorn},
  {Krebs}, {Jhabvala}, {Gezari}, {Fixsen}, {Flores}, {Shakoorzadeh}, {Jungo},
  {Hakun}, {Workman}, {Karpati}, {Kichak}, {Whitley}, {Mann}, {Tollestrup},
  {Eisenhardt}, {Stern}, {Gorjian}, {Bhattacharya}, {Carey}, {Nelson},
  {Glaccum}, {Lacy}, {Lowrance}, {Laine}, {Reach}, {Stauffer}, {Surace},
  {Wilson}, {Wright}, {Hoffman}, {Domingo}, \& {Cohen}}]{Fazio2004}
{Fazio}, G.~G., {Hora}, J.~L., {Allen}, L.~E., {Ashby}, M.~L.~N., {Barmby}, P.,
  {Deutsch}, L.~K., {Huang}, J., {Kleiner}, S., {Marengo}, M., {Megeath},
  S.~T., {Melnick}, G.~J., {Pahre}, M.~A., {Patten}, B.~M., {Polizotti}, J.,
  {Smith}, H.~A., {Taylor}, R.~S., {Wang}, Z., {Willner}, S.~P., {Hoffmann},
  W.~F., {Pipher}, J.~L., {Forrest}, W.~J., {McMurty}, C.~W., {McCreight},
  C.~R., {McKelvey}, M.~E., {McMurray}, R.~E., {Koch}, D.~G., {Moseley}, S.~H.,
  {Arendt}, R.~G., {Mentzell}, J.~E., {Marx}, C.~T., {Losch}, P., {Mayman}, P.,
  {Eichhorn}, W., {Krebs}, D., {Jhabvala}, M., {Gezari}, D.~Y., {Fixsen},
  D.~J., {Flores}, J., {Shakoorzadeh}, K., {Jungo}, R., {Hakun}, C., {Workman},
  L., {Karpati}, G., {Kichak}, R., {Whitley}, R., {Mann}, S., {Tollestrup},
  E.~V., {Eisenhardt}, P., {Stern}, D., {Gorjian}, V., {Bhattacharya}, B.,
  {Carey}, S., {Nelson}, B.~O., {Glaccum}, W.~J., {Lacy}, M., {Lowrance},
  P.~J., {Laine}, S., {Reach}, W.~T., {Stauffer}, J.~A., {Surace}, J.~A.,
  {Wilson}, G., {Wright}, E.~L., {Hoffman}, A., {Domingo}, G., \& {Cohen}, M.
  2004, \apjs, 154, 10

\bibitem[{{Gautier} {et~al.}(2007){Gautier}, {Rieke}, {Stansberry}, {Bryden},
  {Stapelfeldt}, {Werner}, {Beichman}, {Chen}, {Su}, {Trilling}, {Patten}, \&
  {Roellig}}]{Gautier2007}
{Gautier}, III, T.~N., {Rieke}, G.~H., {Stansberry}, J., {Bryden}, G.~C.,
  {Stapelfeldt}, K.~R., {Werner}, M.~W., {Beichman}, C.~A., {Chen}, C., {Su},
  K., {Trilling}, D., {Patten}, B.~M., \& {Roellig}, T.~L. 2007, \apj, 667, 527

\bibitem[{{Gazeas} \& {St{\c e}pie{\'n}}(2008)}]{Gazeas2008}
{Gazeas}, K. \& {St{\c e}pie{\'n}}, K. 2008, \mnras, 390, 1577

\bibitem[{{Grillmair} {et~al.}(2010){Grillmair}, {Laher}, {Surace},
  {Mattingly}, {Hacopians}, {Jackson}, {van Eyken}, {McCollum}, {Groom}, {Mi},
  \& {Teplitz}}]{ptfpipeline}
{Grillmair}, C.~C., {Laher}, R., {Surace}, J., {Mattingly}, S., {Hacopians},
  E., {Jackson}, E., {van Eyken}, J.~C., {McCollum}, B., {Groom}, S.~L., {Mi},
  W., \& {Teplitz}, H.~I. 2010, in ASP~Conf.~Ser., Vol. 434, Astronomical Data
  Analysis Software and Systems XIX, ed. Y.~{Mizumoto}, I.~I. {Morita}, \&
  M.~{Ohishi} (Ann Arbor, MI: ASP), 28

\bibitem[{{Guieu} {et~al.}(2010){Guieu}, {Rebull}, {Stauffer}, {Vrba},
  {Noriega-Crespo}, {Spuck}, {Roelofsen Moody}, {Sepulveda}, {Weehler},
  {Maranto}, {Cole}, {Flagey}, {Laher}, {Penprase}, {Ramirez}, \&
  {Stolovy}}]{GuieuIC2118}
{Guieu}, S., {Rebull}, L.~M., {Stauffer}, J.~R., {Vrba}, F.~J.,
  {Noriega-Crespo}, A., {Spuck}, T., {Roelofsen Moody}, T., {Sepulveda}, B.,
  {Weehler}, C., {Maranto}, A., {Cole}, D.~M., {Flagey}, N., {Laher}, R.,
  {Penprase}, B., {Ramirez}, S., \& {Stolovy}, S. 2010, \apj, 720, 46

\bibitem[{{Hebb} {et~al.}(2010){Hebb}, {Stempels}, {Aigrain},
  {Collier-Cameron}, {Hodgkin}, {Irwin}, {Maxted}, {Pollacco}, {Street},
  {Wilson}, \& {Stassun}}]{Hebb2010}
{Hebb}, L., {Stempels}, H.~C., {Aigrain}, S., {Collier-Cameron}, A., {Hodgkin},
  S.~T., {Irwin}, J.~M., {Maxted}, P.~F.~L., {Pollacco}, D., {Street}, R.~A.,
  {Wilson}, D.~M., \& {Stassun}, K.~G. 2010, \aap, 522, A37+

\bibitem[{{Herbst} {et~al.}(1994){Herbst}, {Herbst}, {Grossman}, \&
  {Weinstein}}]{Herbst1994}
{Herbst}, W., {Herbst}, D.~K., {Grossman}, E.~J., \& {Weinstein}, D. 1994, \aj,
  108, 1906

\bibitem[{{Hillenbrand}(2008)}]{Hillenbrand2008}
{Hillenbrand}, L.~A. 2008, Physica Scripta Volume T, 130, 014024

\bibitem[{{Irwin} {et~al.}(2007){Irwin}, {Aigrain}, {Hodgkin}, {Stassun},
  {Hebb}, {Irwin}, {Moraux}, {Bouvier}, {Alapini}, {Alexander}, {Bramich},
  {Holtzman}, {Mart{\'{\i}}n}, {McCaughrean}, {Pont}, {Verrier}, \& {Zapatero
  Osorio}}]{Irwin2007}
{Irwin}, J., {Aigrain}, S., {Hodgkin}, S., {Stassun}, K.~G., {Hebb}, L.,
  {Irwin}, M., {Moraux}, E., {Bouvier}, J., {Alapini}, A., {Alexander}, R.,
  {Bramich}, D.~M., {Holtzman}, J., {Mart{\'{\i}}n}, E.~L., {McCaughrean},
  M.~J., {Pont}, F., {Verrier}, P.~E., \& {Zapatero Osorio}, M.~R. 2007,
  \mnras, 380, 541

\bibitem[{{Jeffries} {et~al.}(2007){Jeffries}, {Oliveira}, {Naylor}, {Mayne},
  \& {Littlefair}}]{Jeffries2007}
{Jeffries}, R.~D., {Oliveira}, J.~M., {Naylor}, T., {Mayne}, N.~J., \&
  {Littlefair}, S.~P. 2007, \mnras, 376, 580

\bibitem[{{Kenyon} \& {Hartmann}(1995)}]{KenyonHartmann1995}
{Kenyon}, S.~J. \& {Hartmann}, L. 1995, \apjs, 101, 117

\bibitem[{{Kharchenko} {et~al.}(2005){Kharchenko}, {Piskunov}, {R{\"o}ser},
  {Schilbach}, \& {Scholz}}]{Kharchenko2005}
{Kharchenko}, N.~V., {Piskunov}, A.~E., {R{\"o}ser}, S., {Schilbach}, E., \&
  {Scholz}, R. 2005, \aap, 440, 403

\bibitem[{{Kraus} {et~al.}(2007){Kraus}, {Craine}, {Giampapa}, {Scharlach}, \&
  {Tucker}}]{MOTESS-GNAT}
{Kraus}, A.~L., {Craine}, E.~R., {Giampapa}, M.~S., {Scharlach}, W.~W.~G., \&
  {Tucker}, R.~A. 2007, \aj, 134, 1488

\bibitem[{{Kraus} {et~al.}(2011){Kraus}, {Tucker}, {Thompson}, {Craine}, \&
  {Hillenbrand}}]{Kraus2011}
{Kraus}, A.~L., {Tucker}, R.~A., {Thompson}, M.~I., {Craine}, E.~R., \&
  {Hillenbrand}, L.~A. 2011, \apj, 728, 48

\bibitem[{{Lamm} {et~al.}(2004){Lamm}, {Bailer-Jones}, {Mundt}, {Herbst}, \&
  {Scholz}}]{Lamm2004}
{Lamm}, M.~H., {Bailer-Jones}, C.~A.~L., {Mundt}, R., {Herbst}, W., \&
  {Scholz}, A. 2004, \aap, 417, 557

\bibitem[{{Landolt}(1992)}]{Landolt}
{Landolt}, A.~U. 1992, \aj, 104, 340

\bibitem[{{Landsman}(1993)}]{IDLastro}
{Landsman}, W.~B. 1993, in Astronomical Society of the Pacific Conference
  Series, Vol.~52, Astronomical Data Analysis Software and Systems II, ed.
  {R.~J.~Hanisch, R.~J.~V.~Brissenden, \& J.~Barnes}, 246

\bibitem[{{Law} {et~al.}(2010){Law}, {Dekany}, {Rahmer}, {Hale}, {Smith},
  {Quimby}, {Ofek}, {Kasliwal}, {Zolkower}, {Velur}, {Henning}, {Bui},
  {McKenna}, {Nugent}, {Jacobsen}, {Walters}, {Bloom}, {Surace}, {Grillmair},
  {Laher}, {Mattingly}, \& {Kulkarni}}]{Law2010}
{Law}, N.~M., {Dekany}, R.~G., {Rahmer}, G., {Hale}, D., {Smith}, R., {Quimby},
  R., {Ofek}, E.~O., {Kasliwal}, M., {Zolkower}, J., {Velur}, V., {Henning},
  J., {Bui}, K., {McKenna}, D., {Nugent}, P., {Jacobsen}, J., {Walters}, R.,
  {Bloom}, J., {Surace}, J., {Grillmair}, C., {Laher}, R., {Mattingly}, S., \&
  {Kulkarni}, S. 2010, \procspie, 7735, 7735M

\bibitem[{{Law} {et~al.}(2009){Law}, {Kulkarni}, {Dekany}, {Ofek}, {Quimby},
  {Nugent}, {Surace}, {Grillmair}, {Bloom}, {Kasliwal}, {Bildsten}, {Brown},
  {Cenko}, {Ciardi}, {Croner}, {Djorgovski}, {van Eyken}, {Filippenko}, {Fox},
  {Gal-Yam}, {Hale}, {Hamam}, {Helou}, {Henning}, {Howell}, {Jacobsen},
  {Laher}, {Mattingly}, {McKenna}, {Pickles}, {Poznanski}, {Rahmer}, {Rau},
  {Rosing}, {Shara}, {Smith}, {Starr}, {Sullivan}, {Velur}, {Walters}, \&
  {Zolkower}}]{PTFtechnical}
{Law}, N.~M., {Kulkarni}, S.~R., {Dekany}, R.~G., {Ofek}, E.~O., {Quimby},
  R.~M., {Nugent}, P.~E., {Surace}, J., {Grillmair}, C.~C., {Bloom}, J.~S.,
  {Kasliwal}, M.~M., {Bildsten}, L., {Brown}, T., {Cenko}, S.~B., {Ciardi}, D.,
  {Croner}, E., {Djorgovski}, S.~G., {van Eyken}, J., {Filippenko}, A.~V.,
  {Fox}, D.~B., {Gal-Yam}, A., {Hale}, D., {Hamam}, N., {Helou}, G., {Henning},
  J., {Howell}, D.~A., {Jacobsen}, J., {Laher}, R., {Mattingly}, S., {McKenna},
  D., {Pickles}, A., {Poznanski}, D., {Rahmer}, G., {Rau}, A., {Rosing}, W.,
  {Shara}, M., {Smith}, R., {Starr}, D., {Sullivan}, M., {Velur}, V.,
  {Walters}, R., \& {Zolkower}, J. 2009, \pasp, 121, 1395

\bibitem[{{Mathieu} {et~al.}(2007){Mathieu}, {Baraffe}, {Simon}, {Stassun}, \&
  {White}}]{MathieuPPV}
{Mathieu}, R.~D., {Baraffe}, I., {Simon}, M., {Stassun}, K.~G., \& {White}, R.
  2007, {Protostars and Planets V} (Tucson, AZ: Univ. Arizona Press), 411--425

\bibitem[{{McGehee}(2006)}]{McGehee2006}
{McGehee}, P.~M. 2006, \aj, 131, 2959

\bibitem[{{Miller} {et~al.}(2008){Miller}, {Irwin}, {Aigrain}, {Hodgkin}, \&
  {Hebb}}]{Miller2008}
{Miller}, A.~A., {Irwin}, J., {Aigrain}, S., {Hodgkin}, S., \& {Hebb}, L. 2008,
  \mnras, 387, 349

\bibitem[{{Monet} {et~al.}(2003){Monet}, {Levine}, {Canzian}, {Ables}, {Bird},
  {Dahn}, {Guetter}, {Harris}, {Henden}, {Leggett}, {Levison}, {Luginbuhl},
  {Martini}, {Monet}, {Munn}, {Pier}, {Rhodes}, {Riepe}, {Sell}, {Stone},
  {Vrba}, {Walker}, {Westerhout}, {Brucato}, {Reid}, {Schoening}, {Hartley},
  {Read}, \& {Tritton}}]{USNOB}
{Monet}, D.~G., {Levine}, S.~E., {Canzian}, B., {Ables}, H.~D., {Bird}, A.~R.,
  {Dahn}, C.~C., {Guetter}, H.~H., {Harris}, H.~C., {Henden}, A.~A., {Leggett},
  S.~K., {Levison}, H.~F., {Luginbuhl}, C.~B., {Martini}, J., {Monet},
  A.~K.~B., {Munn}, J.~A., {Pier}, J.~R., {Rhodes}, A.~R., {Riepe}, B., {Sell},
  S., {Stone}, R.~C., {Vrba}, F.~J., {Walker}, R.~L., {Westerhout}, G.,
  {Brucato}, R.~J., {Reid}, I.~N., {Schoening}, W., {Hartley}, M., {Read},
  M.~A., \& {Tritton}, S.~B. 2003, \aj, 125, 984

\bibitem[{{Morales-Calder{\'o}n} {et~al.}(2011){Morales-Calder{\'o}n},
  {Stauffer}, {Hillenbrand}, {Gutermuth}, {Song}, {Rebull}, {Plavchan},
  {Carpenter}, {Whitney}, {Covey}, {Alves de Oliveira}, {Winston},
  {McCaughrean}, {Bouvier}, {Guieu}, {Vrba}, {Holtzman}, {Marchis}, {Hora},
  {Wasserman}, {Terebey}, {Megeath}, {Guinan}, {Forbrich}, {Hu{\'e}lamo},
  {Riviere-Marichalar}, {Barrado}, {Stapelfeldt}, {Hern{\'a}ndez}, {Allen},
  {Ardila}, {Bayo}, {Favata}, {James}, {Werner}, \& {Wood}}]{Morales2011}
{Morales-Calder{\'o}n}, M., {Stauffer}, J.~R., {Hillenbrand}, L.~A.,
  {Gutermuth}, R., {Song}, I., {Rebull}, L.~M., {Plavchan}, P., {Carpenter},
  J.~M., {Whitney}, B.~A., {Covey}, K., {Alves de Oliveira}, C., {Winston}, E.,
  {McCaughrean}, M.~J., {Bouvier}, J., {Guieu}, S., {Vrba}, F.~J., {Holtzman},
  J., {Marchis}, F., {Hora}, J.~L., {Wasserman}, L.~H., {Terebey}, S.,
  {Megeath}, T., {Guinan}, E., {Forbrich}, J., {Hu{\'e}lamo}, N.,
  {Riviere-Marichalar}, P., {Barrado}, D., {Stapelfeldt}, K., {Hern{\'a}ndez},
  J., {Allen}, L.~E., {Ardila}, D.~R., {Bayo}, A., {Favata}, F., {James}, D.,
  {Werner}, M., \& {Wood}, K. 2011, \apj, 733, 50

\bibitem[{{Plavchan} {et~al.}(2008){Plavchan}, {Jura}, {Kirkpatrick}, {Cutri},
  \& {Gallagher}}]{PlavchanPeriodogram}
{Plavchan}, P., {Jura}, M., {Kirkpatrick}, J.~D., {Cutri}, R.~M., \&
  {Gallagher}, S.~C. 2008, \apjs, 175, 191

\bibitem[{{Pribulla} \& {Rucinski}(2006)}]{Pribulla2006}
{Pribulla}, T. \& {Rucinski}, S.~M. 2006, \aj, 131, 2986

\bibitem[{{Rau} {et~al.}(2009){Rau}, {Kulkarni}, {Law}, {Bloom}, {Ciardi},
  {Djorgovski}, {Fox}, {Gal-Yam}, {Grillmair}, {Kasliwal}, {Nugent}, {Ofek},
  {Quimby}, {Reach}, {Shara}, {Bildsten}, {Cenko}, {Drake}, {Filippenko},
  {Helfand}, {Helou}, {Howell}, {Poznanski}, \& {Sullivan}}]{PTFsurvey}
{Rau}, A., {Kulkarni}, S.~R., {Law}, N.~M., {Bloom}, J.~S., {Ciardi}, D.,
  {Djorgovski}, G.~S., {Fox}, D.~B., {Gal-Yam}, A., {Grillmair}, C.~C.,
  {Kasliwal}, M.~M., {Nugent}, P.~E., {Ofek}, E.~O., {Quimby}, R.~M., {Reach},
  W.~T., {Shara}, M., {Bildsten}, L., {Cenko}, S.~B., {Drake}, A.~J.,
  {Filippenko}, A.~V., {Helfand}, D.~J., {Helou}, G., {Howell}, D.~A.,
  {Poznanski}, D., \& {Sullivan}, M. 2009, \pasp, 121, 1334

\bibitem[{{Rieke} {et~al.}(2004){Rieke}, {Young}, {Engelbracht}, {Kelly},
  {Low}, {Haller}, {Beeman}, {Gordon}, {Stansberry}, {Misselt}, {Cadien},
  {Morrison}, {Rivlis}, {Latter}, {Noriega-Crespo}, {Padgett}, {Stapelfeldt},
  {Hines}, {Egami}, {Muzerolle}, {Alonso-Herrero}, {Blaylock}, {Dole}, {Hinz},
  {Le Floc'h}, {Papovich}, {P{\'e}rez-Gonz{\'a}lez}, {Smith}, {Su}, {Bennett},
  {Frayer}, {Henderson}, {Lu}, {Masci}, {Pesenson}, {Rebull}, {Rho}, {Keene},
  {Stolovy}, {Wachter}, {Wheaton}, {Werner}, \& {Richards}}]{Rieke2004}
{Rieke}, G.~H., {Young}, E.~T., {Engelbracht}, C.~W., {Kelly}, D.~M., {Low},
  F.~J., {Haller}, E.~E., {Beeman}, J.~W., {Gordon}, K.~D., {Stansberry},
  J.~A., {Misselt}, K.~A., {Cadien}, J., {Morrison}, J.~E., {Rivlis}, G.,
  {Latter}, W.~B., {Noriega-Crespo}, A., {Padgett}, D.~L., {Stapelfeldt},
  K.~R., {Hines}, D.~C., {Egami}, E., {Muzerolle}, J., {Alonso-Herrero}, A.,
  {Blaylock}, M., {Dole}, H., {Hinz}, J.~L., {Le Floc'h}, E., {Papovich}, C.,
  {P{\'e}rez-Gonz{\'a}lez}, P.~G., {Smith}, P.~S., {Su}, K.~Y.~L., {Bennett},
  L., {Frayer}, D.~T., {Henderson}, D., {Lu}, N., {Masci}, F., {Pesenson}, M.,
  {Rebull}, L., {Rho}, J., {Keene}, J., {Stolovy}, S., {Wachter}, S.,
  {Wheaton}, W., {Werner}, M.~W., \& {Richards}, P.~L. 2004, \apjs, 154, 25

\bibitem[{{Robin} {et~al.}(2003){Robin}, {Reyl{\'e}}, {Derri{\`e}re}, \&
  {Picaud}}]{BesanconModel}
{Robin}, A.~C., {Reyl{\'e}}, C., {Derri{\`e}re}, S., \& {Picaud}, S. 2003,
  \aap, 409, 523

\bibitem[{{Roeser} {et~al.}(2010){Roeser}, {Demleitner}, \&
  {Schilbach}}]{PPMXL}
{Roeser}, S., {Demleitner}, M., \& {Schilbach}, E. 2010, \aj, 139, 2440

\bibitem[{{Roxburgh}(1966)}]{Roxburgh1966}
{Roxburgh}, I.~W. 1966, \apj, 143, 111

\bibitem[{{Rucinski}(2004)}]{Rucinski2004WUmaDistances}
{Rucinski}, S.~M. 2004, \nar, 48, 703

\bibitem[{{Rucinski}(2007)}]{Rucinski2007ShortPeriodWUMas}
---. 2007, \mnras, 382, 393

\bibitem[{{Rucinski} {et~al.}(2007){Rucinski}, {Pribulla}, \& {van
  Kerkwijk}}]{Rucinski2007}
{Rucinski}, S.~M., {Pribulla}, T., \& {van Kerkwijk}, M.~H. 2007, \aj, 134,
  2353

\bibitem[{{Schlegel} {et~al.}(1998){Schlegel}, {Finkbeiner}, \&
  {Davis}}]{SchlegelDustMaps}
{Schlegel}, D.~J., {Finkbeiner}, D.~P., \& {Davis}, M. 1998, \apj, 500, 525

\bibitem[{{Siess} {et~al.}(2000){Siess}, {Dufour}, \&
  {Forestini}}]{SiessModels}
{Siess}, L., {Dufour}, E., \& {Forestini}, M. 2000, \aap, 358, 593

\bibitem[{{Skrutskie} {et~al.}(2006){Skrutskie}, {Cutri}, {Stiening},
  {Weinberg}, {Schneider}, {Carpenter}, {Beichman}, {Capps}, {Chester},
  {Elias}, {Huchra}, {Liebert}, {Lonsdale}, {Monet}, {Price}, {Seitzer},
  {Jarrett}, {Kirkpatrick}, {Gizis}, {Howard}, {Evans}, {Fowler}, {Fullmer},
  {Hurt}, {Light}, {Kopan}, {Marsh}, {McCallon}, {Tam}, {Van Dyk}, \&
  {Wheelock}}]{2MASS}
{Skrutskie}, M.~F., {Cutri}, R.~M., {Stiening}, R., {Weinberg}, M.~D.,
  {Schneider}, S., {Carpenter}, J.~M., {Beichman}, C., {Capps}, R., {Chester},
  T., {Elias}, J., {Huchra}, J., {Liebert}, J., {Lonsdale}, C., {Monet}, D.~G.,
  {Price}, S., {Seitzer}, P., {Jarrett}, T., {Kirkpatrick}, J.~D., {Gizis},
  J.~E., {Howard}, E., {Evans}, T., {Fowler}, J., {Fullmer}, L., {Hurt}, R.,
  {Light}, R., {Kopan}, E.~L., {Marsh}, K.~A., {McCallon}, H.~L., {Tam}, R.,
  {Van Dyk}, S., \& {Wheelock}, S. 2006, \aj, 131, 1163

\bibitem[{{Stassun} {et~al.}(2008){Stassun}, {Mathieu}, {Cargile}, {Aarnio},
  {Stempels}, \& {Geller}}]{Stassun2008}
{Stassun}, K.~G., {Mathieu}, R.~D., {Cargile}, P.~A., {Aarnio}, A.~N.,
  {Stempels}, E., \& {Geller}, A. 2008, \nat, 453, 1079

\bibitem[{{Stassun} {et~al.}(2006){Stassun}, {Mathieu}, \&
  {Valenti}}]{Stassun2006}
{Stassun}, K.~G., {Mathieu}, R.~D., \& {Valenti}, J.~A. 2006, \nat, 440, 311

\bibitem[{{Stassun} {et~al.}(2007){Stassun}, {Mathieu}, \&
  {Valenti}}]{Stassun2007}
---. 2007, \apj, 664, 1154

\bibitem[{{Stassun} {et~al.}(2004){Stassun}, {Mathieu}, {Vaz}, {Stroud}, \&
  {Vrba}}]{Stassun2004}
{Stassun}, K.~G., {Mathieu}, R.~D., {Vaz}, L.~P.~R., {Stroud}, N., \& {Vrba},
  F.~J. 2004, \apjs, 151, 357

\bibitem[{{Stauffer} {et~al.}(2007){Stauffer}, {Hartmann}, {Fazio}, {Allen},
  {Patten}, {Lowrance}, {Hurt}, {Rebull}, {Cutri}, {Ramirez}, {Young}, {Rieke},
  {Gorlova}, {Muzerolle}, {Slesnick}, \& {Skrutskie}}]{Stauffer2007}
{Stauffer}, J.~R., {Hartmann}, L.~W., {Fazio}, G.~G., {Allen}, L.~E., {Patten},
  B.~M., {Lowrance}, P.~J., {Hurt}, R.~L., {Rebull}, L.~M., {Cutri}, R.~M.,
  {Ramirez}, S.~V., {Young}, E.~T., {Rieke}, G.~H., {Gorlova}, N.~I.,
  {Muzerolle}, J.~C., {Slesnick}, C.~L., \& {Skrutskie}, M.~F. 2007, \apjs,
  172, 663

\bibitem[{{St{\c e}pie{\'n}}(1995)}]{Stepien1995}
{St{\c e}pie{\'n}}, K. 1995, \mnras, 274, 1019

\bibitem[{{St{\c e}pie{\'n}}(2006)}]{Stepien2006EvStat}
---. 2006, \actaa, 56, 199

\bibitem[{{Stempels} {et~al.}(2008){Stempels}, {Hebb}, {Stassun}, {Holtzman},
  {Dunstone}, {Glowienka}, \& {Frandsen}}]{Stempels2008}
{Stempels}, H.~C., {Hebb}, L., {Stassun}, K.~G., {Holtzman}, J., {Dunstone},
  N., {Glowienka}, L., \& {Frandsen}, S. 2008, \aap, 481, 747

\bibitem[{{Stetson}(1987)}]{DAOPHOT}
{Stetson}, P.~B. 1987, \pasp, 99, 191

\bibitem[{{Tody}(1986)}]{IRAF1}
{Tody}, D. 1986, in Society of Photo-Optical Instrumentation Engineers (SPIE)
  Conference Series, Vol. 627, Society of Photo-Optical Instrumentation
  Engineers (SPIE) Conference Series, ed. {D.~L.~Crawford}, 733

\bibitem[{{Tody}(1993)}]{IRAF2}
{Tody}, D. 1993, in Astronomical Society of the Pacific Conference Series,
  Vol.~52, Astronomical Data Analysis Software and Systems II, ed.
  {R.~J.~Hanisch, R.~J.~V.~Brissenden, \& J.~Barnes}, 173

\bibitem[{{Tokovinin} {et~al.}(2006){Tokovinin}, {Thomas}, {Sterzik}, \&
  {Udry}}]{Tokovinin2006}
{Tokovinin}, A., {Thomas}, S., {Sterzik}, M., \& {Udry}, S. 2006, \aap, 450,
  681

\bibitem[{{Werner} {et~al.}(2004){Werner}, {Roellig}, {Low}, {Rieke}, {Rieke},
  {Hoffmann}, {Young}, {Houck}, {Brandl}, {Fazio}, {Hora}, {Gehrz}, {Helou},
  {Soifer}, {Stauffer}, {Keene}, {Eisenhardt}, {Gallagher}, {Gautier}, {Irace},
  {Lawrence}, {Simmons}, {Van Cleve}, {Jura}, {Wright}, \&
  {Cruikshank}}]{Werner2004}
{Werner}, M.~W., {Roellig}, T.~L., {Low}, F.~J., {Rieke}, G.~H., {Rieke}, M.,
  {Hoffmann}, W.~F., {Young}, E., {Houck}, J.~R., {Brandl}, B., {Fazio}, G.~G.,
  {Hora}, J.~L., {Gehrz}, R.~D., {Helou}, G., {Soifer}, B.~T., {Stauffer}, J.,
  {Keene}, J., {Eisenhardt}, P., {Gallagher}, D., {Gautier}, T.~N., {Irace},
  W., {Lawrence}, C.~R., {Simmons}, L., {Van Cleve}, J.~E., {Jura}, M.,
  {Wright}, E.~L., \& {Cruikshank}, D.~P. 2004, \apjs, 154, 1

\end{thebibliography}
